\documentclass{article}
\usepackage{PRIMEarxiv}
\usepackage{fancyhdr}
\usepackage{graphicx}
\usepackage{amsmath,amssymb}
\usepackage{xcolor}
\usepackage{lscape}
\usepackage{authblk}

\usepackage{setspace}
\usepackage{enumitem}
\usepackage[backend=biber, minnames = 3, maxnames = 6, bibstyle=ije, citestyle=numeric-comp, sorting=none]{biblatex}
\usepackage{hyperref}
\usepackage[noabbrev,capitalise,nameinlink]{cleveref}
\usepackage{booktabs}
\usepackage[ngerman,english]{babel}
\usepackage{csquotes}
\usepackage[font=small]{caption}
\usepackage{longtable}
\usepackage[font=sf]{floatrow}
\definecolor{orangered}{rgb}{1,0.27,0}
\definecolor{violetred4}{rgb}{0.55,0.13,0.32}

\makeatletter
\newcommand{\pfnsymbol}[1]{%
  \textsuperscript{\@fnsymbol{#1}}%
}
\makeatother

\makeatletter

\title{Can synthetic data reproduce real-world findings in epidemiology?\\
A replication study using adversarial random forests}

\author[1,2,3]{Jan Kapar}
\author[1]{Kathrin Günther}
\author[3]{Lori Ann Vallis}
\author[4]{Klaus Berger}
\author[5,6]{Nadine Binder}
\author[7]{Hermann Brenner}
\author[8]{Stefanie Castell}
\author[9]{Beate Fischer}
\author[10]{Volker Harth}
\author[11]{Bernd Holleczek}
\author[1]{Timm Intemann}
\author[12]{Till Ittermann}
\author[4]{André Karch}
\author[13,14,15]{Thomas Keil}
\author[13]{Lilian Krist}
\author[8]{Berit Lange}
\author[9]{Michael F. Leitzmann}
\author[16]{Katharina Nimptsch}
\author[10]{Nadia Obi}
\author[1,2]{Iris Pigeot}
\author[16,17,18]{Tobias Pischon}
\author[19,20]{Tamara Schikowski}
\author[21]{Börge Schmidt}
\author[12]{Carsten Oliver Schmidt}
\author[22,23,9]{Anja M. Sedlmair}
\author[1]{Justine Tanoey}
\author[24]{Harm Wienbergen}
\author[25]{Andreas Wienke}
\author[19]{Claudia Wigmann}
\author[1,2,26]{Marvin N. Wright}

\affil[1]{Leibniz Institute for Prevention Research and Epidemiology—BIPS, Bremen, Germany}
\affil[2]{Faculty of Mathematics and Computer Science, University of Bremen, Bremen, Germany}
\affil[3]{Department of Human Health Sciences, University of Guelph, Guelph, Ontario, Canada}
\affil[4]{Institute of Epidemiology and Social Medicine, University of Münster, Münster, Germany}
\affil[5]{Institute of General Practice/Family Medicine, Faculty of Medicine and Medical Center, University of Freiburg, Freiburg Germany}
\affil[6]{Freiburg Center for Data Analysis, Modeling and AI, University of Freiburg, Freiburg, Germany}
\affil[7]{Cancer Prevention Graduate School, German Cancer Research Center (DKFZ), Heidelberg, Germany}
\affil[8]{Department of Epidemiology, Helmholtz Centre for Infection Research (HZI), Brunswick, Germany}
\affil[9]{Department of Epidemiology and Preventive Medicine, University of Regensburg, Regensburg, Germany}
\affil[10]{Institute for Occupational and Maritime Medicine (ZfAM), University Medical Center Hamburg-Eppendorf, Hamburg, Germany}
\affil[11]{Saarland Cancer Registry, Saarbrücken, Germany}
\affil[12]{Institute for Community Medicine, University Medicine Greifswald, Greifswald, Germany}
\affil[13]{Institute of Social Medicine, Epidemiology and Health Economics, Charité - Universitätsmedizin Berlin, Berlin, Germany}
\affil[14]{Institute of Clinical Epidemiology and Biometry, University of Würzburg, Würzburg, Germany}
\affil[15]{State Institute of Health I, Bavarian Health and Food Safety Authority, Erlangen, Germany}
\affil[16]{Molecular Epidemiology Research Group, Max Delbrück Center for Molecular Medicine in the Helmholtz Association (MDC), Berlin, Germany}
\affil[17]{Biobank Technology Platform, Max Delbrück Center for Molecular Medicine in the Helmholtz Association (MDC), Berlin, Germany}
\affil[18]{Charité - Universitätsmedizin Berlin, corporate member of Freie Universität Berlin and Humboldt-Universität zu Berlin, Berlin, Germany}
\affil[19]{IUF – Leibniz Research Institute for Environmental Medicine, Düsseldorf, Germany}
\affil[20]{Department of Environment and Health, School of Public Health, University of Bielefeld, Bielefeld, Germany}
\affil[21]{Institute for Medical Informatics, Biometry and Epidemiology, University Hospital of Essen, University of Duisburg-Essen, Germany}
\affil[22]{Center for Translational Oncology, University Hospital Regensburg, Regensburg, Germany}
\affil[23]{Bavarian Cancer Research Center (BZKF), Regensburg, Germany}
\affil[24]{Bremen Institute for Heart and Circulation Research (BIHKF), Bremen, Germany}
\affil[25]{Institute of Medical Epidemiology, Biometry, and Informatics, Profile Center Health Sciences, Faculty of Medicine, Martin-Luther-University Halle-Wittenberg, Halle, Germany}
\affil[26]{Department of Public Health, University of Copenhagen, Copenhagen, Denmark}

\affil[ ]{}
\affil[ ]{\texttt{wright@leibniz-bips.de}\vspace{-0.6cm}}

\begin{document}

\maketitle
\begingroup
\renewcommand\thefootnote{}
\footnotetext{%
  \makebox[0pt][l]{%
    \hspace*{-1.8em}%
    \parbox[t]{\textwidth}{%
      Published as an open access article in:
      \textit{International Journal of Epidemiology}, 55(5):dyag164, 2026.
      DOI: \url{https://doi.org/10.1093/ije/dyag164}.
    }%
  }%
}
\endgroup

\begin{abstract}
Synthetic data hold substantial potential to address practical challenges in epidemiology due to restricted data access and privacy concerns. However, many current methods suffer from limited quality, high computational demands, and complexity for non-experts. Furthermore, common evaluation strategies for synthetic data often fail to directly reflect statistical utility and measure privacy risks sufficiently. Against this background, a critical underexplored question is whether synthetic data can reliably reproduce key findings from epidemiological research while preserving privacy.
We propose adversarial random forests (ARF) as an efficient and convenient method for synthesizing tabular epidemiological data. To evaluate its performance, we replicated statistical analyses from six epidemiological publications covering blood pressure, anthropometry, myocardial infarction, accelerometry, loneliness, and diabetes, from the German National Cohort (NAKO Gesundheitsstudie), the Bremen STEMI Registry U45 Study, and the Guelph Family Health Study. We further assessed how dataset dimensionality and variable complexity affect the quality of synthetic data, and contextualized ARF’s performance by comparison with commonly used tabular data synthesizers in terms of utility, privacy, generalization, and runtime.
Across all replicated studies, results on ARF-generated synthetic data consistently aligned with original findings. Even for datasets with relatively low sample size-to-dimensionality ratios, replication outcomes closely matched the original results across descriptive and inferential analyses. Reduced dimensionality and variable complexity further enhanced synthesis quality. ARF demonstrated favourable performance regarding utility, privacy preservation, and generalization relative to other synthesizers and superior computational efficiency.
In summary, ARF reliably generates high-quality synthetic data that replicate diverse epidemiological analyses while offering a competitive privacy–utility trade-off.

\keywords{generative artificial intelligence \and generative machine learning \and adversarial random forests \and synthetic data quality \and statistical utility \and epidemiological study replication \and tabular data}
\end{abstract}

\textbf{Code --- } {\url{https://github.com/bips-hb/ARF_SynthEpiRep}}

\section{Introduction}

Researchers in epidemiology often encounter specific challenges before conducting statistical analyses: health-related data are highly protected by privacy laws, such as Health Insurance Portability and Accountability \cite{centers2003HIPAA} in the United States and General Data Protection Regulation \cite{eu2016GDPR} in the European Union, and access requests can take several months from the initial application to final data access.

Generative artificial intelligence (AI) \cite{foster2022generativeAI}, famously applied to text and image data (e.g. ChatGPT \cite{OpenAI2025ChatGPT}, DALL-E \cite{openAI2025dalle}), could help address these obstacles by producing realistic, privacy-preserving synthetic health data for sharing, previewing, or substituting original data.

Despite successes in medical image synthesis \cite{jeong2022medimageGAN, kazerouni2023medimageDiffusion}, generating and evaluating high-quality tabular health data—rows as individual observations, columns as variables—remains challenging \cite{latner2024generating}. Existing approaches are still developing, no gold-standard method exists, and many new methods are not yet directly applicable in epidemiology, as they often lack support for missing values and are not available in well-maintained packages. With only a few exceptions \cite{kuhnel2024rep1, el2024rep2}, most generative methods are evaluated on standard ML datasets. Benchmarks typically rely on metrics that are difficult to interpret in applied research and rarely or insufficiently assess privacy \cite{xu2019CTGAN, qian2024synthcity, yao2025dcr}, which is essential in epidemiological applications.

We propose adversarial random forests (ARF) \cite{watson2023arf}, a generative AI method specifically designed for tabular data. ARF is available as an \texttt{R} package \cite{wright2024arfR}, easy to use for researchers with limited machine learning experience, computationally efficient, and competitive in performance. We demonstrate its utility in real-world epidemiological settings by replicating analyses from six studies \cite{schikowski2020blutdruckmessung, fischer2020anthropometrisch, wienbergen2022infarction, breau2022cutpoint, berger2021COVID, tanoey2022diabetes} based on the German National Cohort (NAKO) \cite{peters2022NAKO}, the Bremen STEMI Registry U45 Study (BSR-U45) \cite{wienbergen2022infarction}, and the Guelph Family Health Study (GFHS) \cite{haines2018GFHS, breau2022cutpoint} using ARF-generated synthetic data. Moreover, we examine how dataset dimensionality and variable complexity affect the quality and stability of synthetic data to assess how ARF performs across diverse scenarios. In addition, we perform experiments to systematically evaluate privacy preservation, generalization to out-of-sample data, and hyperparameter effects. We further contextualize ARF’s performance by comparing it with commonly used tabular data synthesizers \cite{nowok2016synthpop, koller2009BN, zhang2017PrivBayes, xu2019CTGAN} along these aspects of synthetic data quality as well as in terms of runtime. This comprehensive assessment positions ARF as a valuable approach for generating high-quality synthetic epidemiological data while providing insight into its strengths, limitations, and applicability in real-world research contexts.

\section{Data and methods}

The selected datasets, along with the synthesis and evaluation approach, were chosen with a focus on their applicability to epidemiological research.

\subsection{Data}
We selected the included data sources and original studies with an emphasis on variety regarding study subject, data origin, size and dimensionality, and the complexity of statistical analyses. This way, we cover different challenging scenarios for successful data synthesis, such as small data size, high dimensionality, mixed variable types and irregular distribution shapes.

\subsubsection{Data sources}
The original studies chosen for this replication analysis were based on NAKO, BSR-U45, and GFHS data.
The NAKO is Germany's largest cohort study, following over 200\,000 participants from age 19 to 74 over the long term. It encompasses socioeconomic, demographic, genetic and lifestyle information as well as medical history and examination data in order to investigate the development of major diseases, such as cancer, cardiovascular conditions, diabetes, and mental health disorders \cite{peters2022NAKO}.
The BSR-U45 is a specialized substudy within the Bremen Heart Registry that focuses on patients aged 45 or younger who experienced myocardial infarction, combining retrospective and prospective data on demographic, lifestyle, clinical, and laboratory factors to investigate early-onset MI risk \cite{wienbergen2022infarction}.
The GFHS is a long-term cohort study that involves over 246 families in Wellington County, Ontario, Canada. It aims to identify early life risk factors for obesity and chronic diseases and help families establish healthy routines related to nutrition, physical activity, sleep and screen time \cite{haines2018GFHS}.

\subsubsection{Original studies}

\begin{table}[t]
    \footnotesize
    \centering
    \setlength{\tabcolsep}{5.4pt}
    \begin{tabular}{lllrrrrr}
\toprule
Publication & Data source & Research topic & $n$ & $d$ (num/cat/lvls) & $n_{\text{NA}}$ & $d_{\text{NA}}$ & $\%_{\text{NA}}$ \\
\midrule
Schikowski et al.\@~\cite{schikowski2020blutdruckmessung} & NAKO & Blood pressure & 98\,332 & 7 (5/2/4) & 0 & 0 & 0 \\
Fischer et al.\@~\cite{fischer2020anthropometrisch} & NAKO & Anthropometry & 101\,557 & 12 (10/2/20) & 97\,903 & 9 & 34.43 \\
Wienbergen et al.\@~\cite{wienbergen2022infarction} & NAKO, BSR-U45 & Myocardial infarction & 1713 & 12 (3/9/24) & 176 & 1 & 0.86 \\
Breau et al.\@~\cite{breau2022cutpoint} & GFHS & Accelerometry & 262 & 28 (25/3/34) & 16 & 2 & 0.23 \\
Berger et al.\@~\cite{berger2021COVID} & NAKO & Loneliness/COVID-19 & 113\,527 & 47 (3/44/230) & 0 & 0 & 0 \\
Tanoey et al.\@~\cite{tanoey2022diabetes} & NAKO & Type 1 diabetes & 101\,570 & 19 (7/12/42) & 100\,621 & 6 & 12.66 \\
\bottomrule
\bottomrule
\end{tabular}
    \caption{Overview of publications and data used for synthetic data replications. The values for $n$ partially deviate from the numbers given in the original publications due to participants withdrawing their consent for data usage in the respective cohort studies over time. $n$, number of rows/participants; $d$, dimensionality; num, number of numeric variables; cat, number of categorical variables; lvls, total number of categories (i.e., levels) across all categorical variables; $n_{\text{NA}}$, number of rows containing missing values; $d_{\text{NA}}$, number of variables containing missing values; $\%_{\text{NA}}$, percentage of missing values; NAKO, German National Cohort (NAKO Gesundheitsstudie); BSR-U45, Bremen STEMI Registry U45 Study; GFHS, Guelph Family Health Study}
    \label{tab:dataset_info}

\end{table}

Five of the six selected original studies were based on NAKO data \cite{berger2021COVID, fischer2020anthropometrisch, schikowski2020blutdruckmessung, tanoey2022diabetes, wienbergen2022infarction}, one of which also used BSR-U45 data \cite{wienbergen2022infarction}, and one was based on GFHS data \cite{breau2022cutpoint}. \cref{tab:dataset_info} summarizes all datasets used for these studies, including data source, size, dimensionality, variable types, and missing values.
Schikowski et al.\@~\cite{schikowski2020blutdruckmessung} examined methodological differences in blood pressure measurement, with NAKO conducting two readings compared to three in other German population studies. Age- and sex-specific mean blood pressure values and hypertension frequencies were compared.
Fischer et al.\@~\cite{fischer2020anthropometrisch} presented descriptive analyses of anthropometric measures, comparing overweight and obesity frequencies alongside mean adipose tissue thickness by sex and age.
In a case-control study, Wienbergen et al.~\cite{wienbergen2022infarction} used logistic regression to determine associations of lifestyle and metabolic factors with early-onset myocardial infarction risk.
Breau et al.\@~\cite{breau2022cutpoint} examined the effect of movement intensity cutpoints for accelerometer data of young children on physical activity metrics.
Berger et al.\@~\cite{berger2021COVID} analysed loneliness frequency and its association with depression and anxiety during the first SARS-CoV-2 wave, using linear regression to identify factors influencing loneliness.
In Tanoey et al.\@~\cite{tanoey2022diabetes}, the relationships between birth order, delivery mode, and daycare attendance with the risk of type 1 diabetes were examined and stratified by childhood- and adult-onset type 1 diabetes using Cox regression.

\subsection{Methods}
The generative AI method ARF was used for data synthesis. The statistical analyses of the included studies were replicated using synthetic data, and their results then compared to the original findings to assess the statistical utility of the synthetic data.

\subsubsection{Tree-based generative artificial intelligence for tabular data: adversarial random forests}
Generative AI is concerned with modelling the underlying joint distribution of given data in order to generate realistic synthetic samples \cite{foster2022generativeAI}. Deep learning approaches dominate this discipline, especially for image and text data. Tabular data, however, present unique challenges, such as mixed variable types (e.g. continuous or categorical), a lack of natural positional ordering of variables, and irregular distributions, all of which substantially complicate the learning task \cite{borisov2022deepAItab, davila2025navigating, latner2024generating}. Recent advances of generative deep learning methods for tabular data often come at increased computational cost and demanding hardware requirements, and using these methods is often non-trivial for non-experts \cite{davila2025navigating}.

Tree-based methods \cite{quinlan1986DTs} are a well-suited and less resource-demanding alternative to deep learning for tabular learning tasks \cite{grinsztajn2022tree, qian2024synthcity, drechsler2011empirical}. The tree-based method ARF is competitive to state-of-the-art deep learning models for tabular data synthesis while often operating orders of magnitude faster and requiring no GPU calculations \cite{watson2023arf, qian2024synthcity}. Synthetic data can be created with one line of \texttt{R} code: \texttt{synthetic\_data~<-~rarf(original\_data)}.

ARF utilizes random forests (RFs) \cite{breiman2001random} as its foundation. It iteratively learns data dependencies starting with an RF classification, where the original dataset is distinguished from a constructed version built from original entries with permuted, and thus independent, variables. Hereby, a partitioning of the data manifold is determined in which original and constructed entries are indistinguishable for the RF. This is utilized to assume variable independence for the original data locally and perform variable-wise univariate density estimation within the partitioning units, which are then combined in a global mixture distribution \cite{robbins1948mixture}. ARF generates data by drawing samples from this estimated mixture \cite{watson2023arf}. It further supports joint and conditional density estimation, training and synthesis with missing data, and conditional sampling, which enables applications such as what-if analyses, missing data imputation, and data balancing \cite{dandl2024countarfactuals, golchian2025missARF, burakov2025balancing}.

\subsubsection{Replication and downstream utility evaluation procedure}
The replication and evaluation procedure is split into three main parts: original dataset preparation before synthesis, generating synthetic data with ARF, and performing the original statistical analyses both on original and synthetic data to compare the results.

\paragraph{Original dataset preparation}
Data cleaning, missing data handling, and variable derivations followed the original authors’ protocols where available. To show the effects of dimensionality and complexity induced by derived variables, a dataset with a task-specific variable subset was prepared for selected analyses. We refer to the corresponding synthesis approaches as \emph{general synthesis} and \emph{task-specific synthesis}, respectively: for general synthesis, all variables present in the original dataset were included without derivations; for task-specific synthesis, only the variable subset used for the analysis was included, with derived variables replacing original ones when applicable.

\paragraph{Synthesis} Data synthesis with ARF contains two sources of randomness: model training and data sampling. To account for that, we trained 100 ARF models for each original dataset and sampled 20 times from each model, resulting in a total of 2000 synthetic datasets per original dataset.

\paragraph{Evaluation} For each statistical result of the original studies, we reported the estimate together with a 95\% bootstrap percentile confidence interval (CI) \cite{efron1994bootstrap} based on 2000 bootstrap resamples to quantify the uncertainty of the original analyses.
Across the 2000 synthetic datasets, we reported the median estimate and the corresponding empirical 95\% percentile CI.
To quantify deviations between original and synthetic results across all reported statistics within each analysis table or plot, we additionally reported summary metrics, including the mean absolute standardized difference (MASD) and the mean confidence interval overlap (CIO) \cite{drechsler2022CIO}. Further details on the evaluation procedure and the employed metrics for quantification are provided in \cref{sec:appx_A}.

\subsubsection{Additional evaluation of privacy, generalization, and comparative performance}

While the main evaluation focuses on the reproducibility of published analyses using ARF-generated synthetic data, it does not assess other important aspects of synthetic data quality, such as privacy risk and generalization to unseen data. In practice, privacy preservation and utility typically form a trade-off \cite{stadler2022groundhog}. We therefore conducted additional experiments evaluating these aspects.

To contextualize ARF’s performance, we further compared it with several commonly used tabular data synthesizers: synthpop \cite{nowok2016synthpop}, Bayesian networks \cite{koller2009BN}, PrivBayes \cite{zhang2017PrivBayes}, CTGAN \cite{xu2019CTGAN} and TVAE \cite{xu2019CTGAN}. As the minimum node size in ARF controls the granularity of tree partitions and thereby affects the utility–privacy trade-off, we additionally evaluated performance across different values of this hyperparameter. Further methodological details are provided in \cref{appx:C}.

\section{Results}

We first present general synthesis results to assess the utility of ARF-generated data across datasets and analyses, followed by the evaluation of task-specific synthesis, highlighting the effects of dimensionality and derived variables. The complete set of replication results for all analyses in the original publications including numerical tables for figures is available in the \cref{appx:B}. The section concludes with a summary of the additional evaluation results regarding privacy, generalization, and runtime.

\subsection{Downstream utility (general synthesis)} 

\subsubsection{Replication results for Schikowski et al. \texorpdfstring{\cite{schikowski2020blutdruckmessung}}{[Schikowski et al., 2020]}}

\begin{figure}[t]
    \centering
    \includegraphics[width=\columnwidth]{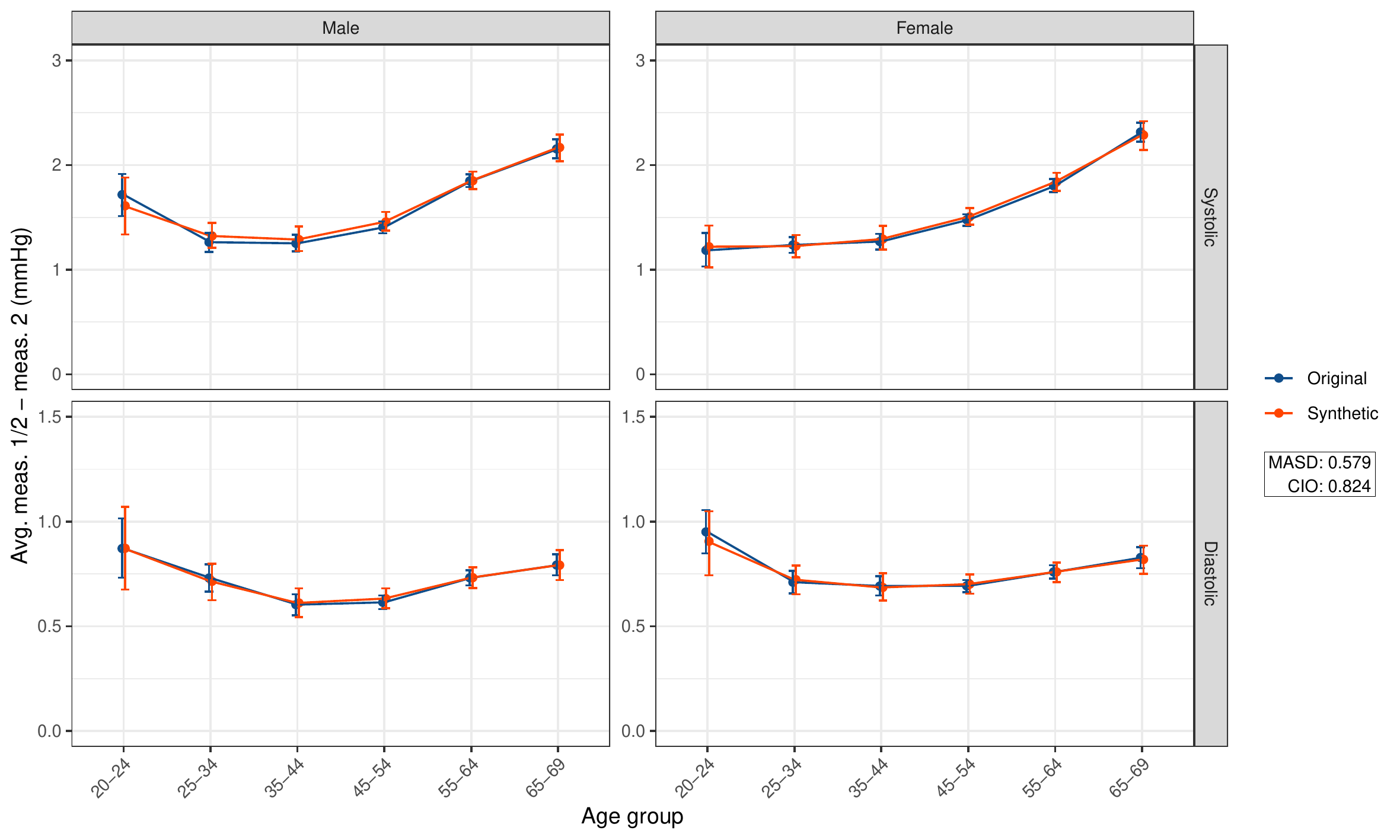}
    \caption{Replication of Figure~2 in Schikowski et al.\@~\cite{schikowski2020blutdruckmessung}: differences of mean blood pressure values (in mmHg) using the mean of first and second measurement or the second measurement only by sex and age group (in years). Percentile-based 95\% bootstrap confidence intervals are reported for original data results. Median and percentile-based 95\% confidence intervals of synthetic data results are reported. avg., average; meas., measurement; MASD, mean absolute standardized difference; CIO, mean confidence interval overlap}
    \label{fig:Schikwoski2}
\end{figure}

The dataset for Schikowski et al.\@~\cite{schikowski2020blutdruckmessung} contained 98\,332 participants, seven variables, and no missing values. \cref{fig:Schikwoski2} (replicating Figure~2 of Schikowski et al.\@~\cite{schikowski2020blutdruckmessung}) shows the interplay of six of these variables, following derivations as in the original study: for systolic and diastolic blood pressure, the difference between using the mean of the first and second measurements versus only the second measurement was calculated. All median values of the synthetic estimates closely resembled those from the original data, while the 95\% CIs were consistently well-aligned (MASD = 0.85; CIO = 0.823). Similar conclusions were drawn for the remaining analyses of this publication (\cref{appx:B1}). Given the strong similarity of results, all conclusions in Schikowski et al.\@~\cite{schikowski2020blutdruckmessung} drawn from original data were equally supported by ARF-generated data.

\subsubsection{Replication results for Fischer et al. \texorpdfstring{\cite{fischer2020anthropometrisch}}{[Fischer et al., 2020]}}

\begin{figure}[t]
    \centering
    \includegraphics[width=\columnwidth]{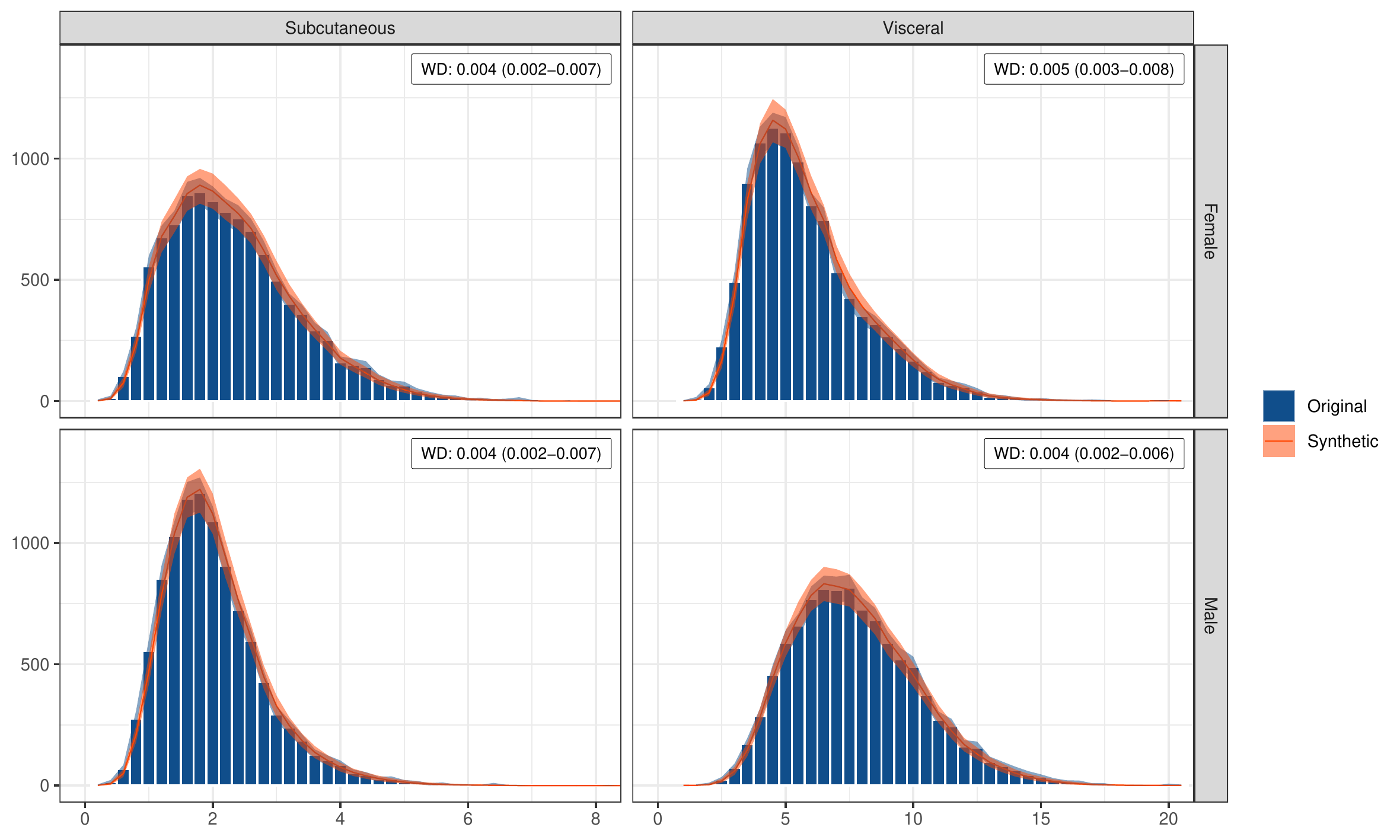}
    \caption{Replication of Figure~5 in Fischer et al.\@~\cite{fischer2020anthropometrisch}: subcutaneous and visceral abdominal adipose tissue thickness by sex. Percentile-based 95\% bootstrap confidence intervals are reported for original data results. Median and percentile-based 95\% confidence intervals of synthetic data results are reported. In order to provide a clear comparability of the distributions despite the variability in the distribution of missing values in the synthetic data, the counts for the synthetic results were rescaled to match the total count of the real ones; WD, Wasserstein distance}
    \label{fig:Fischer5}
\end{figure}

With 101\,557 participants and 12 variables, the data for Fischer et al.\@~\cite{fischer2020anthropometrisch} was of similar size and dimensionality as those from Schikowski et al.\@~\cite{schikowski2020blutdruckmessung} but contained over one third of missing values. The histograms shown in \cref{fig:Fischer5} (replicating Figure~5 in Fischer et al.\@~\cite{fischer2020anthropometrisch}) were calculated using five variables. The sex-specific means of both measurements for subcutaneous and visceral adipose tissue thickness were calculated, with 80\% missingness reflecting NAKO’s scan coverage plan. For comparability between original and synthetic distributions, the synthetic results were rescaled sex-independently to match the total count of non-missing original values. This accounts for variability in the distribution of missing values in the synthetic data. Despite the high missingness rate, the synthetic data histograms recreated the original ones closely, which is reflected by low Wasserstein distance \cite{villani2021WD} values (WD $\leq$ 0.005 for all subgroups). Also the further synthetic results for this study (\cref{appx:B2}) consistently mirrored the original outcomes.

\subsubsection{Replication results for Wienbergen et al. \texorpdfstring{\cite{wienbergen2022infarction}}{[Wienbergen et al., 2022]}}

\begin{figure}[t]
    \centering
    \includegraphics[width=\columnwidth]{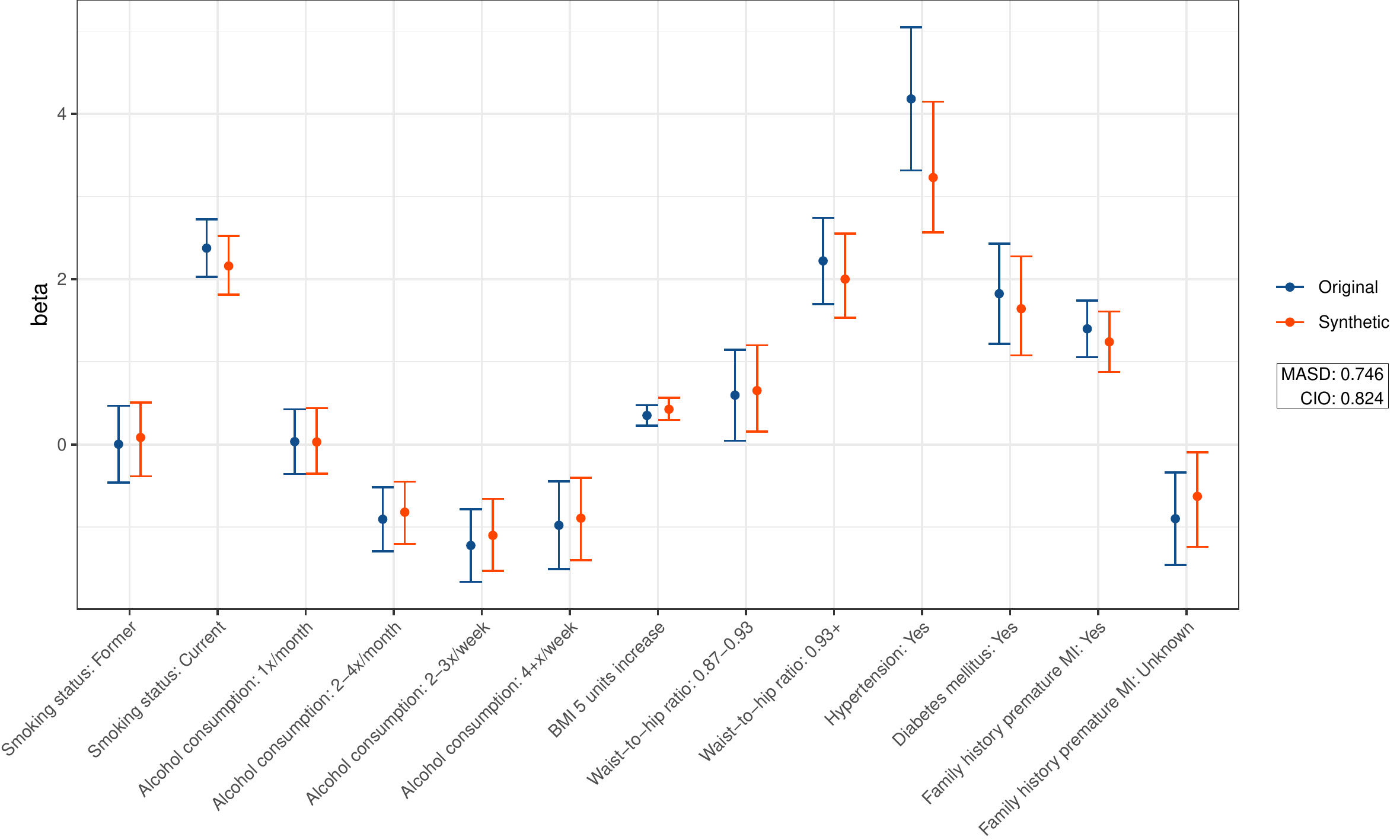}
    \caption{Replication of separate logistic regressions per variable, each adjusted for age, sex, country of birth, and years of school education, Table~2 in Wienbergen et al.\@~\cite{wienbergen2022infarction}:
    associations between lifestyle and metabolic factors, as well as family history of premature MI and risk of early–onset MI. Median beta estimates and percentile-based 95\% confidence intervals computed from the beta coefficient distribution across synthesis repetitions are reported for synthetic data. BMI, body mass index; MI, myocardial infarction; MASD, mean absolute standardized difference; mean CIO, confidence interval overlap}
    \label{fig:Wienbergen_reg}
\end{figure}

Wienbergen et al.\@~\cite{wienbergen2022infarction} used a dataset with 12 variables and 1713 participants. \cref{tab:Wienbergen1,fig:Wienbergen_reg} (replicating Table~1 and the basic adjustment level logistic regressions of Table~2 in Wienbergen et al.\@~\cite{wienbergen2022infarction}) show the performance of ARF-generated data replicating a typical "Table 1" and logistic regression results. The synthetic data medians and 95\% CIs closely resembled the original values in \cref{tab:Wienbergen1} (MASD = 0.664; CIO = 0.863). The largest relative median deviation was observed for the hypertension cases in the original control group of only six participants, but the synthetic data 95\% CI still covered the original value. These findings align with \cref{fig:Wienbergen_reg}, where synthetic data results were close to the original ones (MASD = 0.746; CIO = 0.824), while the hypertension variable posed the biggest difficulty as well. All results for this publication are presented in \cref{appx:B3}.

\begin{table}[p]
    \scriptsize
    \centering
    \begin{tabular}{llllll}
        \hline
Variable & Value cases & Proportion cases & Value controls & Proportion controls \\ 
  \hline
Number of participants (n) & 522 (485-560) &  & 1191 (1153-1228) &  \\ 
   & \textcolor{orangered}{522 (486-560)} & \textcolor{orangered}{} & \textcolor{orangered}{1191 (1153-1227)} & \textcolor{orangered}{} \\ 
  Age (years) (median) & 42 (42-43) &  & 42 (41-42) &  \\ 
   & \textcolor{orangered}{41 (41-42)} & \textcolor{orangered}{} & \textcolor{orangered}{41 (40-41)} & \textcolor{orangered}{} \\ 
  Age (years) (1st quartile) & 39 (38-39) &  & 38 (37-38) &  \\ 
   & \textcolor{orangered}{38 (38-39)} & \textcolor{orangered}{} & \textcolor{orangered}{37 (37-38)} & \textcolor{orangered}{} \\ 
  Age (years) (3rd quartile) & 44 (44-44) &  & 44 (43-44) &  \\ 
   & \textcolor{orangered}{43 (43-44)} & \textcolor{orangered}{} & \textcolor{orangered}{43 (43-43)} & \textcolor{orangered}{} \\ 
  Sex: Male (n) & 434 (399-470) & 0.83 (0.80-0.86) & 927 (888-967) & 0.78 (0.75-0.80) \\ 
   & \textcolor{orangered}{435 (401-469)} & \textcolor{orangered}{0.83 (0.80-0.86)} & \textcolor{orangered}{927 (889-965)} & \textcolor{orangered}{0.78 (0.76-0.80)} \\ 
  Sex: Female (n) & 88 (71-106) & 0.17 (0.14-0.20) & 264 (235-295) & 0.22 (0.20-0.25) \\ 
   & \textcolor{orangered}{87 (71-105)} & \textcolor{orangered}{0.17 (0.14-0.20)} & \textcolor{orangered}{264 (236-292)} & \textcolor{orangered}{0.22 (0.20-0.24)} \\ 
  Country of birth: Other countries (n) & 102 (83-122) & 0.20 (0.16-0.23) & 230 (204-258) & 0.19 (0.17-0.22) \\ 
   & \textcolor{orangered}{104 (84-124)} & \textcolor{orangered}{0.20 (0.16-0.23)} & \textcolor{orangered}{229 (203-257)} & \textcolor{orangered}{0.19 (0.17-0.21)} \\ 
  Country of birth: Germany (n) & 420 (386-455) & 0.80 (0.77-0.84) & 961 (920-1001) & 0.81 (0.78-0.83) \\ 
   & \textcolor{orangered}{419 (384-455)} & \textcolor{orangered}{0.80 (0.77-0.84)} & \textcolor{orangered}{961 (921-1000)} & \textcolor{orangered}{0.81 (0.79-0.83)} \\ 
  School education (years): $<$10 (n) & 38 (27-51) & 0.07 (0.05-0.10) & 76 (60-94) & 0.06 (0.05-0.08) \\ 
   & \textcolor{orangered}{38 (27-52)} & \textcolor{orangered}{0.07 (0.05-0.10)} & \textcolor{orangered}{75 (59-93)} & \textcolor{orangered}{0.06 (0.05-0.08)} \\ 
  School education (years): 10-11 (n) & 382 (348-414) & 0.73 (0.69-0.77) & 241 (212-268) & 0.20 (0.18-0.22) \\ 
   & \textcolor{orangered}{371 (337-408)} & \textcolor{orangered}{0.71 (0.67-0.76)} & \textcolor{orangered}{251 (223-282)} & \textcolor{orangered}{0.21 (0.19-0.24)} \\ 
  School education (years): 12+ (n) & 102 (82-121) & 0.20 (0.16-0.23) & 874 (835-918) & 0.73 (0.71-0.76) \\ 
   & \textcolor{orangered}{113 (92-134)} & \textcolor{orangered}{0.21 (0.18-0.25)} & \textcolor{orangered}{864 (820-902)} & \textcolor{orangered}{0.73 (0.70-0.75)} \\ 
  Smoking status: Never (n) & 53 (40-67) & 0.10 (0.08-0.13) & 580 (543-617) & 0.49 (0.46-0.52) \\ 
   & \textcolor{orangered}{61 (47-78)} & \textcolor{orangered}{0.12 (0.09-0.15)} & \textcolor{orangered}{572 (532-612)} & \textcolor{orangered}{0.48 (0.45-0.51)} \\ 
  Smoking status: Former (n) & 39 (27-51) & 0.07 (0.05-0.10) & 324 (294-357) & 0.27 (0.25-0.30) \\ 
   & \textcolor{orangered}{46 (33-60)} & \textcolor{orangered}{0.09 (0.06-0.11)} & \textcolor{orangered}{317 (285-349)} & \textcolor{orangered}{0.27 (0.24-0.29)} \\ 
  Smoking status: Current (n) & 430 (395-465) & 0.82 (0.79-0.86) & 287 (257-318) & 0.24 (0.22-0.27) \\ 
   & \textcolor{orangered}{415 (382-452)} & \textcolor{orangered}{0.80 (0.76-0.83)} & \textcolor{orangered}{302 (271-334)} & \textcolor{orangered}{0.25 (0.23-0.28)} \\ 
  Alcohol consumption: Never (n) & 112 (92-133) & 0.21 (0.18-0.25) & 121 (101-142) & 0.10 (0.08-0.12) \\ 
   & \textcolor{orangered}{110 (91-130)} & \textcolor{orangered}{0.21 (0.18-0.25)} & \textcolor{orangered}{124 (103-144)} & \textcolor{orangered}{0.10 (0.09-0.12)} \\ 
  Alcohol consumption: Once a month (n) & 176 (153-201) & 0.34 (0.30-0.38) & 193 (169-219) & 0.16 (0.14-0.18) \\ 
   & \textcolor{orangered}{171 (147-197)} & \textcolor{orangered}{0.33 (0.29-0.37)} & \textcolor{orangered}{197 (172-224)} & \textcolor{orangered}{0.17 (0.15-0.19)} \\ 
  Alcohol consumption: 2-4 times a month (n) & 130 (108-151) & 0.25 (0.21-0.29) & 441 (405-477) & 0.37 (0.34-0.40) \\ 
   & \textcolor{orangered}{133 (113-155)} & \textcolor{orangered}{0.26 (0.22-0.29)} & \textcolor{orangered}{437 (404-473)} & \textcolor{orangered}{0.37 (0.34-0.39)} \\ 
  Alcohol consumption: 2-3 times a week (n) & 67 (52-84) & 0.13 (0.10-0.16) & 302 (273-332) & 0.25 (0.23-0.28) \\ 
   & \textcolor{orangered}{70 (54-86)} & \textcolor{orangered}{0.13 (0.10-0.16)} & \textcolor{orangered}{299 (269-329)} & \textcolor{orangered}{0.25 (0.23-0.28)} \\ 
  Alcohol consumption: 4+ times a week (n) & 37 (26-49) & 0.07 (0.05-0.09) & 134 (112-156) & 0.11 (0.09-0.13) \\ 
   & \textcolor{orangered}{38 (26-50)} & \textcolor{orangered}{0.07 (0.05-0.09)} & \textcolor{orangered}{133 (112-155)} & \textcolor{orangered}{0.11 (0.09-0.13)} \\ 
  Body mass index (kg/$\text{m}^2$) (median) & 28.40 (27.76-28.88) &  & 25.50 (25.30-25.80) &  \\ 
   & \textcolor{orangered}{28.62 (28.07-29.20)} & \textcolor{orangered}{} & \textcolor{orangered}{25.68 (25.38-25.97)} & \textcolor{orangered}{} \\ 
  Body mass index (kg/$\text{m}^2$) (1st quartile) & 25.10 (24.84-25.51) &  & 23.00 (22.80-23.50) &  \\ 
   & \textcolor{orangered}{25.46 (24.98-25.96)} & \textcolor{orangered}{} & \textcolor{orangered}{23.25 (22.98-23.55)} & \textcolor{orangered}{} \\ 
  Body mass index (kg/$\text{m}^2$) (3rd quartile) & 31.79 (31.23-32.51) &  & 28.35 (28.00-28.70) &  \\ 
   & \textcolor{orangered}{32.17 (31.52-32.86)} & \textcolor{orangered}{} & \textcolor{orangered}{28.63 (28.23-29.03)} & \textcolor{orangered}{} \\ 
  Body mass index: $<$25.0 (n) & 123 (103-144) & 0.24 (0.20-0.27) & 514 (475-549) & 0.43 (0.40-0.46) \\ 
   & \textcolor{orangered}{111 (92-133)} & \textcolor{orangered}{0.21 (0.18-0.25)} & \textcolor{orangered}{511 (471-549)} & \textcolor{orangered}{0.43 (0.40-0.46)} \\ 
  Body mass index: 25.0-29.9 (n) & 208 (180-233) & 0.40 (0.36-0.44) & 482 (447-517) & 0.40 (0.38-0.43) \\ 
   & \textcolor{orangered}{205 (178-234)} & \textcolor{orangered}{0.39 (0.35-0.44)} & \textcolor{orangered}{472 (432-511)} & \textcolor{orangered}{0.40 (0.37-0.43)} \\ 
  Body mass index: 30.0+ (n) & 191 (165-217) & 0.37 (0.33-0.41) & 195 (170-220) & 0.16 (0.14-0.18) \\ 
   & \textcolor{orangered}{205 (178-234)} & \textcolor{orangered}{0.39 (0.35-0.44)} & \textcolor{orangered}{209 (180-236)} & \textcolor{orangered}{0.18 (0.15-0.20)} \\ 
  Waist-to-hip ratio (median) & 0.98 (0.97-0.99) &  & 0.90 (0.89-0.90) &  \\ 
   & \textcolor{orangered}{0.98 (0.97-0.99)} & \textcolor{orangered}{} & \textcolor{orangered}{0.90 (0.89-0.90)} & \textcolor{orangered}{} \\ 
  Waist-to-hip ratio (1st quartile) & 0.93 (0.91-0.95) &  & 0.85 (0.84-0.86) &  \\ 
   & \textcolor{orangered}{0.92 (0.91-0.94)} & \textcolor{orangered}{} & \textcolor{orangered}{0.85 (0.84-0.85)} & \textcolor{orangered}{} \\ 
  Waist-to-hip ratio (3rd quartile) & 1.03 (1.02-1.04) &  & 0.95 (0.94-0.96) &  \\ 
   & \textcolor{orangered}{1.03 (1.02-1.04)} & \textcolor{orangered}{} & \textcolor{orangered}{0.95 (0.94-0.96)} & \textcolor{orangered}{} \\ 
  Hypertension: No (n) & 391 (358-425) & 0.75 (0.71-0.79) & 1185 (1147-1223) & 0.99 (0.99-1.00) \\ 
   & \textcolor{orangered}{398 (365-433)} & \textcolor{orangered}{0.76 (0.73-0.80)} & \textcolor{orangered}{1178 (1140-1214)} & \textcolor{orangered}{0.99 (0.98-0.99)} \\ 
  Hypertension: Yes (n) & 131 (111-153) & 0.25 (0.21-0.29) & 6 (2-11) & 0.01 (0.00-0.01) \\ 
   & \textcolor{orangered}{124 (103-146)} & \textcolor{orangered}{0.24 (0.20-0.27)} & \textcolor{orangered}{13 (6-21)} & \textcolor{orangered}{0.01 (0.01-0.02)} \\ 
  Diabetes mellitus: No (n) & 461 (425-497) & 0.88 (0.85-0.91) & 1171 (1133-1209) & 0.98 (0.98-0.99) \\ 
   & \textcolor{orangered}{464 (429-500)} & \textcolor{orangered}{0.89 (0.86-0.92)} & \textcolor{orangered}{1168 (1131-1204)} & \textcolor{orangered}{0.98 (0.97-0.99)} \\ 
  Diabetes mellitus: Yes (n) & 61 (46-77) & 0.12 (0.09-0.15) & 20 (12-29) & 0.02 (0.01-0.02) \\ 
   & \textcolor{orangered}{58 (44-74)} & \textcolor{orangered}{0.11 (0.08-0.14)} & \textcolor{orangered}{22 (13-32)} & \textcolor{orangered}{0.02 (0.01-0.03)} \\ 
  Family history of premature MI: No (n) & 358 (326-391) & 0.69 (0.65-0.72) & 1012 (970-1052) & 0.85 (0.83-0.87) \\ 
   & \textcolor{orangered}{361 (328-393)} & \textcolor{orangered}{0.69 (0.65-0.73)} & \textcolor{orangered}{1008 (968-1048)} & \textcolor{orangered}{0.85 (0.83-0.87)} \\ 
  Family history of premature MI: Yes (n) & 144 (122-166) & 0.28 (0.24-0.32) & 97 (79-116) & 0.08 (0.07-0.10) \\ 
   & \textcolor{orangered}{138 (116-162)} & \textcolor{orangered}{0.27 (0.23-0.30)} & \textcolor{orangered}{103 (83-124)} & \textcolor{orangered}{0.09 (0.07-0.10)} \\ 
  Family history of premature MI: Unknown (n) & 20 (12-29) & 0.04 (0.02-0.06) & 82 (65-100) & 0.07 (0.05-0.08) \\ 
   & \textcolor{orangered}{23 (14-32)} & \textcolor{orangered}{0.04 (0.03-0.06)} & \textcolor{orangered}{79 (63-97)} & \textcolor{orangered}{0.07 (0.05-0.08)} \\ 
   \hline

        \textbf{MASD: $0.664$; CIO: $0.863$}
 \\
        \hline
    \end{tabular}
    \caption{Replication of Table~1 in Wienbergen et al.\@~\cite{wienbergen2022infarction}: Distribution of sociodemographic, lifestyle, and metabolic factors and family history of premature myocardial infarction according to cases and controls. Percentile-based 95\% bootstrap confidence intervals are reported for original data results. Median and percentile-based 95\% confidence intervals of synthetic data results are printed in orange. n, number of participants; MI, myocaridal infarction; MASD, mean absolute standardized difference; CIO, mean confidence interval overlap}
    \label{tab:Wienbergen1}
\end{table}

\subsubsection{Replication results for Breau et al. \texorpdfstring{\cite{breau2022cutpoint}}{[Breau et al., 2022]}}

\begin{figure}[!ht]
    \centering
    \includegraphics[width=\columnwidth]{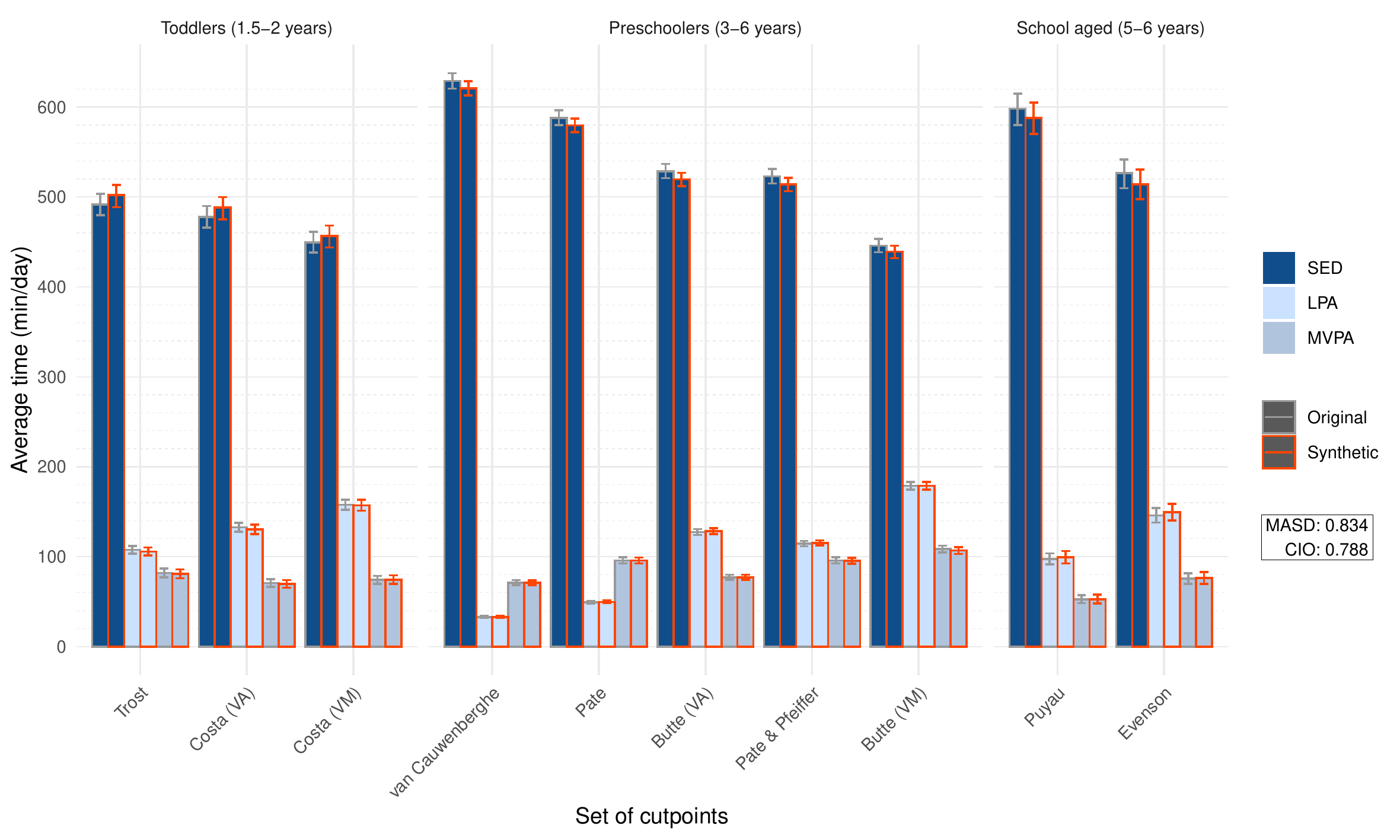}
    \caption{Replication of Figure~2 in Breau et al.\@~\cite{breau2022cutpoint}:
    calculated average valid wear time minutes per day spent in SED, LPA, and MVPA according to age-appropriate ActiGraph cutpoint sets by age group. Percentile-based 95\% bootstrap confidence intervals are reported for original data results. Median and percentile-based 95\% confidence intervals of synthetic data results are reported. SED, sedentary behaviour; LPA, light physical activity; MVPA, moderate to vigorous physical activity; VA, vertical axis; VM, vector magnitude; MASD, mean absolute standardized difference; CIO, mean confidence interval overlap}
    \label{fig:Breau2}
\end{figure}

Containing 262 participants and 28 variables, three of which were categorical with a total of 34 categories, the data characteristics in Breau et al.\@~\cite{breau2022cutpoint} present a challenge for generative machine learning methods. However, the synthetic data results shown in \cref{fig:Breau2} and \cref{fig:Breau_fig1} (replicating Figures 2 and 1 in Breau et al.\@~\cite{breau2022cutpoint}, respectively) matched the original findings well (MASD = 0.834; CIO = 0.788). Similarly, the replication results in \cref{fig:Breau_fig3} (replicating Figure 3) exhibited strong resemblance, allowing all main conclusions to remain valid with synthetic data. While the synthetic data results in \cref{tab:Breau_tab3} (replicating Table~3 in Breau et al.\@~\cite{breau2022cutpoint}) yielded closely matching mean estimates, they showed lower precision for standard deviations and subgroup sizes, likely due to the relatively low participants-to-dimensionality ratio and the presence of derived variables (age group and body mass index). All evaluation results are provided in \cref{appx:B4}.

\subsection{Effect of dimensionality and derived variables (task-specific synthesis)}

\subsubsection{Replication results for Berger et al. \texorpdfstring{\cite{berger2021COVID}}{[Berger et al., 2021]}}

\begin{figure}[t!]
    \centering
    \includegraphics[width=\columnwidth]{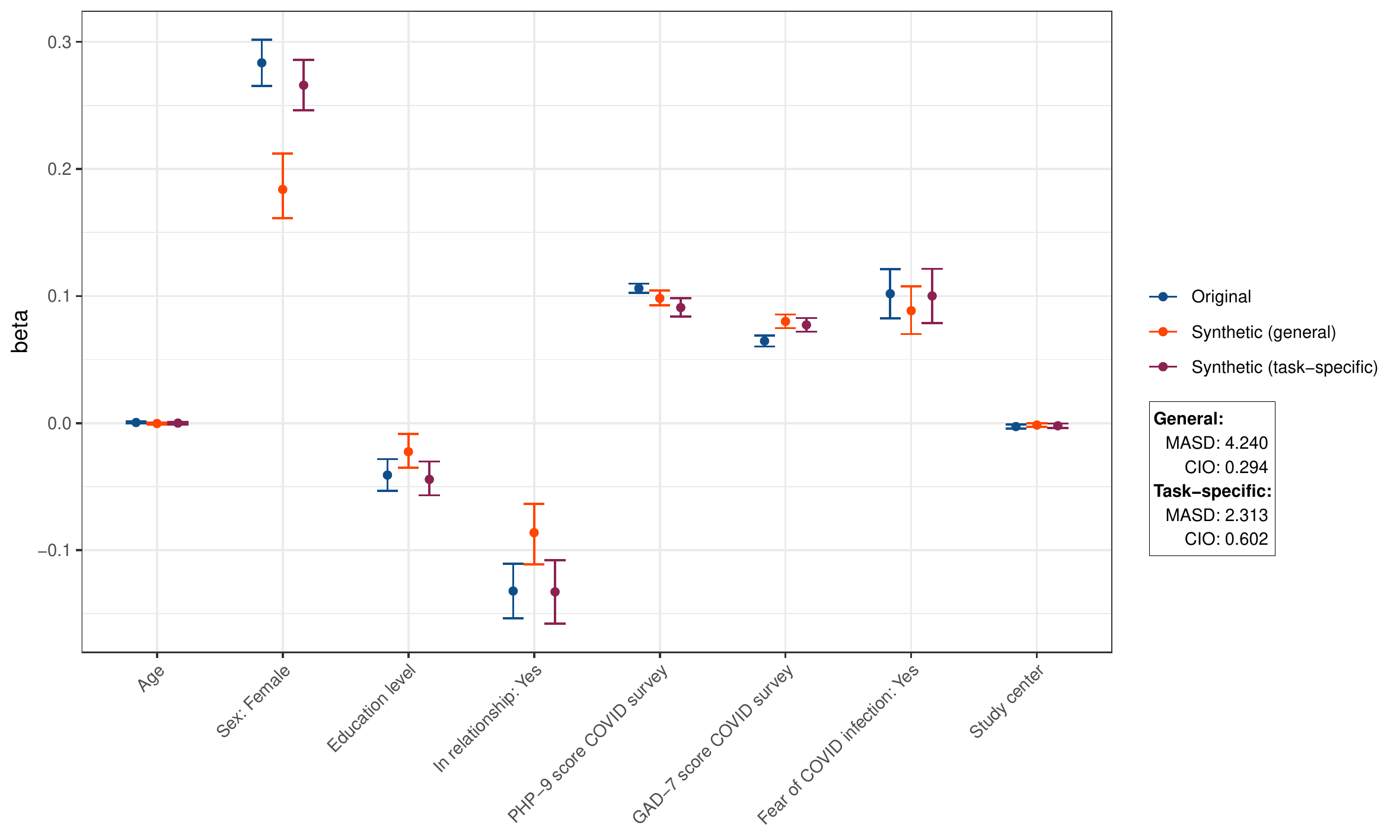}
    \caption{Replication of multivariable linear regression, Table 3 in Berger et al.\@~\cite{berger2021COVID}, with general and task-specific synthesis:
    relationship between perceived loneliness and sociodemographic factors as well as symptoms of depression and anxiety among German National Cohort (NAKO Gesundheitsstudie) participants in May 2020. Median beta estimates and percentile-based 95\% confidence intervals computed from the beta coefficient distribution across synthesis repetitions are reported for synthetic data. PHQ-9, nine-item Patient Health Questionnaire; GAD-7, Generalized Anxiety Disorder seven-item scale; MASD, mean absolute standardized difference; CIO, mean confidence interval overlap}
    \label{fig:berger_reg}
\end{figure}

The data used in Berger et al.\@~\cite{berger2021COVID} contained 113\,527 participants and 47 variables, 44 of which were categorical with a total count of 230 categories. After aggregating PHQ-9 \cite{kroenke2001phq}, GAD-7 \cite{spitzer2006gad}, and loneliness questionnaire items to a sum score each, nine final variables were used in the study's linear regression analysis. \cref{fig:berger_reg} (replicating Table~3 in Berger et al.\@~\cite{berger2021COVID}) shows the resulting $\beta$-estimates for the association between perceived loneliness as the dependent variable and the eight independent variables. The effect of the reduced dimensionality and complexity on the synthesis quality was apparent comparing general and task-specific synthesis results:. While general synthesis results (MASD = 4.24; CIO = 0.294) in many cases overlapped with the original results but struggled especially for the sex and relationship variable, task-specific synthesis results (MASD = 2.313; CIO = 0.602) were consistently closer. Aligning observations were made in \cref{tab:Berger_tab1_taskspec} (replicating Table~1 in Berger et al.\@~\cite{berger2021COVID}), where task-specific synthesis (MASD = 1.792; CIO = 0.747) yielded estimates closer to the original results compared to general synthesis (MASD = 9.161; CIO = 0.438) for the derived age group variable and for the standard deviations of PHQ-9 and GAD-7 sum scores. All results for this publication are reported in \cref{appx:B5}.

\begin{figure}[t!]
    \centering
    \includegraphics[width=\columnwidth]{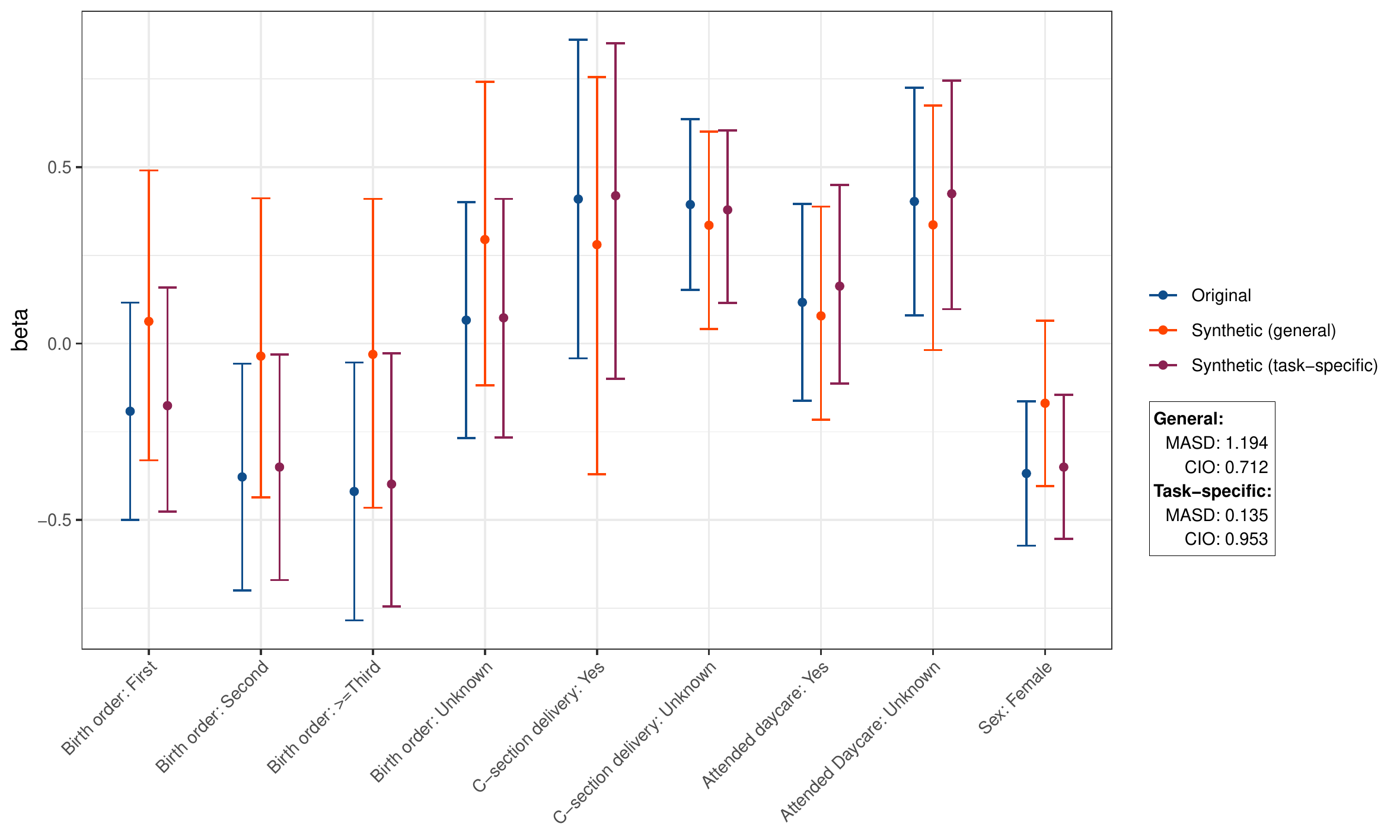}
    \caption{Replication of univariable Cox regressions, Table~2 in Tanoey et al.\@~\cite{tanoey2022diabetes}, with general and task-specific synthesis:
    type 1 diabetes univariable Cox regression estimates. Median beta estimates and percentile-based 95\% confidence intervals computed from the beta coefficient distribution across synthesis repetitions are reported for synthetic data. C-section, caesarean section; MASD, mean absolute standardized difference; CIO, mean confidence interval overlap}
    \label{fig:tanoey_reg}
\end{figure}

\subsubsection{Replication results for Tanoey et al. \texorpdfstring{\cite{tanoey2022diabetes}}{[Tanoey et al., 2022]}}

The Cox regression results in \cref{fig:tanoey_reg} (replicating the univariable Cox regression on type 1 diabetes risk given in Table~2 in Tanoey et al.\@~\cite{tanoey2022diabetes}) support the previous findings. A complex identification algorithm for type 1 diabetes combined five variables and depended on the presence of missing values in some of these. A relatively low share of type 1 diabetes cases (371 out of 101\,570 participants) was identified. The results performed on the full original dataset with 19 variables (12 categorical variables with a total count of 42 categories) were comparable to the original ones (MASD = 1.194; CIO = 0.712) but improved using a task-specific dataset of eight derived variables (MASD = 0.135; CIO = 0.953). The further replication results (\cref{appx:B6}) showed a similar pattern. The effect of variable derivations and the small subset size was most evident in \cref{fig:Tanoey_fig2_taskspec}, where only type 1 diabetes cases were considered (general synthesis: MASD = 2.448, CIO = 0.436; task-specific synthesis: MASD = 0.422, CIO = 0.901).

\subsection{Competitive performance: privacy, generalization, and runtime}

Across datasets, ARF consistently ranked among the synthesizers with the highest utility, with only minor differences for minimum node sizes of 2, 5, and 10. Lower node sizes generally improved utility but, especially in the case of the mainly categorical Berger et al.\@~\cite{berger2021COVID}, increased privacy risks, reflecting the expected utility–privacy trade-off. ARF matched or exceeded all alternatives in the utility–privacy trade-off across four of six datasets, ranking second-best in the others.

These results remained consistent across the generalization analysis, confirming that ARF did not merely reproduce the training distribution.

Runtime comparisons further highlighted the efficiency of ARF. For the largest dataset, Berger et al.\@~\cite{berger2021COVID}, ARF completed training and sampling within 35 seconds, whereas alternative approaches required substantially longer runtimes (six to 36 minutes). Full results are provided \cref{appx:C}.

\section{Discussion}

Summarizing the results, ARF consistently generated synthetic datasets that closely replicated original study findings across a range of data sizes and dimensionalities without tuning. Various descriptive and inferential analyses were repeated using synthetic ARF data with similar outcomes. In most of the cases, median results on synthetic data were close and the percentile-based 95\% CIs well aligned. These findings correspond to existing benchmark studies reporting competitive ARF performance on non-health datasets using different evaluation metrics \cite{watson2023arf, qian2024synthcity}.

We found that a higher participants-to-dimensionality ratio generally led to closer resemblance. Whenever possible, we therefore recommend restricting the data to variables used in subsequent analyses and performing any necessary variable derivation before synthesis. While such task-specific preprocessing is feasible for applications such as data balancing or augmentation, it may not always be possible for privacy-preserving data sharing scenarios where preprocessing must be performed by the data provider.

The extended evaluation further showed that ARF typically achieves a favourable balance between utility and privacy risk compared with commonly used tabular data synthesizers while operating orders of magnitude faster. The minimum node size was found to influence this trade-off: smaller values generally improved utility but increased privacy risks. In practice, slightly larger values (e.g. 5 or 10 instead of the default of 2) may therefore provide a more balanced default in privacy-sensitive applications.

A key strength of this study is its epidemiology-focused validation across multiple real-world datasets and analysis types. Unlike typical machine-learning papers, which rarely perform epidemiology-relevant analyses and often report metrics that are difficult to interpret, we provide application-oriented assessments; moreover, privacy aspects—especially crucial in epidemiological contexts—are frequently omitted, as their evaluation is non-trivial.

Future directions could include developing a differentially private \cite{dwork2014DP} variant of ARF or an extension for extremely high-dimensional data, such as omics datasets.

\section{Conclusion}
This study demonstrated that ARF is a practical and valuable tool for synthesizing tabular epidemiological data, producing synthetic datasets that closely replicate descriptive and inferential findings across diverse datasets and analyses. In the abscence of a universal gold standard for tabular data synthesis, ARF offers a competitive balance between utility and privacy risk compared with commonly used synthesizers, while remaining computationally efficient.

\section*{Ethics approval}
The German National Cohort (NAKO Gesundheitsstudie, NAKO) study is performed with the approval of the relevant local ethics committees, and is in accordance with national law and with the Declaration of Helsinki of 1975 (in the current, revised version). The Bremen STEMI Registry U45 Study (BSR-U45) was approved by the ethics committee of the Ärztekammer Bremen, Germany. All participants gave written informed consent. The study was performed in accordance to the Declaration of Helsinki, with approved protocols conducted by trained investigators subject to independent ethical review and under oversight by a properly convened committee. The Guelph Familiy Health Study (GFHS) was conducted in accordance with the Declaration of Helsinki and approved by the University of Guelph Research Ethics Board (REB 17-07-003).

\section*{Acknowledgements}
This work was conducted with data (Application Numbers NAKO-384, NAKO-812) from the German National Cohort (NAKO Gesundheitsstudie, NAKO) (www.nako.de). The NAKO is funded by the Federal Ministry of Education and Research (BMBF) [project funding reference numbers: 01ER1301A/B/C, 01ER1511D, 01ER1801A/B/C/D, and 01ER2301A/B/C], federal states of Germany and the Helmholtz Association, the participating universities, and the institutes of the Leibniz Association. We thank all participants who took part in the NAKO study and the staff of this research initiative. We incorporated data from the Bremen STEMI Registry U45 Study (BSR-U45). We thank all participants who took part in the BSR-U45 Study, as well as all staff involved. Finally, we acknowledge the Guelph Family Health Study (GFHS) for providing data. The GFHS is funded by the Canada Foundation for Innovation (John R. Evans Leaders Fund) [grant number 36619] and the Canadian Institutes of Health Research [project grant number 376067]. We thank David Ma, director of the GFHS, Jess Haines, co-director of the GFHS, Angela Annis, and Maddy Nixon, who lead the data collection, and all families enrolled.

\section*{Author contributions}
J.K. and M.N.W. created the evaluation concept. J.K., K.G., and L.A.V. selected the original publications for the replications. J.K., K.G., and M.N.W. created and submitted the German National Cohort data request. J.K. and L.A.V. created and submitted the Guelph Family Health Study data request. J.K. performed data preparation, synthesis, and analyses. J.K. and M.N.W. interpreted the results. J.K. wrote the first draft. All authors critically revised the manuscript for important intellectual content and gave final approval and agreed to be accountable for all aspects of the work, ensuring integrity and accuracy.

\section*{Conflict of interest}
None declared.

\section*{Funding}
This work was supported by the U Bremen Research Alliance/AI Center for Health Care, funded by the Federal State of Bremen, and by the European Union’s ERASMUS+ programme. L.A.V. contributed to this work as part of the Guelph Family Health Study, which is funded by the Canada Foundation for Innovation (John R. Evans Leaders Fund) [grant number 36619] and the Canadian Institutes of Health Research [project grant number 376067]. N.B. was supported by the German Research Foundation (Deutsche Forschungsgemeinschaft, DFG) [project number 499552394 – SFB 1597]. T.In. and I.P. contributed to this work as part of the NFDI4Health (National Research Data Infrastructure for Personal Health Data) Consortium, which is publicly funded by the DFG [project number 442326535]. C.W. contributed to this work as part of the IUF – Leibniz Research Institute for Environmental Medicine, which is funded by the Ministry of Culture and Science of North Rhine-Westphalia and the Federal Ministry of Research, Technology and Space of Germany.

\section*{Data availability}
The German National Cohort (NAKO Gesundheitsstudie, NAKO) data are not openly available due to data protection measures. However, scientists can apply for data access following the official usage regulations and upon formal request to the NAKO use and access committee (https://transfer.nako.de). Participant data from the Bremen STEMI Registry U45 Study (BSR-U45) are not publicly available. Due to University of Guelph Research Ethics Board restrictions and participant confidentiality, no Guelph Family Health Study (GFHS) participant data are publicly available. The GFHS welcomes outside collaborators. Interested investigators can contact GFHS investigators to explore this option, which preserves participant confidentiality and meets the requirements of the University of Guelph Research Ethics Board, to protect human subjects.

\section*{Use of artificial intelligence (AI) tools}
The AI tool ChatGPT (model GPT-4o) was solely used to improve readability and review English grammar of the manuscript. No AI tools were used for any other purpose.

\newpage

\printbibliography

\newpage

\appendix
\phantomsection\label{sec:appendix}
\counterwithin{table}{subsection}
\counterwithin{figure}{subsection}

\tiny

\section{Evaluation details}
\label{sec:appx_A}

{\normalsize

In this section, we provide additional details on the evaluation of synthetic data generated by ARF. The evaluation assesses bias in the reproduction of the original analytical results and examines the extent to which statistical uncertainty was preserved, while also quantifying variability introduced by stochastic model training.

We measured the uncertainty in the original analyses using 95\% bootstrap percentile confidence intervals \cite{efron1994bootstrap} for every estimate based on 2000 bootstrap resamples. These were reported alongside the original estimates. For synthetic data, we repeated ARF training 100 times, sampled 20 synthetic datasets of the same size as the original data from each trained ARF model, and reported the median of the estimates across all 2000 synthetic datasets, together with the corresponding 95\% percentile confidence intervals.

Summary measures were reported to capture deviations between the original and synthetic estimates across all reported statistics within a table or plot. Bias was quantified using the mean absolute standardized difference (MASD) \cite{drechsler2022CIO} between the original estimates and the median synthetic estimates:

\[
    \mathrm{MASD} = \frac{1}{n_\text{stat}} \sum_{i=1}^{n_\text{stat}} \frac{\left|\mathrm{median}\left({\smash{\widehat{\text{stat}}}\vphantom{stat}}_\text{synthetic}^i\right) - {\smash{\widehat{\text{stat}}}\vphantom{stat}}_{\text{original}}^i\right|}{\mathrm{SE}\vphantom{\widehat{stat}}\left({\smash{\widehat{\text{stat}}}\vphantom{stat}}_\text{original}^i\right)} \in \mathbb{R}_{0}^{+}
\]

The standard error of the original estimates were obtained from the bootstrap distribution, ensuring a uniform and comparable measure of uncertainty across all statistics, which might be heterogeneous and not amenable to simple analytical variance formulas. MASD values are non-negative, with an optimum at 0.

Agreement in uncertainty was assessed using the mean confidence interval overlap (CIO) \cite{drechsler2022CIO}:

\[
 \mathrm{CIO} = \frac{1}{n_\text{stat}}\sum_{i=1}^{n_\text{stat}}\frac{1}{2} \left(\frac{\|\bigcap\{{\text{CI}_\text{original}^i}, {\text{CI}_\text{synthetic}^i}\}\|}{\|\text{CI}_\text{original}^i\|} + \frac{\|\bigcap\{{\text{CI}_\text{original}^i}, {\text{CI}_\text{synthetic}^i}\}\|}{\|\text{CI}_\text{synthetic}^i\|}\right) \in [0,1]
\],

This metric measures the overlap between two confidence intervals as the average proportion of their intersection relative to each interval. Non-overlapping intervals yield $\mathrm{CIO}=0$, identical intervals $\mathrm{CIO} = 1$.

MASD and CIO were reported for all replicated tables and plots. An exception is the subgroup distribution analysis in \cref{fig:Fischer5} (Fischer et al. \cite{fischer2020anthropometrisch}), where deviations between original and synthetic distributions were quantified using the Wasserstein distance (WD) \cite{villani2021WD}. Following normalization to the original data ranges, WD values range from 0 to 1, with lower values indicating higher distributional similarity.

}

\newpage


\section{Complete set of replication results}
\label{appx:B}


\subsection{Schikowski et al. \texorpdfstring{\cite{schikowski2020blutdruckmessung}}{[Schikowski et al., 2020]}}
\label{appx:B1}

\subsubsection{Replication of Figure 2}

\begin{longtable}[l]{llll}
\hline
Sex & Age group & Blood pressure measure & Avg. meas. 1/2 - meas. 2 (mean) \\ 
  \hline
Male & 20-24 & Systolic & 1.72 (1.51-1.91) \\ 
   &  &  & \textcolor{orangered}{1.61 (1.34-1.88)} \\ 
   &  & Diastolic & 0.87 (0.73-1.02) \\ 
   &  &  & \textcolor{orangered}{0.87 (0.68-1.07)} \\ 
   & 25-34 & Systolic & 1.26 (1.17-1.35) \\ 
   &  &  & \textcolor{orangered}{1.32 (1.21-1.45)} \\ 
   &  & Diastolic & 0.73 (0.67-0.80) \\ 
   &  &  & \textcolor{orangered}{0.71 (0.62-0.80)} \\ 
   & 35-44 & Systolic & 1.25 (1.17-1.33) \\ 
   &  &  & \textcolor{orangered}{1.29 (1.18-1.41)} \\ 
   &  & Diastolic & 0.60 (0.55-0.65) \\ 
   &  &  & \textcolor{orangered}{0.61 (0.54-0.68)} \\ 
   & 45-54 & Systolic & 1.41 (1.35-1.46) \\ 
   &  &  & \textcolor{orangered}{1.46 (1.37-1.55)} \\ 
   &  & Diastolic & 0.61 (0.58-0.65) \\ 
   &  &  & \textcolor{orangered}{0.63 (0.59-0.68)} \\ 
   & 55-64 & Systolic & 1.85 (1.79-1.91) \\ 
   &  &  & \textcolor{orangered}{1.85 (1.77-1.94)} \\ 
   &  & Diastolic & 0.73 (0.70-0.77) \\ 
   &  &  & \textcolor{orangered}{0.73 (0.68-0.78)} \\ 
   & 65-69 & Systolic & 2.15 (2.07-2.24) \\ 
   &  &  & \textcolor{orangered}{2.17 (2.04-2.29)} \\ 
   &  & Diastolic & 0.79 (0.74-0.84) \\ 
   &  &  & \textcolor{orangered}{0.79 (0.72-0.86)} \\ 
  Female & 20-24 & Systolic & 1.19 (1.03-1.35) \\ 
   &  &  & \textcolor{orangered}{1.22 (1.02-1.42)} \\ 
   &  & Diastolic & 0.95 (0.85-1.05) \\ 
   &  &  & \textcolor{orangered}{0.91 (0.74-1.05)} \\ 
   & 25-34 & Systolic & 1.24 (1.16-1.31) \\ 
   &  &  & \textcolor{orangered}{1.22 (1.12-1.33)} \\ 
   &  & Diastolic & 0.71 (0.66-0.76) \\ 
   &  &  & \textcolor{orangered}{0.72 (0.65-0.79)} \\ 
   & 35-44 & Systolic & 1.27 (1.19-1.34) \\ 
   &  &  & \textcolor{orangered}{1.29 (1.19-1.42)} \\ 
   &  & Diastolic & 0.69 (0.65-0.74) \\ 
   &  &  & \textcolor{orangered}{0.68 (0.62-0.75)} \\ 
   & 45-54 & Systolic & 1.48 (1.42-1.53) \\ 
   &  &  & \textcolor{orangered}{1.50 (1.43-1.59)} \\ 
   &  & Diastolic & 0.69 (0.66-0.72) \\ 
   &  &  & \textcolor{orangered}{0.70 (0.66-0.75)} \\ 
   & 55-64 & Systolic & 1.80 (1.74-1.87) \\ 
   &  &  & \textcolor{orangered}{1.84 (1.75-1.92)} \\ 
   &  & Diastolic & 0.76 (0.73-0.79) \\ 
   &  &  & \textcolor{orangered}{0.76 (0.71-0.80)} \\ 
   & 65-69 & Systolic & 2.31 (2.22-2.40) \\ 
   &  &  & \textcolor{orangered}{2.29 (2.14-2.42)} \\ 
   &  & Diastolic & 0.83 (0.78-0.88) \\ 
   &  &  & \textcolor{orangered}{0.82 (0.75-0.88)} \\ 
   \hline

\textbf{MASD: $0.579$; CIO: $0.824$}
\\
\hline
\caption{Differences of mean blood pressure values (in mmHg) using the mean of first and second measurement or the second measurement only by sex and age group (in years). Percentile-based 95\% bootstrap confidence intervals are reported for original data results. Median and percentile-based 95\% confidence intervals of synthetic data results are printed in orange. avg., average; meas., measurement; MASD, mean absolute standardized difference; CIO, confidence interval overlap}
\label{tab:Schikowski_fig2}
\end{longtable}

\newpage

\begin{figure}[ht]
    \centering
    \includegraphics[width=\columnwidth]{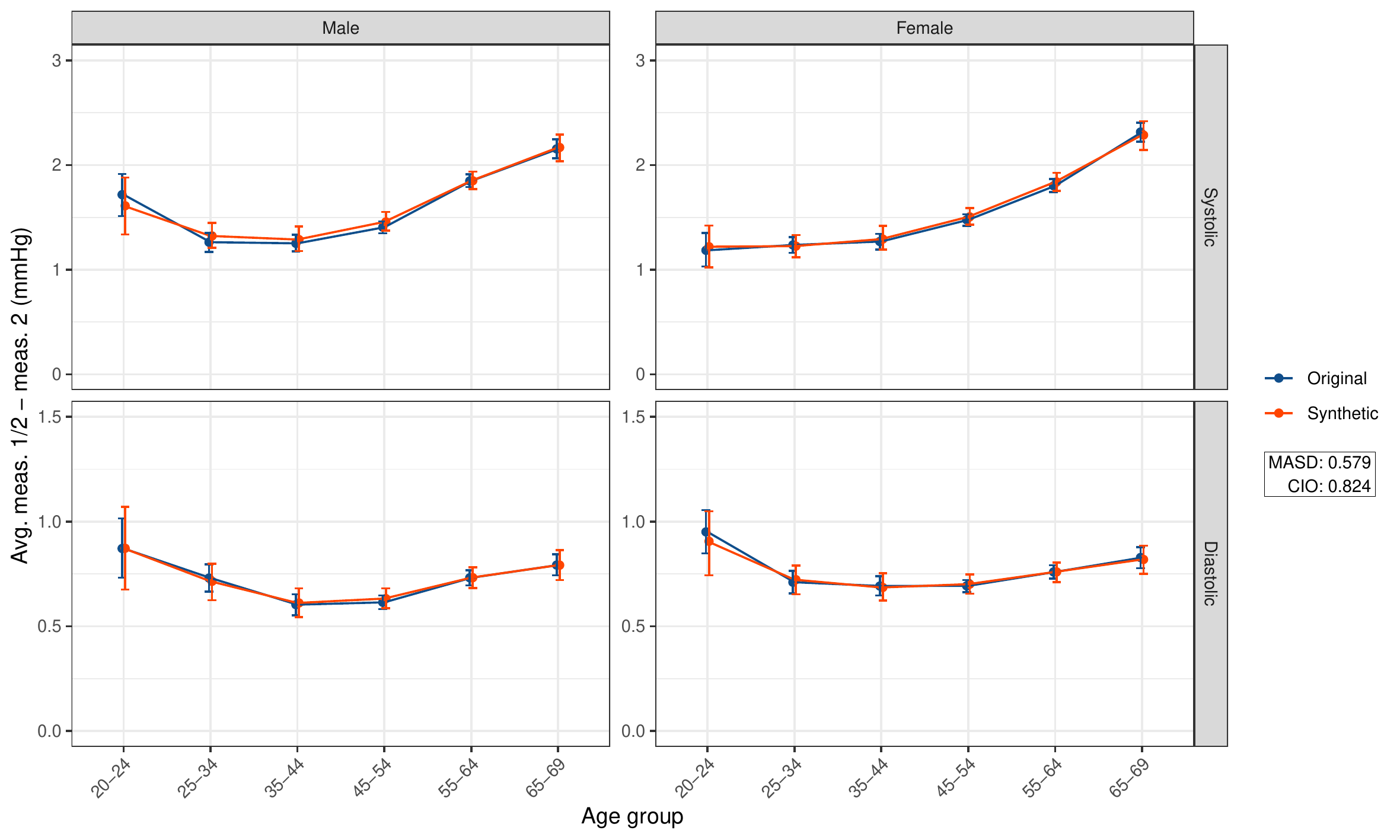}
    \caption{Differences of mean blood pressure values (in mmHg) using the mean of first and second measurement or the second measurement only by sex and age group (in years). Percentile-based 95\% bootstrap confidence intervals are reported for original data results. Median and percentile-based 95\% confidence intervals of synthetic data results are reported. avg., average; meas., measurement; MASD, mean absolute standardized difference; CIO, confidence interval overlap}
    \label{fig:Schikowski_fig2}
\end{figure}

\newpage

\subsubsection{Replication of Figures 3 and 4}

\begin{longtable}[l]{lllll}
\hline
Sex & Age group & Blood pressure measure & Aggregation & Value \\ 
  \hline
Male & 20-24 & Systolic & Mean & 126.22 (125.50-126.91) \\ 
   &  &  &  & \textcolor{orangered}{126.01 (125.09-126.97)} \\ 
   &  &  & SD & 11.47 (10.91-12.00) \\ 
   &  &  &  & \textcolor{orangered}{11.22 (10.59-11.90)} \\ 
   &  & Diastolic & Mean & 72.14 (71.64-72.65) \\ 
   &  &  &  & \textcolor{orangered}{72.39 (71.69-73.24)} \\ 
   &  &  & SD & 8.10 (7.69-8.48) \\ 
   &  &  &  & \textcolor{orangered}{7.93 (7.44-8.42)} \\ 
   & 25-34 & Systolic & Mean & 127.29 (126.96-127.64) \\ 
   &  &  &  & \textcolor{orangered}{127.09 (126.68-127.55)} \\ 
   &  &  & SD & 11.61 (11.32-11.91) \\ 
   &  &  &  & \textcolor{orangered}{11.57 (11.25-11.98)} \\ 
   &  & Diastolic & Mean & 75.49 (75.24-75.75) \\ 
   &  &  &  & \textcolor{orangered}{75.63 (75.23-76.21)} \\ 
   &  &  & SD & 8.42 (8.21-8.63) \\ 
   &  &  &  & \textcolor{orangered}{8.50 (8.24-8.83)} \\ 
   & 35-44 & Systolic & Mean & 128.61 (128.29-128.94) \\ 
   &  &  &  & \textcolor{orangered}{128.67 (128.12-129.29)} \\ 
   &  &  & SD & 12.39 (12.12-12.69) \\ 
   &  &  &  & \textcolor{orangered}{12.64 (12.28-13.06)} \\ 
   &  & Diastolic & Mean & 80.30 (80.05-80.53) \\ 
   &  &  &  & \textcolor{orangered}{80.00 (79.56-80.41)} \\ 
   &  &  & SD & 9.27 (9.08-9.48) \\ 
   &  &  &  & \textcolor{orangered}{9.26 (9.04-9.50)} \\ 
   & 45-54 & Systolic & Mean & 131.83 (131.59-132.07) \\ 
   &  &  &  & \textcolor{orangered}{132.04 (131.70-132.52)} \\ 
   &  &  & SD & 14.40 (14.17-14.62) \\ 
   &  &  &  & \textcolor{orangered}{14.62 (14.33-15.04)} \\ 
   &  & Diastolic & Mean & 82.96 (82.79-83.12) \\ 
   &  &  &  & \textcolor{orangered}{82.81 (82.51-83.08)} \\ 
   &  &  & SD & 9.64 (9.51-9.76) \\ 
   &  &  &  & \textcolor{orangered}{9.74 (9.57-9.90)} \\ 
   & 55-64 & Systolic & Mean & 135.47 (135.19-135.75) \\ 
   &  &  &  & \textcolor{orangered}{135.76 (135.32-136.50)} \\ 
   &  &  & SD & 16.68 (16.43-16.91) \\ 
   &  &  &  & \textcolor{orangered}{16.84 (16.52-17.28)} \\ 
   &  & Diastolic & Mean & 82.51 (82.34-82.67) \\ 
   &  &  &  & \textcolor{orangered}{82.56 (82.32-82.92)} \\ 
   &  &  & SD & 9.96 (9.83-10.09) \\ 
   &  &  &  & \textcolor{orangered}{10.06 (9.89-10.26)} \\ 
   & 65-69 & Systolic & Mean & 138.60 (138.22-139.01) \\ 
   &  &  &  & \textcolor{orangered}{138.57 (137.80-139.17)} \\ 
   &  &  & SD & 17.85 (17.51-18.20) \\ 
   &  &  &  & \textcolor{orangered}{17.59 (17.13-18.03)} \\ 
   &  & Diastolic & Mean & 80.59 (80.35-80.80) \\ 
   &  &  &  & \textcolor{orangered}{80.63 (80.14-80.98)} \\ 
   &  &  & SD & 10.13 (9.95-10.31) \\ 
   &  &  &  & \textcolor{orangered}{10.00 (9.76-10.27)} \\ 
  Female & 20-24 & Systolic & Mean & 114.78 (114.23-115.30) \\ 
   &  &  &  & \textcolor{orangered}{114.78 (113.93-115.61)} \\ 
   &  &  & SD & 10.22 (9.76-10.70) \\ 
   &  &  &  & \textcolor{orangered}{10.04 (9.47-10.65)} \\ 
   &  & Diastolic & Mean & 70.97 (70.59-71.37) \\ 
   &  &  &  & \textcolor{orangered}{71.05 (70.50-71.62)} \\ 
   &  &  & SD & 7.64 (7.33-7.98) \\ 
   &  &  &  & \textcolor{orangered}{7.50 (7.10-7.93)} \\ 
   & 25-34 & Systolic & Mean & 114.74 (114.44-115.02) \\ 
   &  &  &  & \textcolor{orangered}{114.92 (114.47-115.49)} \\ 
   &  &  & SD & 10.72 (10.43-11.02) \\ 
   &  &  &  & \textcolor{orangered}{10.84 (10.55-11.39)} \\ 
   &  & Diastolic & Mean & 72.70 (72.46-72.92) \\ 
   &  &  &  & \textcolor{orangered}{72.70 (72.35-73.04)} \\ 
   &  &  & SD & 8.53 (8.32-8.73) \\ 
   &  &  &  & \textcolor{orangered}{8.43 (8.21-8.68)} \\ 
   & 35-44 & Systolic & Mean & 117.41 (117.10-117.69) \\ 
   &  &  &  & \textcolor{orangered}{117.67 (116.79-118.62)} \\ 
   &  &  & SD & 12.83 (12.57-13.11) \\ 
   &  &  &  & \textcolor{orangered}{12.95 (12.50-13.44)} \\ 
   &  & Diastolic & Mean & 75.64 (75.42-75.85) \\ 
   &  &  &  & \textcolor{orangered}{75.47 (74.86-75.94)} \\ 
   &  &  & SD & 9.41 (9.23-9.58) \\ 
   &  &  &  & \textcolor{orangered}{9.32 (9.07-9.54)} \\ 
   & 45-54 & Systolic & Mean & 123.17 (122.93-123.41) \\ 
   &  &  &  & \textcolor{orangered}{123.57 (123.21-124.06)} \\ 
   &  &  & SD & 15.44 (15.23-15.65) \\ 
   &  &  &  & \textcolor{orangered}{15.64 (15.35-16.03)} \\ 
   &  & Diastolic & Mean & 78.60 (78.45-78.75) \\ 
   &  &  &  & \textcolor{orangered}{78.60 (78.37-78.89)} \\ 
   &  &  & SD & 9.82 (9.69-9.94) \\ 
   &  &  &  & \textcolor{orangered}{9.88 (9.74-10.03)} \\ 
   & 55-64 & Systolic & Mean & 129.91 (129.64-130.19) \\ 
   &  &  &  & \textcolor{orangered}{130.15 (129.67-131.01)} \\ 
   &  &  & SD & 17.53 (17.29-17.75) \\ 
   &  &  &  & \textcolor{orangered}{17.73 (17.40-18.14)} \\ 
   &  & Diastolic & Mean & 79.50 (79.34-79.65) \\ 
   &  &  &  & \textcolor{orangered}{79.61 (79.37-80.00)} \\ 
   &  &  & SD & 9.76 (9.64-9.88) \\ 
   &  &  &  & \textcolor{orangered}{9.87 (9.71-10.11)} \\ 
   & 65-69 & Systolic & Mean & 135.72 (135.32-136.14) \\ 
   &  &  &  & \textcolor{orangered}{135.65 (134.69-136.35)} \\ 
   &  &  & SD & 18.59 (18.25-18.93) \\ 
   &  &  &  & \textcolor{orangered}{18.41 (17.99-18.88)} \\ 
   &  & Diastolic & Mean & 78.84 (78.63-79.08) \\ 
   &  &  &  & \textcolor{orangered}{78.89 (78.33-79.26)} \\ 
   &  &  & SD & 9.86 (9.69-10.03) \\ 
   &  &  &  & \textcolor{orangered}{9.77 (9.56-9.99)} \\ 
   \hline

\textbf{MASD: $1.095$; CIO: $0.729$}
\\
\hline
\caption{Distribution of systolic and diastolic blood pressure (second measurement; in mmHg) by sex and age group (in years). Percentile-based 95\% bootstrap confidence intervals are reported for original data results. Median and percentile-based 95\% confidence intervals of synthetic data results are printed in orange. SD, standard deviation; MASD, mean absolute standardized difference; CIO, confidence interval overlap}
\label{tab:Schikowski_fig3-4}
\end{longtable}

\begin{figure}[ht]
    \centering
    \includegraphics[width=\columnwidth]{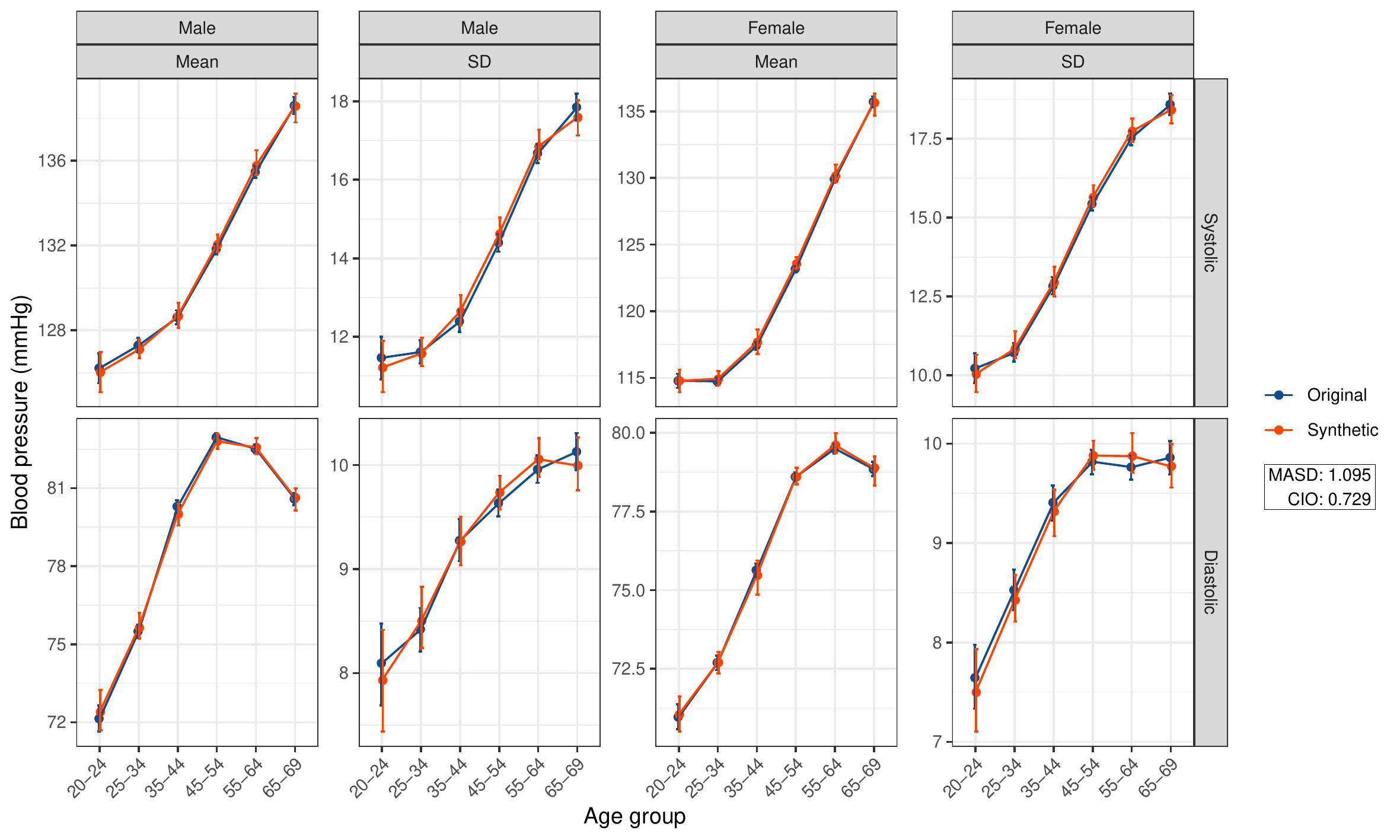}
    \caption{Distribution of systolic and diastolic blood pressure (second measurement; in mmHg) by sex and age group (in years). Percentile-based 95\% bootstrap confidence intervals are reported for original data results. Median and percentile-based 95\% confidence intervals of synthetic data results are reported. SD, standard deviation; MASD, mean absolute standardized difference; CIO, confidence interval overlap}
    \label{fig:Schikowski_fig3-4}
\end{figure}

\newpage

\subsubsection{Replication of Figures 5 and 6}

\begin{longtable}[l]{lllll}
\hline
Sex & Age group & Hypertension status & Proportion using avg. meas. 1/2 & Proportion using meas. 2 \\ 
  \hline
Male & 20-24 & Normotension & 0.82 (0.79-0.84) & 0.83 (0.81-0.85) \\ 
   &  &  & \textcolor{orangered}{0.83 (0.80-0.86)} & \textcolor{orangered}{0.84 (0.81-0.87)} \\ 
   &  & Hypertension known controlled & 0.03 (0.02-0.04) & 0.03 (0.02-0.04) \\ 
   &  &  & \textcolor{orangered}{0.03 (0.02-0.04)} & \textcolor{orangered}{0.03 (0.02-0.04)} \\ 
   &  & Hypertension known uncontrolled & 0.02 (0.01-0.02) & 0.01 (0.01-0.02) \\ 
   &  &  & \textcolor{orangered}{0.01 (0.01-0.02)} & \textcolor{orangered}{0.01 (0.01-0.02)} \\ 
   &  & Hypertension unknown uncontrolled & 0.14 (0.12-0.16) & 0.12 (0.10-0.14) \\
\end{longtable}

\newpage


\begin{landscape}

\subsection{Fischer et al. \texorpdfstring{\cite{fischer2020anthropometrisch}}{[Fischer et al., 2020]}}
\label{appx:B2}

\subsubsection{Replication of Table 1}

\begin{longtable}[l]{lllllllll}
\hline
Variable & Sex & Age group & n & Mean & SD & 10th percentile & 50th perecentile & 90th percentile \\ 
  \hline
Height & Male & 20-29 & 2899 (2798-2999) & 180.75 (180.50-181.01) & 7.14 (6.90-7.38) & 171.90 (171.50-172.40) & 180.70 (180.40-181.10) & 189.90 (189.50-190.20) \\ 
   &  &  & \textcolor{orangered}{2210 (2044-2462)} & \textcolor{orangered}{180.19 (179.58-180.71)} & \textcolor{orangered}{7.31 (6.96-7.76)} & \textcolor{orangered}{170.98 (169.76-171.80)} & \textcolor{orangered}{180.30 (179.70-180.90)} & \textcolor{orangered}{189.38 (188.70-189.95)} \\ 
   &  & 30-39 & 4168 (4048-4296) & 180.66 (180.45-180.88) & 7.15 (7.00-7.30) & 171.50 (171.10-171.80) & 180.60 (180.30-180.80) & 189.60 (189.21-190.00) \\ 
   &  &  & \textcolor{orangered}{4933 (4572-5155)} & \textcolor{orangered}{179.95 (179.16-180.40)} & \textcolor{orangered}{7.47 (7.18-8.05)} & \textcolor{orangered}{170.40 (168.90-171.20)} & \textcolor{orangered}{180.00 (179.40-180.50)} & \textcolor{orangered}{189.30 (188.90-189.70)} \\ 
   &  & 40-49 & 9431 (9241-9621) & 179.83 (179.69-179.97) & 7.14 (7.01-7.29) & 170.80 (170.60-171.10) & 179.90 (179.70-180.00) & 188.80 (188.50-189.00) \\ 
   &  &  & \textcolor{orangered}{9799 (93312-10253)} & \textcolor{orangered}{179.17 (178.34-179.62)} & \textcolor{orangered}{7.45 (7.17-8.01)} & \textcolor{orangered}{169.70 (168.10-170.50)} & \textcolor{orangered}{179.30 (178.60-179.70)} & \textcolor{orangered}{188.50 (188.20-188.90)} \\ 
   &  & 50-59 & 12750 (12541-12961) & 178.46 (178.34-178.59) & 7.11 (6.99-7.28) & 169.60 (169.44-169.90) & 178.30 (178.20-178.40) & 187.50 (187.30-187.70) \\
\end{longtable}

\end{landscape}

\newpage

\subsubsection{Replication of Figure 2}

\begin{longtable}[l]{lll}
\hline
Study centre & Height men (mean) & Height women (mean) \\ 
  \hline
Augsburg & 176.83 (176.63-177.03) & 163.66 (163.48-163.84) \\ 
   & \textcolor{orangered}{176.64 (176.19-176.96)} & \textcolor{orangered}{164.34 (164.03-164.90)} \\ 
  Berlin Mitte & 178.17 (177.90-178.45) & 165.25 (165.02-165.47) \\ 
   & \textcolor{orangered}{177.67 (177.05-178.12)} & \textcolor{orangered}{165.53 (165.19-166.06)} \\ 
  Berlin Nord & 177.60 (177.30-177.89) & 164.70 (164.47-164.92) \\ 
   & \textcolor{orangered}{177.34 (176.70-177.77)} & \textcolor{orangered}{165.20 (164.84-165.86)} \\ 
  Berlin Süd & 178.53 (178.24-178.84) & 165.58 (165.32-165.83) \\
\end{longtable}

\newpage

\subsubsection{Replication of Figure 3}

\begin{longtable}[l]{llll}
\hline
Sex & Age group & BMI group & Proportion \\ 
  \hline
Male & 20-29 & Overweight & 0.30 (0.28-0.31) \\ 
   &  &  & \textcolor{orangered}{0.30 (0.28-0.32)} \\ 
   &  & Obese & 0.10 (0.09-0.11) \\ 
   &  &  & \textcolor{orangered}{0.09 (0.08-0.11)} \\ 
   & 30-39 & Overweight & 0.39 (0.38-0.41) \\ 
   &  &  & \textcolor{orangered}{0.38 (0.37-0.40)} \\ 
   &  & Obese & 0.14 (0.13-0.15) \\
\end{longtable}

\newpage

\begin{figure}[ht]
    \centering
    \includegraphics[width=\columnwidth]{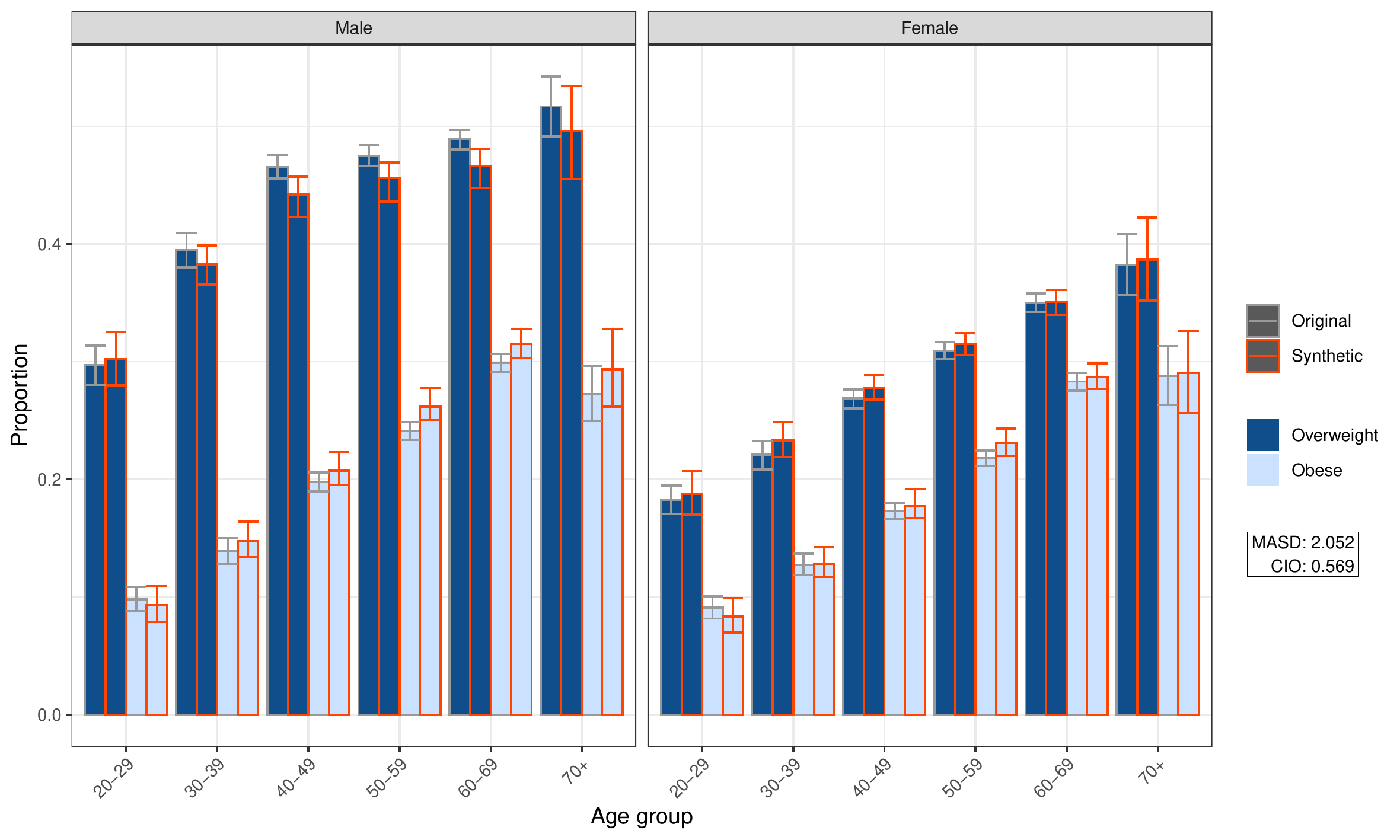}
    \caption{Proportion of overweight and obese study participants by sex and age group (in years). Percentile-based 95\% bootstrap confidence intervals are reported for original data results. Median and percentile-based 95\% confidence intervals of synthetic data results are reported. BMI, body mass index; MASD, mean absolute standardized difference; CIO, confidence interval overlap}
    \label{fig:Fischer_fig3}
\end{figure}

\newpage

\subsubsection{Replication of Figure 4}

\begin{longtable}[l]{lll}
\hline
Study centre & BMI men (mean) & BMI women (mean) \\ 
  \hline
Augsburg & 27.91 (27.78-28.05) & 26.81 (26.66-26.97) \\ 
   & \textcolor{orangered}{27.93 (27.77-28.09)} & \textcolor{orangered}{26.84 (26.66-27.03)} \\ 
  Berlin Mitte & 26.74 (26.59-26.90) & 25.45 (25.27-25.62) \\ 
   & \textcolor{orangered}{26.90 (26.70-27.12)} & \textcolor{orangered}{25.57 (25.36-25.81)} \\ 
  Berlin Nord & 27.71 (27.53-27.90) & 26.41 (26.23-26.60) \\ 
   & \textcolor{orangered}{27.65 (27.45-27.86)} & \textcolor{orangered}{26.44 (26.23-26.66)} \\ 
  Berlin Süd & 26.93 (26.75-27.11) & 25.46 (25.29-25.65) \\ 
   & \textcolor{orangered}{27.03 (26.80-27.25)} & \textcolor{orangered}{25.61 (25.41-25.85)} \\ 
  Bremen & 26.74 (26.58-26.92) & 25.62 (25.44-25.80) \\ 
   & \textcolor{orangered}{26.82 (26.63-27.02)} & \textcolor{orangered}{25.72 (25.51-25.94)} \\ 
  Düsseldorf & 27.21 (26.92-27.56) & 26.13 (25.86-26.41) \\ 
   & \textcolor{orangered}{27.21 (26.93-27.50)} & \textcolor{orangered}{26.20 (25.89-26.51)} \\ 
  Essen & 27.95 (27.75-28.16) & 26.68 (26.43-26.93) \\ 
   & \textcolor{orangered}{27.99 (27.74-28.24)} & \textcolor{orangered}{26.77 (26.50-27.05)} \\ 
  Freiburg & 26.66 (26.48-26.83) & 25.16 (24.95-25.37) \\ 
   & \textcolor{orangered}{26.67 (26.47-26.91)} & \textcolor{orangered}{25.21 (24.95-25.49)} \\ 
  Halle & 27.53 (27.34-27.70) & 26.68 (26.44-26.88) \\ 
   & \textcolor{orangered}{27.64 (27.43-27.85)} & \textcolor{orangered}{26.75 (26.50-27.01)} \\ 
  Hamburg & 27.04 (26.80-27.27) & 26.06 (25.78-26.35) \\ 
   & \textcolor{orangered}{27.21 (26.93-27.49)} & \textcolor{orangered}{26.17 (25.88-26.47)} \\ 
  Hannover & 26.75 (26.59-26.93) & 25.75 (25.52-26.00) \\ 
   & \textcolor{orangered}{26.92 (26.72-27.16)} & \textcolor{orangered}{25.81 (25.59-26.06)} \\ 
  Kiel & 27.54 (27.34-27.74) & 26.51 (26.26-26.75) \\ 
   & \textcolor{orangered}{27.68 (27.42-27.93)} & \textcolor{orangered}{26.50 (26.22-26.78)} \\ 
  Leipzig & 27.36 (27.22-27.51) & 26.48 (26.31-26.65) \\ 
   & \textcolor{orangered}{27.47 (27.29-27.66)} & \textcolor{orangered}{26.57 (26.36-26.78)} \\ 
  Mannheim & 27.33 (27.12-27.56) & 26.26 (26.01-26.50) \\ 
   & \textcolor{orangered}{27.44 (27.20-27.69)} & \textcolor{orangered}{26.28 (26.02-26.54)} \\ 
  Münster & 26.65 (26.48-26.81) & 25.28 (25.08-25.47) \\ 
   & \textcolor{orangered}{26.74 (26.55-26.94)} & \textcolor{orangered}{25.47 (25.23-25.73)} \\ 
  Neubrandenburg & 28.34 (28.23-28.47) & 27.36 (27.21-27.51) \\ 
   & \textcolor{orangered}{28.32 (28.14-28.48)} & \textcolor{orangered}{27.32 (27.14-27.51)} \\ 
  Regensburg & 27.84 (27.64-28.03) & 26.44 (26.21-26.69) \\ 
   & \textcolor{orangered}{27.73 (27.51-27.98)} & \textcolor{orangered}{26.47 (26.21-26.73)} \\ 
  Saarbrücken & 27.86 (27.67-28.06) & 26.58 (26.36-26.81) \\ 
   & \textcolor{orangered}{27.82 (27.56-28.06)} & \textcolor{orangered}{26.57 (26.32-26.83)} \\ 
   \hline

\textbf{MASD: $0.818$; CIO: $0.801$}
\\
\hline
\caption{Mean BMI (in kg/m²) of participants by study centre and sex. Percentile-based 95\% bootstrap confidence intervals are reported for original data results. Median and percentile-based 95\% confidence intervals of synthetic data results are printed in orange. BMI, body mass index; MASD, mean absolute standardized difference; CIO, confidence interval overlap}
\label{tab:Fischer_fig4}
\end{longtable}

\newpage

\subsubsection{Replication of Figure 5}

\begin{figure}[ht]
    \centering
    \includegraphics[width=\columnwidth]{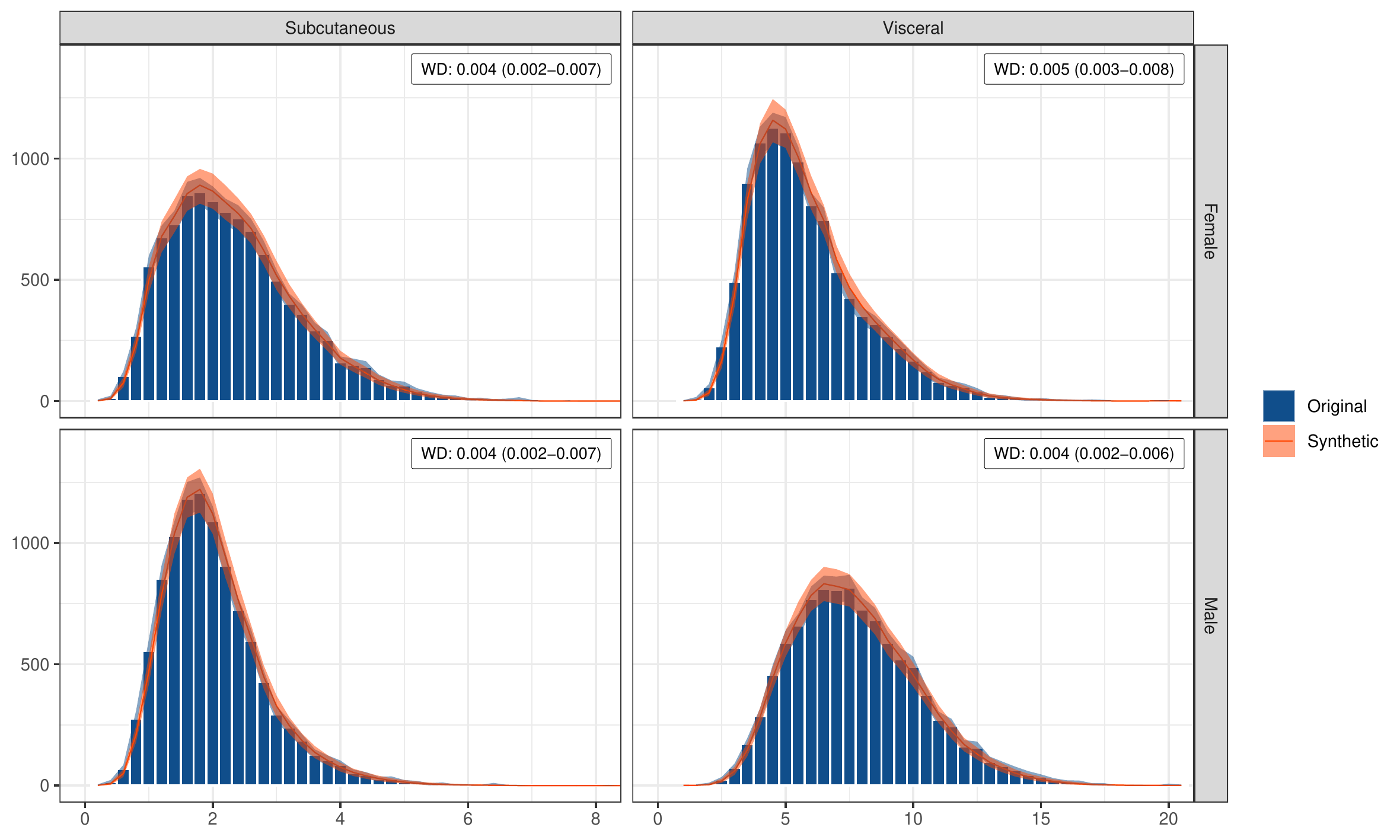}
    \caption{Subcutaneous and visceral abdominal adipose tissue thickness by sex. Percentile-based 95\% bootstrap confidence intervals are reported for original data results. Median and percentile-based 95\% confidence intervals of synthetic data results are reported. In order to provide a clear comparability of the distributions despite the variability in the distribution of missing values in the synthetic data, the counts for the synthetic results were rescaled to match the total count of the real ones; WD, Wasserstein distance}
    \label{fig:Fischer_fig5}
\end{figure}

\newpage


\subsection{Wienbergen et al. \texorpdfstring{\cite{wienbergen2022infarction}}{[Wienbergen et al., 2022]}}
\label{appx:B3}

\subsubsection{Replication of Table 1}

\begin{longtable}[l]{llllll}
\hline

\textbf{MASD: $0.664$; CIO: $0.863$}
\\
\hline
\caption{Distribution of sociodemographic, lifestyle, and metabolic factors and family history of premature myocardial infarction according to cases and controls. Percentile-based 95\% bootstrap confidence intervals are reported for original data results. Median and percentile-based 95\% confidence intervals of synthetic data results are printed in orange. n, number of participants; MI, myocaridal infarction; MASD, mean absolute standardized difference; CIO, confidence interval overlap}
\label{tab:Wienbergen_tab1}
\end{longtable}

\newpage

\subsubsection{Replication of Table 2}

\begin{longtable}[l]{lllll}
\hline
Adj. level & Variable & Odds ratio & 2.5\% & 97.5\% \\ 
  \hline
1 & Smoking status: Former & 1.00 (0.62-1.59) & 0.63 (0.38-1.00) & 1.60 (1.01-2.54) \\ 
   &  & \textcolor{orangered}{1.09 (0.68-1.66)} & \textcolor{orangered}{0.70 (0.43-1.07)} & \textcolor{orangered}{1.68 (1.08-2.58)} \\ 
   & Smoking status: Current & 10.75 (7.87-15.80) & 7.60 (5.61-10.79) & 15.21 (10.98-23.13) \\ 
   &  & \textcolor{orangered}{8.66 (6.14-12.45)} & \textcolor{orangered}{6.25 (4.52-8.75)} & \textcolor{orangered}{12.02 (8.37-17.78)} \\ 
   & Alcohol consumption: 1x/month & 1.04 (0.69-1.54) & 0.70 (0.46-1.04) & 1.53 (1.03-2.30) \\ 
   &  & \textcolor{orangered}{1.03 (0.70-1.55)} & \textcolor{orangered}{0.70 (0.48-1.06)} & \textcolor{orangered}{1.52 (1.02-2.28)} \\ 
   & Alcohol consumption: 2-4x/month & 0.40 (0.27-0.59) & 0.28 (0.18-0.40) & 0.60 (0.40-0.87) \\ 
   &  & \textcolor{orangered}{0.44 (0.30-0.64)} & \textcolor{orangered}{0.30 (0.21-0.44)} & \textcolor{orangered}{0.64 (0.44-0.93)} \\ 
   & Alcohol consumption: 2-3x/week & 0.29 (0.19-0.45) & 0.19 (0.12-0.29) & 0.46 (0.30-0.69) \\ 
   &  & \textcolor{orangered}{0.33 (0.22-0.52)} & \textcolor{orangered}{0.22 (0.14-0.34)} & \textcolor{orangered}{0.51 (0.33-0.79)} \\ 
   & Alcohol consumption: 4+x/week & 0.38 (0.22-0.63) & 0.22 (0.12-0.37) & 0.64 (0.37-1.06) \\ 
   &  & \textcolor{orangered}{0.41 (0.25-0.67)} & \textcolor{orangered}{0.25 (0.14-0.41)} & \textcolor{orangered}{0.69 (0.42-1.12)} \\ 
   & BMI: 25.0-29.9 & 1.59 (1.19-2.12) & 1.18 (0.88-1.56) & 2.15 (1.61-2.88) \\ 
   &  & \textcolor{orangered}{1.76 (1.30-2.37)} & \textcolor{orangered}{1.31 (0.96-1.74)} & \textcolor{orangered}{2.38 (1.75-3.22)} \\ 
   & BMI: 30.0+ & 2.72 (1.94-3.84) & 1.96 (1.39-2.75) & 3.77 (2.69-5.36) \\ 
   &  & \textcolor{orangered}{3.11 (2.22-4.36)} & \textcolor{orangered}{2.25 (1.61-3.14)} & \textcolor{orangered}{4.30 (3.06-6.10)} \\ 
   & BMI 5 units increase & 1.42 (1.25-1.63) & 1.26 (1.11-1.43) & 1.61 (1.41-1.85) \\ 
   &  & \textcolor{orangered}{1.53 (1.35-1.76)} & \textcolor{orangered}{1.35 (1.19-1.54)} & \textcolor{orangered}{1.74 (1.52-2.00)} \\ 
   & Waist-to-hip ratio: 0.87–0.93 & 1.82 (1.03-3.27) & 1.05 (0.60-1.82) & 3.15 (1.76-5.94) \\ 
   &  & \textcolor{orangered}{1.92 (1.17-3.32)} & \textcolor{orangered}{1.17 (0.72-1.95)} & \textcolor{orangered}{3.14 (1.88-5.64)} \\ 
   & Waist-to-hip ratio: 0.93+ & 9.21 (5.41-16.75) & 5.47 (3.36-9.33) & 15.50 (8.76-30.19) \\ 
   &  & \textcolor{orangered}{7.39 (4.62-12.83)} & \textcolor{orangered}{4.64 (3.01-7.61)} & \textcolor{orangered}{11.72 (7.09-21.46)} \\ 
   & Hypertension: Yes & 65.37 (32.46-218.27) & 27.51 (16.16-52.56) & 155.34 (65.36-907.00) \\ 
   &  & \textcolor{orangered}{25.26 (12.98-63.28)} & \textcolor{orangered}{13.43 (7.65-26.95)} & \textcolor{orangered}{47.59 (22.13-151.25)} \\ 
   & Diabetes mellitus: Yes & 6.20 (3.68-11.19) & 3.38 (2.06-5.61) & 11.34 (6.47-22.21) \\ 
   &  & \textcolor{orangered}{5.17 (2.94-9.73)} & \textcolor{orangered}{2.89 (1.69-5.13)} & \textcolor{orangered}{9.22 (5.02-18.67)} \\ 
   & Family history premature MI: Yes & 4.05 (2.82-6.00) & 2.87 (2.00-4.19) & 5.71 (3.95-8.57) \\ 
   &  & \textcolor{orangered}{3.46 (2.41-5.00)} & \textcolor{orangered}{2.48 (1.73-3.55)} & \textcolor{orangered}{4.83 (3.32-7.05)} \\ 
   & Family history premature MI: Unknown & 0.41 (0.21-0.67) & 0.23 (0.11-0.40) & 0.71 (0.41-1.14) \\ 
   &  & \textcolor{orangered}{0.53 (0.29-0.91)} & \textcolor{orangered}{0.31 (0.16-0.55)} & \textcolor{orangered}{0.91 (0.53-1.51)} \\ 
  2 & Smoking status: Former & 0.97 (0.57-1.63) & 0.60 (0.34-1.00) & 1.59 (0.96-2.67) \\ 
   &  & \textcolor{orangered}{1.05 (0.65-1.66)} & \textcolor{orangered}{0.67 (0.40-1.05)} & \textcolor{orangered}{1.65 (1.04-2.61)} \\ 
   & Smoking status: Current & 12.09 (8.75-18.45) & 8.38 (6.17-12.34) & 17.44 (12.41-27.48) \\ 
   &  & \textcolor{orangered}{9.08 (6.34-13.59)} & \textcolor{orangered}{6.44 (4.56-9.33)} & \textcolor{orangered}{12.79 (8.77-19.76)} \\ 
   & Alcohol consumption: 1x/month & 0.99 (0.63-1.57) & 0.62 (0.39-0.98) & 1.56 (1.01-2.50) \\ 
   &  & \textcolor{orangered}{1.03 (0.65-1.62)} & \textcolor{orangered}{0.66 (0.42-1.04)} & \textcolor{orangered}{1.59 (1.01-2.53)} \\ 
   & Alcohol consumption: 2-4x/month & 0.35 (0.22-0.55) & 0.22 (0.14-0.35) & 0.55 (0.35-0.87) \\ 
   &  & \textcolor{orangered}{0.44 (0.28-0.67)} & \textcolor{orangered}{0.28 (0.18-0.44)} & \textcolor{orangered}{0.67 (0.44-1.02)} \\ 
   & Alcohol consumption: 2-3x/week & 0.29 (0.16-0.47) & 0.17 (0.10-0.28) & 0.48 (0.28-0.79) \\ 
   &  & \textcolor{orangered}{0.36 (0.22-0.59)} & \textcolor{orangered}{0.22 (0.13-0.37)} & \textcolor{orangered}{0.58 (0.36-0.95)} \\ 
   & Alcohol consumption: 4+x/week & 0.27 (0.14-0.48) & 0.15 (0.07-0.26) & 0.49 (0.26-0.89) \\ 
   &  & \textcolor{orangered}{0.35 (0.20-0.63)} & \textcolor{orangered}{0.20 (0.11-0.36)} & \textcolor{orangered}{0.63 (0.36-1.09)} \\ 
   & BMI: 25.0-29.9 & 1.65 (1.16-2.32) & 1.16 (0.81-1.62) & 2.35 (1.66-3.34) \\ 
   &  & \textcolor{orangered}{1.79 (1.28-2.50)} & \textcolor{orangered}{1.27 (0.91-1.76)} & \textcolor{orangered}{2.51 (1.79-3.55)} \\ 
   & BMI: 30.0+ & 3.05 (2.05-4.59) & 2.07 (1.39-3.09) & 4.49 (3.03-6.88) \\ 
   &  & \textcolor{orangered}{3.20 (2.17-4.73)} & \textcolor{orangered}{2.22 (1.51-3.23)} & \textcolor{orangered}{4.64 (3.13-6.95)} \\ 
   & BMI 5 units increase & 1.47 (1.27-1.74) & 1.28 (1.10-1.49) & 1.70 (1.47-2.02) \\ 
   &  & \textcolor{orangered}{1.54 (1.32-1.81)} & \textcolor{orangered}{1.34 (1.15-1.55)} & \textcolor{orangered}{1.78 (1.52-2.09)} \\ 
   & Waist-to-hip ratio: 0.87–0.93 & 1.49 (0.80-2.89) & 0.81 (0.44-1.50) & 2.75 (1.48-5.60) \\ 
   &  & \textcolor{orangered}{1.61 (0.93-3.00)} & \textcolor{orangered}{0.94 (0.56-1.71)} & \textcolor{orangered}{2.77 (1.58-5.30)} \\ 
   & Waist-to-hip ratio: 0.93+ & 7.27 (4.31-14.96) & 4.08 (2.49-7.60) & 12.97 (7.53-28.98) \\ 
   &  & \textcolor{orangered}{5.79 (3.48-10.40)} & \textcolor{orangered}{3.50 (2.19-5.96)} & \textcolor{orangered}{9.60 (5.51-18.19)} \\ 
   & Hypertension: Yes & 82.75 (35.58-365.81) & 31.26 (15.29-77.73) & 219.10 (80.73-1730.39) \\ 
   &  & \textcolor{orangered}{21.54 (9.66-59.86)} & \textcolor{orangered}{10.58 (5.32-23.97)} & \textcolor{orangered}{43.16 (17.44-151.58)} \\ 
   & Diabetes mellitus: Yes & 5.66 (3.12-11.54) & 2.73 (1.54-5.13) & 11.73 (6.30-26.64) \\ 
   &  & \textcolor{orangered}{4.39 (2.19-9.20)} & \textcolor{orangered}{2.24 (1.15-4.33)} & \textcolor{orangered}{8.67 (4.17-20.01)} \\ 
   & Family history premature MI: Yes & 2.92 (1.88-4.70) & 1.96 (1.26-3.09) & 4.36 (2.79-7.15) \\ 
   &  & \textcolor{orangered}{2.58 (1.73-3.95)} & \textcolor{orangered}{1.76 (1.19-2.67)} & \textcolor{orangered}{3.77 (2.51-5.83)} \\ 
   & Family history premature MI: Unknown & 0.24 (0.11-0.45) & 0.13 (0.05-0.24) & 0.46 (0.22-0.83) \\ 
   &  & \textcolor{orangered}{0.39 (0.19-0.74)} & \textcolor{orangered}{0.21 (0.09-0.42)} & \textcolor{orangered}{0.71 (0.37-1.31)} \\ 
  3 & Smoking status: Former & 0.70 (0.37-1.25) & 0.40 (0.20-0.73) & 1.21 (0.69-2.15) \\ 
   &  & \textcolor{orangered}{0.90 (0.54-1.48)} & \textcolor{orangered}{0.55 (0.32-0.92)} & \textcolor{orangered}{1.46 (0.91-2.40)} \\ 
   & Smoking status: Current & 11.76 (8.28-18.42) & 7.94 (5.69-11.97) & 17.40 (12.00-28.20) \\ 
   &  & \textcolor{orangered}{8.45 (5.79-12.89)} & \textcolor{orangered}{5.90 (4.14-8.69)} & \textcolor{orangered}{12.13 (8.17-19.16)} \\ 
   & Alcohol consumption: 1x/month & 1.13 (0.70-1.82) & 0.69 (0.42-1.11) & 1.84 (1.15-3.04) \\ 
   &  & \textcolor{orangered}{1.08 (0.67-1.74)} & \textcolor{orangered}{0.68 (0.42-1.09)} & \textcolor{orangered}{1.71 (1.07-2.74)} \\ 
   & Alcohol consumption: 2-4x/month & 0.38 (0.23-0.62) & 0.24 (0.14-0.38) & 0.62 (0.38-1.01) \\ 
   &  & \textcolor{orangered}{0.46 (0.29-0.73)} & \textcolor{orangered}{0.30 (0.19-0.47)} & \textcolor{orangered}{0.72 (0.46-1.14)} \\ 
   & Alcohol consumption: 2-3x/week & 0.33 (0.19-0.57) & 0.19 (0.10-0.33) & 0.58 (0.33-0.98) \\ 
   &  & \textcolor{orangered}{0.38 (0.23-0.65)} & \textcolor{orangered}{0.23 (0.14-0.40)} & \textcolor{orangered}{0.64 (0.39-1.08)} \\ 
   & Alcohol consumption: 4+x/week & 0.27 (0.13-0.50) & 0.14 (0.07-0.26) & 0.52 (0.27-0.97) \\ 
   &  & \textcolor{orangered}{0.36 (0.19-0.66)} & \textcolor{orangered}{0.20 (0.10-0.37)} & \textcolor{orangered}{0.67 (0.37-1.20)} \\ 
   & BMI: 25.0-29.9 & 1.50 (1.05-2.15) & 1.03 (0.72-1.47) & 2.17 (1.52-3.15) \\ 
   &  & \textcolor{orangered}{1.68 (1.18-2.38)} & \textcolor{orangered}{1.18 (0.83-1.66)} & \textcolor{orangered}{2.40 (1.68-3.43)} \\ 
   & BMI: 30.0+ & 2.43 (1.54-3.82) & 1.60 (1.01-2.49) & 3.69 (2.34-5.93) \\ 
   &  & \textcolor{orangered}{2.58 (1.72-3.84)} & \textcolor{orangered}{1.74 (1.17-2.58)} & \textcolor{orangered}{3.81 (2.52-5.75)} \\ 
   & BMI 5 units increase & 1.34 (1.13-1.60) & 1.15 (0.97-1.36) & 1.57 (1.32-1.89) \\ 
   &  & \textcolor{orangered}{1.41 (1.20-1.67)} & \textcolor{orangered}{1.21 (1.04-1.42)} & \textcolor{orangered}{1.64 (1.40-1.95)} \\ 
   & Waist-to-hip ratio: 0.87–0.93 & 1.56 (0.80-3.13) & 0.82 (0.41-1.55) & 2.99 (1.50-6.31) \\ 
   &  & \textcolor{orangered}{1.57 (0.90-2.96)} & \textcolor{orangered}{0.89 (0.52-1.66)} & \textcolor{orangered}{2.75 (1.54-5.30)} \\ 
   & Waist-to-hip ratio: 0.93+ & 6.27 (3.54-13.23) & 3.40 (1.93-6.69) & 11.54 (6.27-26.50) \\ 
   &  & \textcolor{orangered}{4.96 (2.92-9.32)} & \textcolor{orangered}{2.96 (1.78-5.13)} & \textcolor{orangered}{8.35 (4.74-16.41)} \\ 
   & Hypertension: Yes & 82.75 (35.58-365.81) & 31.26 (15.29-77.73) & 219.10 (80.73-1730.39) \\ 
   &  & \textcolor{orangered}{21.54 (9.66-59.86)} & \textcolor{orangered}{10.58 (5.32-23.97)} & \textcolor{orangered}{43.16 (17.44-151.58)} \\ 
   & Diabetes mellitus: Yes & 5.66 (3.12-11.54) & 2.73 (1.54-5.13) & 11.73 (6.30-26.64) \\ 
   &  & \textcolor{orangered}{4.39 (2.19-9.20)} & \textcolor{orangered}{2.24 (1.15-4.33)} & \textcolor{orangered}{8.67 (4.17-20.01)} \\ 
   & Family history premature MI: Yes & 2.76 (1.74-4.50) & 1.80 (1.14-2.90) & 4.22 (2.67-7.10) \\ 
   &  & \textcolor{orangered}{2.44 (1.61-3.81)} & \textcolor{orangered}{1.64 (1.07-2.53)} & \textcolor{orangered}{3.64 (2.38-5.73)} \\ 
   & Family history premature MI: Unknown & 0.22 (0.10-0.42) & 0.11 (0.04-0.22) & 0.45 (0.22-0.83) \\ 
   &  & \textcolor{orangered}{0.39 (0.18-0.75)} & \textcolor{orangered}{0.20 (0.09-0.42)} & \textcolor{orangered}{0.74 (0.37-1.38)} \\ 
   \hline

\textbf{MASD: $0.786$; CIO: $0.721$}
\\
\hline
\caption{Associations between lifestyle and metabolic factors, as well as family history of premature myocardial infarction and risk of early–onset myocardial infarction. Adj. level 1: adjusted for age, sex, country of birth, and years of school education; Adj. level 2: additionally adjusted for body mass index, smoking, and frequency of alcohol consumption; Adj. level 3: additionally adjusted for self–reported diabetes mellitus and hypertension. Percentile-based 95\% bootstrap confidence intervals are reported for original data results. Median and percentile-based 95\% confidence intervals of synthetic data results are printed in orange. adj., adjustment; 2.5\% and 97.5\%, regression 95\% confidence interval limits; BMI, body mass index; MI, myocardial infarction; MASD, mean absolute standardized difference; CIO, confidence interval overlap}
\label{tab:Wienbergen_tab2}
\end{longtable}

\subsubsection{Replication of separate logistic regressions per variable, each adjusted for age, sex, country of
birth, and years of school education - Table 2}

\begin{figure}[ht]
    \centering
    \includegraphics[width=\columnwidth]{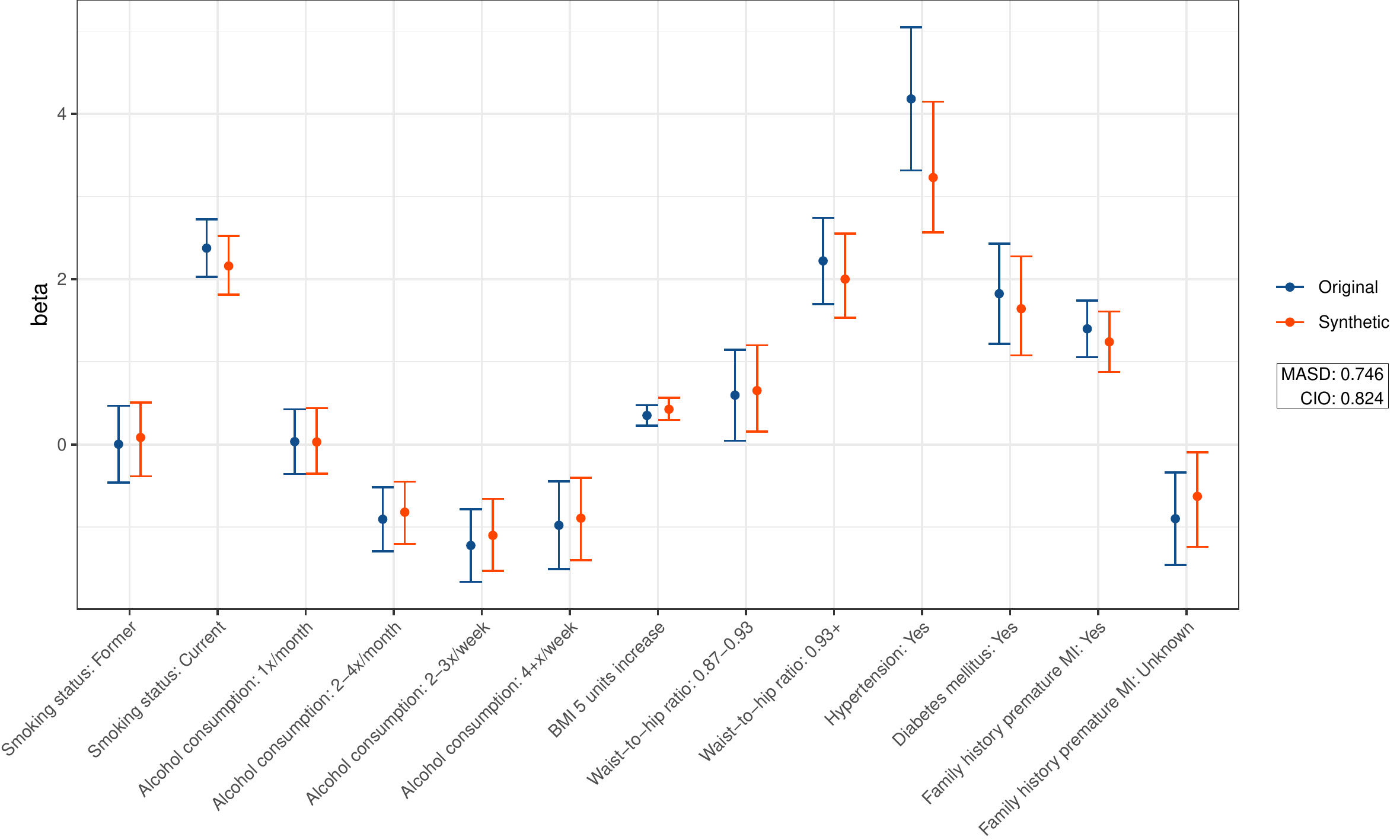}
    \caption{Associations between lifestyle and metabolic factors, as well as family history of
premature MI and risk of early–onset MI. Median beta estimates and percentile-based 95\% confidence intervals computed from the beta coefficient distribution across synthesis repetitions are reported for synthetic data. BMI, body mass index; MI, myocardial infarction; MASD, mean absolute standardized difference; CIO, confidence interval overlap}
    \label{fig:Wienbergen_reg_appx}
\end{figure}

\newpage

\subsubsection{Replication of Table 3}

\begin{longtable}[l]{llllll}
\hline
Subset & Adj. level & No. risk factors & Odds ratio & 2.5\% & 97.5\% \\ 
  \hline
All & 1 & 1 & 5.40 (3.08-12.42) & 2.88 (1.80-5.20) & 10.11 (5.28-29.45) \\ 
   &  &  & \textcolor{orangered}{4.40 (2.58-8.95)} & \textcolor{orangered}{2.48 (1.57-4.28)} & \textcolor{orangered}{7.81 (4.20-19.06)} \\ 
   &  & 2+ & 43.81 (25.52-97.73) & 23.27 (14.85-41.49) & 82.48 (43.44-229.52) \\ 
   &  &  & \textcolor{orangered}{30.91 (17.28-64.17)} & \textcolor{orangered}{17.44 (10.68-30.29)} & \textcolor{orangered}{54.41 (28.09-139.38)} \\ 
   &  & 1 unit increase & 6.41 (5.25-8.16) & 5.12 (4.26-6.40) & 8.01 (6.47-10.39) \\ 
   &  &  & \textcolor{orangered}{5.32 (4.18-6.81)} & \textcolor{orangered}{4.33 (3.47-5.44)} & \textcolor{orangered}{6.53 (5.05-8.52)} \\ 
   & 2 & 1 & 23.04 (9.57-74.53) & 9.14 (3.95-24.88) & 58.09 (22.42-227.19) \\ 
   &  &  & \textcolor{orangered}{11.50 (4.82-31.41)} & \textcolor{orangered}{4.90 (2.16-12.22)} & \textcolor{orangered}{26.92 (10.67-85.09)} \\ 
   &  & 2+ & 582.31 (140.81-3606.48) & 136.45 (34.21-664.72) & 2485.08 (563.87-19064.33) \\ 
   &  &  & \textcolor{orangered}{141.81 (35.23-672.94)} & \textcolor{orangered}{37.54 (10.03-157.69)} & \textcolor{orangered}{533.45 (119.85-2921.45)} \\ 
   &  & 1 unit increase & 21.14 (12.35-44.31) & 11.71 (7.18-22.41) & 38.16 (21.17-89.43) \\ 
   &  &  & \textcolor{orangered}{9.92 (5.59-18.28)} & \textcolor{orangered}{6.09 (3.66-10.48)} & \textcolor{orangered}{16.12 (8.54-32.03)} \\ 
  Family history of premature MI: No & 1 & 1 & 5.49 (3.02-14.18) & 2.74 (1.66-5.23) & 11.00 (5.39-38.91) \\ 
   &  &  & \textcolor{orangered}{4.53 (2.45-10.30)} & \textcolor{orangered}{2.37 (1.42-4.30)} & \textcolor{orangered}{8.58 (4.24-25.40)} \\ 
   &  & 2+ & 43.08 (23.96-108.52) & 21.25 (13.02-40.77) & 87.35 (43.95-293.87) \\ 
   &  &  & \textcolor{orangered}{31.43 (16.57-72.03)} & \textcolor{orangered}{16.41 (9.50-30.48)} & \textcolor{orangered}{60.11 (28.62-176.70)} \\ 
   &  & 1 unit increase & 6.24 (4.96-8.30) & 4.80 (3.89-6.21) & 8.10 (6.32-11.15) \\ 
   &  &  & \textcolor{orangered}{5.38 (4.12-7.01)} & \textcolor{orangered}{4.24 (3.30-5.38)} & \textcolor{orangered}{6.84 (5.14-9.13)} \\ 
   & 2 & 1 & 5.49 (3.02-14.18) & 2.74 (1.66-5.23) & 11.00 (5.39-38.91) \\ 
   &  &  & \textcolor{orangered}{4.53 (2.45-10.30)} & \textcolor{orangered}{2.37 (1.42-4.30)} & \textcolor{orangered}{8.58 (4.24-25.40)} \\ 
   &  & 2+ & 43.08 (23.96-108.52) & 21.25 (13.02-40.77) & 87.35 (43.95-293.87) \\ 
   &  &  & \textcolor{orangered}{31.43 (16.57-72.03)} & \textcolor{orangered}{16.41 (9.50-30.48)} & \textcolor{orangered}{60.11 (28.62-176.70)} \\ 
   &  & 1 unit increase & 6.24 (4.96-8.30) & 4.80 (3.89-6.21) & 8.10 (6.32-11.15) \\ 
   &  &  & \textcolor{orangered}{5.38 (4.12-7.01)} & \textcolor{orangered}{4.24 (3.30-5.38)} & \textcolor{orangered}{6.84 (5.14-9.13)} \\ 
  Family history of premature MI: Yes & 1 & 1 & 4.14 (1.23-44652545.64) & 0.84 (0.00-1.75) & 20.41 (4.65-Inf) \\ 
   &  &  & \textcolor{orangered}{4.03 (1.06-36378208.19)} & \textcolor{orangered}{0.72 (0.00-1.76)} & \textcolor{orangered}{19.36 (3.65-Inf)} \\ 
   &  & 2+ & 42.88 (13.83-542617242.56) & 8.59 (0.00-21.15) & 214.19 (49.90-Inf) \\ 
   &  &  & \textcolor{orangered}{30.00 (7.48-302773299.34)} & \textcolor{orangered}{5.28 (0.00-13.71)} & \textcolor{orangered}{145.45 (25.38-Inf)} \\ 
   &  & 1 unit increase & 7.49 (4.63-16.75) & 4.15 (2.72-7.64) & 13.49 (7.85-37.15) \\ 
   &  &  & \textcolor{orangered}{5.29 (3.11-9.76)} & \textcolor{orangered}{3.14 (2.00-5.16)} & \textcolor{orangered}{8.94 (4.80-18.64)} \\ 
   & 2 & 1 & 4.14 (1.23-44652545.64) & 0.84 (0.00-1.75) & 20.41 (4.65-Inf) \\ 
   &  &  & \textcolor{orangered}{4.03 (1.06-36378208.19)} & \textcolor{orangered}{0.72 (0.00-1.76)} & \textcolor{orangered}{19.36 (3.65-Inf)} \\ 
   &  & 2+ & 42.88 (13.83-542617242.56) & 8.59 (0.00-21.15) & 214.19 (49.90-Inf) \\ 
   &  &  & \textcolor{orangered}{30.00 (7.48-302773299.34)} & \textcolor{orangered}{5.28 (0.00-13.71)} & \textcolor{orangered}{145.45 (25.38-Inf)} \\ 
   &  & 1 unit increase & 7.49 (4.63-16.75) & 4.15 (2.72-7.64) & 13.49 (7.85-37.15) \\ 
   &  &  & \textcolor{orangered}{5.29 (3.11-9.76)} & \textcolor{orangered}{3.14 (2.00-5.16)} & \textcolor{orangered}{8.94 (4.80-18.64)} \\ 
  Sex: Male & 1 & 1 & 4.13 (2.22-11.14) & 2.01 (1.20-3.80) & 8.50 (4.05-32.30) \\ 
   &  &  & \textcolor{orangered}{3.89 (2.07-9.56)} & \textcolor{orangered}{1.97 (1.18-3.67)} & \textcolor{orangered}{7.65 (3.58-25.53)} \\ 
   &  & 2+ & 34.68 (19.14-91.16) & 16.92 (10.47-31.86) & 71.08 (35.07-275.98) \\ 
   &  &  & \textcolor{orangered}{27.99 (14.97-67.03)} & \textcolor{orangered}{14.31 (8.59-26.36)} & \textcolor{orangered}{54.50 (25.97-173.90)} \\ 
   &  & 1 unit increase & 6.38 (5.07-8.39) & 4.98 (4.03-6.39) & 8.17 (6.40-10.98) \\ 
   &  &  & \textcolor{orangered}{5.39 (4.15-7.10)} & \textcolor{orangered}{4.28 (3.38-5.51)} & \textcolor{orangered}{6.79 (5.14-9.15)} \\ 
   & 2 & 1 & 4.21 (2.25-11.05) & 2.02 (1.20-3.82) & 8.75 (4.19-32.93) \\ 
   &  &  & \textcolor{orangered}{3.78 (1.97-9.53)} & \textcolor{orangered}{1.91 (1.12-3.64)} & \textcolor{orangered}{7.49 (3.48-24.91)} \\ 
   &  & 2+ & 34.38 (19.01-92.57) & 16.53 (10.08-31.56) & 71.50 (34.85-279.20) \\ 
   &  &  & \textcolor{orangered}{26.43 (13.88-64.72)} & \textcolor{orangered}{13.39 (7.83-24.79)} & \textcolor{orangered}{51.85 (24.40-168.07)} \\ 
   &  & 1 unit increase & 6.24 (4.94-8.40) & 4.84 (3.90-6.31) & 8.06 (6.27-11.17) \\ 
   &  &  & \textcolor{orangered}{5.22 (4.01-6.91)} & \textcolor{orangered}{4.13 (3.24-5.36)} & \textcolor{orangered}{6.61 (4.97-8.97)} \\ 
  Sex: Female & 1 & 1 & 11.36 (3.96-148177227.37) & 3.12 (0.00-7.01) & 41.32 (11.11-Inf) \\ 
   &  &  & \textcolor{orangered}{6.32 (2.25-37.78)} & \textcolor{orangered}{2.04 (0.69-4.53)} & \textcolor{orangered}{19.03 (5.58-297.28)} \\ 
   &  & 2+ & 78.34 (25.92-1049261545.60) & 19.59 (0.00-52.43) & 313.30 (79.69-Inf) \\ 
   &  &  & \textcolor{orangered}{40.33 (13.80-255.58)} & \textcolor{orangered}{12.13 (4.52-28.78)} & \textcolor{orangered}{130.47 (35.30-2095.27)} \\ 
   &  & 1 unit increase & 6.69 (4.50-13.35) & 3.85 (2.74-6.43) & 11.64 (7.28-27.97) \\ 
   &  &  & \textcolor{orangered}{5.26 (3.35-9.17)} & \textcolor{orangered}{3.29 (2.23-5.12)} & \textcolor{orangered}{8.46 (5.01-16.58)} \\ 
   & 2 & 1 & 10.38 (3.60-152846697.04) & 2.82 (0.00-6.78) & 38.17 (10.25-Inf) \\ 
   &  &  & \textcolor{orangered}{6.14 (2.17-36.91)} & \textcolor{orangered}{1.95 (0.65-4.40)} & \textcolor{orangered}{18.71 (5.43-286.80)} \\ 
   &  & 2+ & 71.82 (24.41-1198973535.71) & 17.77 (0.00-50.01) & 290.20 (77.07-Inf) \\ 
   &  &  & \textcolor{orangered}{39.15 (13.24-254.66)} & \textcolor{orangered}{11.52 (4.22-28.10)} & \textcolor{orangered}{128.01 (34.30-2133.37)} \\ 
   &  & 1 unit increase & 6.73 (4.48-14.32) & 3.80 (2.66-6.59) & 11.91 (7.49-31.43) \\ 
   &  &  & \textcolor{orangered}{5.19 (3.29-9.50)} & \textcolor{orangered}{3.19 (2.14-5.12)} & \textcolor{orangered}{8.43 (4.93-17.53)} \\ 
   \hline

\textbf{MASD: $0.576$; CIO: $0.671$}
\\
\hline
\caption{Joint associations of lifestyle and metabolic factors with risk of early–onset myocardial infarction, overall and by family history of premature myocardial infarction and sex. Percentile-based 95\% bootstrap confidence intervals are reported for original data results. Median and percentile-based 95\% confidence intervals of synthetic data results are printed in orange. adj., adjustment; No., number; 2.5\% and 97.5\%, regression 95\% confidence interval limits; BMI, body mass index; MI, myocardial infarction; MASD, mean absolute standardized difference; CIO, confidence interval overlap}
\label{tab:Wienbergen_tab3}
\end{longtable}

\newpage


\subsection{Breau et al. \texorpdfstring{\cite{breau2022cutpoint}}{[Breau et al., 2022]}}
\label{appx:B4}

\subsubsection{Replication of Table 3}

\begin{longtable}[l]{llllll}
\hline
Variable & Everyone & Toddlers 1.5-2 yrs & Preschoolers 3-4 yrs & Preschoolers 3-6 yrs & School aged 5-6 yrs \\ 
  \hline
Number of participants (n) & 262 (262-262) & 96 (80-111) & 121 (105-136) & 166 (151-182) & 45 (34-57) \\ 
   & \textcolor{orangered}{262 (262-262)} & \textcolor{orangered}{84 (69-100)} & \textcolor{orangered}{143 (126-159.02)} & \textcolor{orangered}{178 (162-193)} & \textcolor{orangered}{35 (24-48)} \\ 
  Age (years) (mean) & 3.62 (3.46-3.77) & 2.26 (2.17-2.34) & 3.96 (3.86-4.06) & 4.40 (4.26-4.54) & 5.58 (5.49-5.68) \\ 
   & \textcolor{orangered}{3.64 (3.50-3.78)} & \textcolor{orangered}{2.39 (2.30-2.47)} & \textcolor{orangered}{3.94 (3.84-4.03)} & \textcolor{orangered}{4.23 (4.12-4.36)} & \textcolor{orangered}{5.44 (5.34-5.54)} \\ 
  Age (years) (SD) & 1.28 (1.20-1.35) & 0.42 (0.38-0.45) & 0.59 (0.54-0.64) & 0.90 (0.83-0.96) & 0.33 (0.27-0.37) \\ 
   & \textcolor{orangered}{1.10 (1.03-1.18)} & \textcolor{orangered}{0.38 (0.33-0.42)} & \textcolor{orangered}{0.56 (0.52-0.60)} & \textcolor{orangered}{0.79 (0.72-0.86)} & \textcolor{orangered}{0.30 (0.23-0.36)} \\ 
  Height (cm) (mean) & 99.29 (97.99-100.61) & 88.29 (87.24-89.31) & 101.76 (100.75-102.79) & 104.75 (103.64-105.92) & 112.73 (111.30-114.14) \\ 
   & \textcolor{orangered}{99.37 (98.18-100.44)} & \textcolor{orangered}{92.74 (90.76-94.73)} & \textcolor{orangered}{100.80 (99.36-102.33)} & \textcolor{orangered}{102.22 (100.81-103.57)} & \textcolor{orangered}{107.72 (104.53-110.78)} \\ 
  Height (cm) (SD) & 10.22 (9.49-10.92) & 4.99 (4.30-5.57) & 5.72 (4.99-6.46) & 7.34 (6.58-8.06) & 4.75 (3.78-5.50) \\ 
   & \textcolor{orangered}{9.11 (8.44-9.81)} & \textcolor{orangered}{7.70 (6.41-9.10)} & \textcolor{orangered}{7.64 (6.80-8.45)} & \textcolor{orangered}{8.11 (7.28-8.94)} & \textcolor{orangered}{7.40 (5.36-9.62)} \\ 
  Weight (kg) (mean) & 15.92 (15.52-16.32) & 13.02 (12.68-13.35) & 16.59 (16.21-16.99) & 17.58 (17.16-18.02) & 20.21 (19.44-21.06) \\ 
   & \textcolor{orangered}{16.04 (15.66-16.42)} & \textcolor{orangered}{14.11 (13.57-14.68)} & \textcolor{orangered}{16.48 (16.01-16.98)} & \textcolor{orangered}{16.95 (16.49-17.41)} & \textcolor{orangered}{18.81 (17.71-19.90)} \\ 
  Weight (kg) (SD) & 3.32 (3.00-3.64) & 1.67 (1.40-1.93) & 2.17 (1.87-2.48) & 2.86 (2.49-3.23) & 2.83 (2.12-3.41) \\ 
   & \textcolor{orangered}{3.00 (2.74-3.27)} & \textcolor{orangered}{2.29 (1.85-2.82)} & \textcolor{orangered}{2.63 (2.28-2.99)} & \textcolor{orangered}{2.86 (2.54-3.20)} & \textcolor{orangered}{2.94 (2.23-3.68)} \\ 
  BMI z score (mean) & 0.56 (0.44-0.68) & 0.86 (0.63-1.11) & 0.44 (0.28-0.59) & 0.40 (0.27-0.54) & 0.31 (0.02-0.62) \\ 
   & \textcolor{orangered}{0.54 (0.29-0.79)} & \textcolor{orangered}{0.54 (0.09-1.00)} & \textcolor{orangered}{0.55 (0.22-0.90)} & \textcolor{orangered}{0.54 (0.24-0.85)} & \textcolor{orangered}{0.49 (-0.14-1.13)} \\ 
  BMI z score (SD) & 0.99 (0.87-1.13) & 1.08 (0.84-1.34) & 0.85 (0.73-0.95) & 0.91 (0.79-1.03) & 1.07 (0.77-1.38) \\ 
   & \textcolor{orangered}{1.92 (1.64-2.19)} & \textcolor{orangered}{1.94 (1.53-2.42)} & \textcolor{orangered}{1.94 (1.61-2.31)} & \textcolor{orangered}{1.90 (1.60-2.22)} & \textcolor{orangered}{1.66 (1.17-2.31)} \\ 
  Valid days (mean) & 7.65 (7.51-7.79) & 7.53 (7.28-7.76) & 7.66 (7.46-7.84) & 7.72 (7.55-7.88) & 7.87 (7.53-8.18) \\ 
   & \textcolor{orangered}{7.56 (7.42-7.68)} & \textcolor{orangered}{7.40 (7.15-7.64)} & \textcolor{orangered}{7.61 (7.43-7.77)} & \textcolor{orangered}{7.63 (7.47-7.78)} & \textcolor{orangered}{7.72 (7.35-8.03)} \\ 
  Valid days (SD) & 1.14 (0.98-1.30) & 1.28 (0.99-1.53) & 1.03 (0.79-1.26) & 1.06 (0.86-1.24) & 1.12 (0.74-1.44) \\ 
   & \textcolor{orangered}{1.06 (0.95-1.18)} & \textcolor{orangered}{1.15 (0.93-1.37)} & \textcolor{orangered}{1.01 (0.86-1.17)} & \textcolor{orangered}{1.01 (0.88-1.16)} & \textcolor{orangered}{0.97 (0.68-1.31)} \\ 
  Valid min/day (mean) & 714.24 (705.47-722.91) & 681.41 (666.11-696.45) & 727.65 (716.75-737.52) & 733.23 (723.60-742.74) & 748.24 (724.64-769.14) \\ 
   & \textcolor{orangered}{713.16 (704.72-721.34)} & \textcolor{orangered}{688.42 (671.78-703.21)} & \textcolor{orangered}{721.33 (710.82-731.14)} & \textcolor{orangered}{725.11 (715.75-734.04)} & \textcolor{orangered}{740.30 (717.66-761.75)} \\ 
  Valid min/day (SD) & 72.16 (64.43-79.98) & 75.64 (62.81-87.18) & 57.16 (47.29-67.62) & 62.86 (54.52-71.89) & 74.78 (56.23-89.77) \\ 
   & \textcolor{orangered}{66.32 (59.03-73.58)} & \textcolor{orangered}{70.28 (56.97-83.04)} & \textcolor{orangered}{59.46 (51.02-69.04)} & \textcolor{orangered}{60.75 (53.18-69.14)} & \textcolor{orangered}{61.12 (46.42-80.15)} \\ 
  Sex: Male (n) & 126 (110-142) & 40 (29-52) & 60 (47-73) & 86 (71-101) & 26 (17-36) \\ 
   & \textcolor{orangered}{126 (110-142)} & \textcolor{orangered}{35 (24-46)} & \textcolor{orangered}{71 (58-86)} & \textcolor{orangered}{91 (76-106)} & \textcolor{orangered}{20 (12-30)} \\ 
  Sex: Female (n) & 136 (120-152) & 56 (43-70) & 61 (47-74) & 80 (65-94) & 19 (11-27) \\ 
   & \textcolor{orangered}{136 (120-152)} & \textcolor{orangered}{49 (37-62)} & \textcolor{orangered}{71 (57-86)} & \textcolor{orangered}{86 (72-102)} & \textcolor{orangered}{15 (8-23)} \\ 
  Ethnicity: White (n) & 192 (178-205) & 78 (63-93) & 85 (71-100) & 114 (98-129) & 29 (20-39) \\ 
   & \textcolor{orangered}{192 (178-206)} & \textcolor{orangered}{64 (50-78)} & \textcolor{orangered}{103 (87-118)} & \textcolor{orangered}{128 (111-143)} & \textcolor{orangered}{24 (16-35)} \\ 
  Ethnicity: Other (n) & 70 (57-84) & 18 (10-26) & 36 (26-47) & 52 (40-65) & 16 (9-24) \\ 
   & \textcolor{orangered}{70 (56-84)} & \textcolor{orangered}{20 (12-28)} & \textcolor{orangered}{40 (28-51)} & \textcolor{orangered}{50 (38-63)} & \textcolor{orangered}{11 (5-18)} \\ 
  HHI: No answer (n) & 10 (4-17) & 2 (0-5) & 6 (2-11) & 8 (3-14) & 2 (0-5) \\ 
   & \textcolor{orangered}{10 (4-16)} & \textcolor{orangered}{3 (0-7)} & \textcolor{orangered}{6 (2-11)} & \textcolor{orangered}{7 (2-12)} & \textcolor{orangered}{1 (0-4)} \\ 
  HHI: $<$\textdollar 100,000 (n) & 122 (106-138) & 45 (33-57) & 53 (41-66) & 77 (63-91) & 24 (15-34) \\ 
   & \textcolor{orangered}{122 (106-137)} & \textcolor{orangered}{38 (28-50)} & \textcolor{orangered}{67 (52-81)} & \textcolor{orangered}{84 (69-99)} & \textcolor{orangered}{16 (9-25)} \\ 
  HHI: \$100,000+ (n) & 130 (114-146) & 49 (37-62) & 62 (49-75) & 81 (66-96) & 19 (11-27) \\ 
   & \textcolor{orangered}{130 (115-146)} & \textcolor{orangered}{43 (31-55)} & \textcolor{orangered}{70 (56-84)} & \textcolor{orangered}{87 (72-102)} & \textcolor{orangered}{17 (10-26)} \\ 
   \hline

\textbf{MASD: $2.132$; CIO: $0.627$}
\\
\hline
\caption{Characteristics of study population for the entire sample population and by age group. Percentile-based 95\% bootstrap confidence intervals are reported for original data results. Median and percentile-based 95\% confidence intervals of synthetic data results are printed in orange. n, number of participants; SD, standard deviation; BMI, body mass index; HHI, household income; MASD, mean absolute standardized difference; CIO, confidence interval overlap}
\label{tab:Breau_tab3}
\end{longtable}

\newpage

\subsubsection{Replication of Figure 1}

\begin{longtable}[l]{llll}
\hline
Set of cutpoints & SED & LPA & MVPA \\ 
  \hline
van Cauwenberghe & 615.34 (607.77-623.08) & 31.81 (30.92-32.68) & 67.09 (64.70-69.44) \\ 
   & \textcolor{orangered}{614.10 (606.94-620.96)} & \textcolor{orangered}{31.88 (30.99-32.73)} & \textcolor{orangered}{67.22 (64.97-69.54)} \\ 
  Pate & 574.95 (567.58-582.27) & 48.54 (47.33-49.75) & 90.75 (87.77-93.69) \\ 
   & \textcolor{orangered}{573.58 (566.89-579.93)} & \textcolor{orangered}{48.67 (47.47-49.84)} & \textcolor{orangered}{90.92 (88.07-93.83)} \\ 
  Puyau & 574.95 (567.58-582.27) & 93.60 (91.17-96.06) & 45.69 (43.94-47.51) \\ 
   & \textcolor{orangered}{573.61 (566.97-580.11)} & \textcolor{orangered}{93.73 (91.37-96.15)} & \textcolor{orangered}{45.82 (44.07-47.61)} \\ 
  Trost & 515.14 (508.14-522.02) & 108.35 (105.93-110.76) & 90.75 (87.77-93.69) \\ 
   & \textcolor{orangered}{513.81 (507.48-520.02)} & \textcolor{orangered}{108.42 (106.15-110.82)} & \textcolor{orangered}{90.91 (87.97-93.86)} \\ 
  Butte (VA) & 515.14 (508.14-522.02) & 126.38 (123.55-129.24) & 72.73 (70.22-75.23) \\ 
   & \textcolor{orangered}{513.77 (507.34-520.01)} & \textcolor{orangered}{126.45 (123.65-129.21)} & \textcolor{orangered}{72.89 (70.40-75.34)} \\ 
  Pate \& Pfeiffer & 509.42 (502.33-516.36) & 114.14 (111.68-116.66) & 90.69 (87.68-93.61) \\ 
   & \textcolor{orangered}{508.20 (501.82-514.43)} & \textcolor{orangered}{114.14 (111.77-116.55)} & \textcolor{orangered}{90.83 (87.98-93.71)} \\ 
  Evenson & 501.33 (494.38-508.17) & 145.78 (142.58-149.04) & 67.13 (64.80-69.51) \\ 
   & \textcolor{orangered}{499.98 (493.70-506.25)} & \textcolor{orangered}{145.88 (142.72-148.93)} & \textcolor{orangered}{67.29 (65.00-69.66)} \\ 
  Costa (VA) & 501.50 (494.46-508.43) & 133.90 (131.02-136.82) & 78.84 (76.14-81.47) \\ 
   & \textcolor{orangered}{500.24 (493.91-506.69)} & \textcolor{orangered}{133.89 (131.08-136.73)} & \textcolor{orangered}{79.01 (76.40-81.63)} \\ 
  Costa (VM) & 463.98 (457.37-470.51) & 163.44 (159.97-166.80) & 86.82 (83.68-89.73) \\ 
   & \textcolor{orangered}{463.08 (456.92-468.87)} & \textcolor{orangered}{163.34 (160.04-166.62)} & \textcolor{orangered}{86.81 (83.95-89.85)} \\ 
  Butte (VM) & 437.20 (430.77-443.48) & 176.36 (172.73-179.96) & 100.68 (97.21-103.91) \\ 
   & \textcolor{orangered}{436.05 (430.09-442.23)} & \textcolor{orangered}{176.33 (172.97-179.87)} & \textcolor{orangered}{100.65 (97.42-103.98)} \\ 
   \hline

\textbf{MASD: $0.172$; CIO: $0.950$}
\\
\hline
\caption{Calculated average valid wear time minutes per day spent in SED, LPA, and MVPA according to the 10 different sets of ActiGraph cutpoints examined.  Cutpoints were applied to the entire cohort. Percentile-based 95\% bootstrap confidence intervals are reported for original data results. Median and percentile-based 95\% confidence intervals of synthetic data results are printed in orange. SED, sedentary behaviour; LPA, light physical activity; MVPA, moderate to vigorous physical activity; VA, vertical axis; VM, vector magnitude; MASD, mean absolute standardized difference; CIO, confidence interval overlap}
\label{tab:Breau_fig1}
\end{longtable}

\begin{figure}[ht]
    \centering
    \includegraphics[width=\columnwidth]{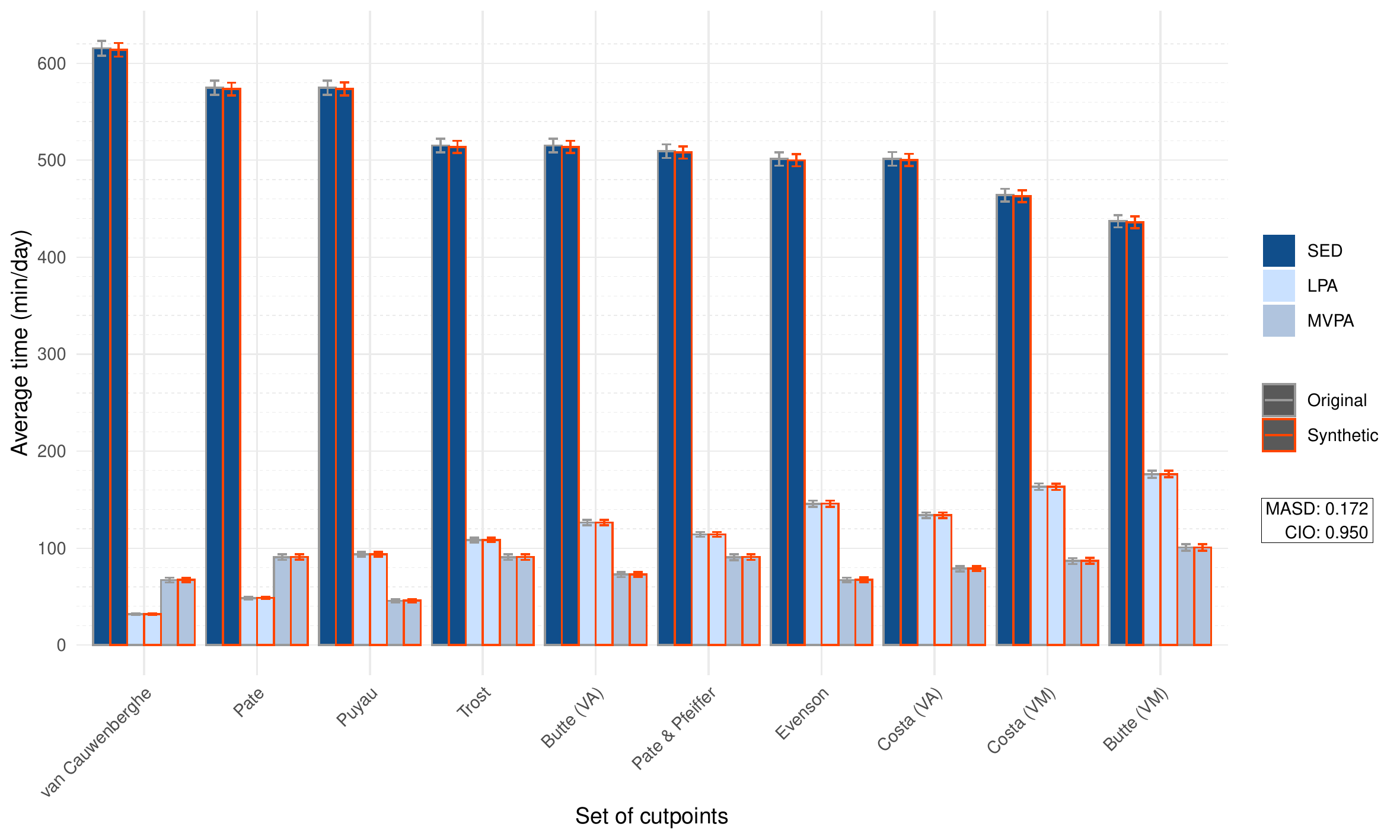}
    \caption{Calculated average valid wear time minutes per day spent in SED, LPA, and MVPA according to the 10 different sets of ActiGraph cutpoints examined.  Cutpoints were applied to the entire cohort. Percentile-based 95\% bootstrap confidence intervals are reported for original data results. Median and percentile-based 95\% confidence intervals of synthetic data results are reported. SED, sedentary behaviour; LPA, light physical activity; MVPA, moderate to vigorous physical activity; VA, vertical axis; VM, vector magnitude; MASD, mean absolute standardized difference; CIO, confidence interval overlap}
    \label{fig:Breau_fig1}
\end{figure}

\newpage

\subsubsection{Replication of Figure 2}

\begin{longtable}[l]{lllll}
\hline
Age group & Set of cutpoints & SED & LPA & MVPA \\ 
  \hline
Toddlers (1.5-2 years) & Trost & 491.66 (479.62-503.62) & 107.73 (103.48-111.87) & 82.01 (77.21-86.77) \\ 
   &  & \textcolor{orangered}{501.96 (488.56-513.23)} & \textcolor{orangered}{105.80 (101.54-110.26)} & \textcolor{orangered}{80.77 (76.03-85.99)} \\ 
   & Costa (VA) & 477.97 (465.90-489.90) & 132.69 (127.62-137.62) & 70.75 (66.49-75.13) \\ 
   &  & \textcolor{orangered}{488.53 (475.31-499.88)} & \textcolor{orangered}{130.24 (125.12-135.73)} & \textcolor{orangered}{69.67 (65.49-74.12)} \\ 
   & Costa (VM) & 449.33 (438.15-461.16) & 157.79 (151.94-163.57) & 74.28 (69.63-78.73) \\ 
   &  & \textcolor{orangered}{456.76 (443.80-468.22)} & \textcolor{orangered}{157.03 (151.06-163.17)} & \textcolor{orangered}{74.46 (69.94-79.37)} \\ 
  Preschoolers (3-6 years) & van Cauwenberghe & 628.95 (620.51-637.51) & 32.99 (32.00-34.02) & 71.29 (68.60-74.05) \\ 
   &  & \textcolor{orangered}{620.87 (612.79-628.62)} & \textcolor{orangered}{33.01 (32.00-34.04)} & \textcolor{orangered}{71.13 (68.36-73.84)} \\ 
   & Pate & 588.03 (579.86-596.40) & 49.40 (47.98-50.86) & 95.81 (92.42-99.29) \\ 
   &  & \textcolor{orangered}{579.69 (572.13-587.05)} & \textcolor{orangered}{49.68 (48.26-51.14)} & \textcolor{orangered}{95.71 (92.43-99.02)} \\ 
   & Butte (VA) & 528.71 (520.88-536.86) & 127.41 (124.16-130.84) & 77.11 (74.29-80.05) \\ 
   &  & \textcolor{orangered}{519.72 (511.95-526.81)} & \textcolor{orangered}{128.38 (125.16-131.66)} & \textcolor{orangered}{77.00 (74.20-79.89)} \\ 
   & Pate \& Pfeiffer & 522.89 (515.09-531.00) & 114.54 (111.73-117.57) & 95.81 (92.42-99.29) \\ 
   &  & \textcolor{orangered}{513.94 (506.50-521.20)} & \textcolor{orangered}{115.45 (112.75-118.21)} & \textcolor{orangered}{95.61 (92.34-98.93)} \\ 
   & Butte (VM) & 445.84 (438.62-453.26) & 178.91 (174.60-183.16) & 108.48 (104.78-112.35) \\ 
   &  & \textcolor{orangered}{439.11 (431.86-445.84)} & \textcolor{orangered}{178.87 (174.77-182.94)} & \textcolor{orangered}{107.00 (103.16-110.65)} \\ 
  School aged (5-6 years) & Puyau & 598.02 (579.82-614.92) & 97.55 (91.59-103.70) & 52.66 (48.23-57.20) \\ 
   &  & \textcolor{orangered}{588.02 (570.17-604.99)} & \textcolor{orangered}{99.42 (92.55-106.31)} & \textcolor{orangered}{52.78 (47.91-57.90)} \\ 
   & Evenson & 526.41 (509.72-541.64) & 145.98 (138.04-154.08) & 75.84 (69.98-81.77) \\ 
   &  & \textcolor{orangered}{514.18 (497.56-530.53)} & \textcolor{orangered}{149.58 (140.40-158.77)} & \textcolor{orangered}{76.33 (69.70-82.95)} \\ 
   \hline

\textbf{MASD: $0.834$; CIO: $0.788$}
\\
\hline
\caption{Calculated average valid wear time minutes per day spent in SED, LPA, and MVPA according to age-appropriate ActiGraph cutpoint sets by age group. Percentile-based 95\% bootstrap confidence intervals are reported for original data results. Median and percentile-based 95\% confidence intervals of synthetic data results are printed in orange. SED, sedentary behaviour; LPA, light physical activity; MVPA, moderate to vigorous physical activity; VA, vertical axis; VM, vector magnitude; MASD, mean absolute standardized difference; CIO, confidence interval overlap}
\label{tab:Breau_fig2}
\end{longtable}

\begin{figure}[ht]
    \centering
    \includegraphics[width=\columnwidth]{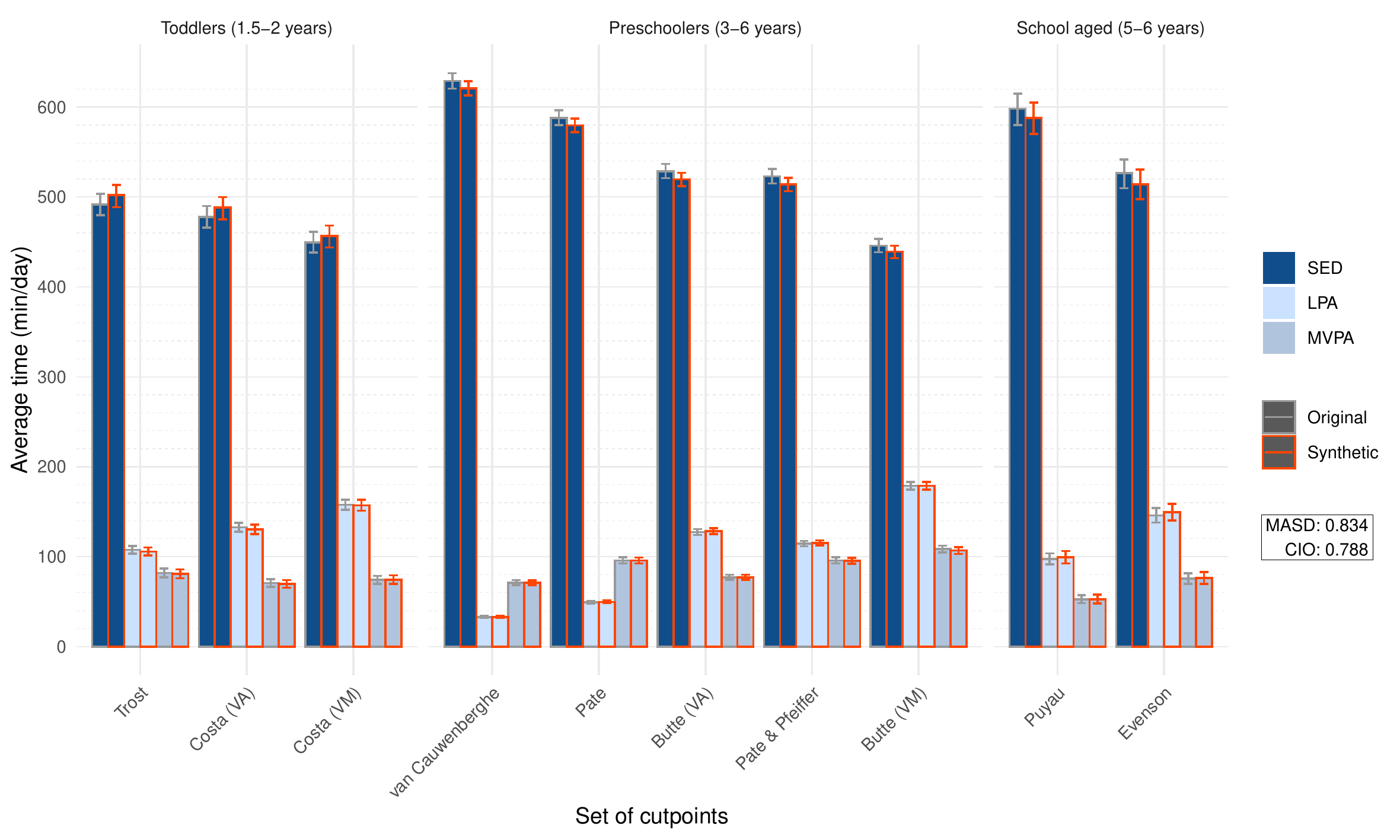}
    \caption{Calculated average valid wear time minutes per day spent in SED, LPA, and MVPA according to age-appropriate ActiGraph cutpoint sets by age group. Percentile-based 95\% bootstrap confidence intervals are reported for original data results. Median and percentile-based 95\% confidence intervals of synthetic data results are reported. SED, sedentary behaviour; LPA, light physical activity; MVPA, moderate to vigorous physical activity; VA, vertical axis; VM, vector magnitude; MASD, mean absolute standardized difference; CIO, confidence interval overlap}
    \label{fig:Breau_fig2}
\end{figure}

\newpage

\subsubsection{Replication of Figure 3}

\begin{longtable}[l]{lll}
\hline
Age group & Set of cutpoints & Proportion meeting PA guideline \\ 
  \hline
Toddlers (1.5-2 years) & Trost & 0.62 (0.53-0.72) \\ 
   &  & \textcolor{orangered}{0.59 (0.48-0.71)} \\ 
   & Costa (VA) & 0.72 (0.62-0.80) \\ 
   &  & \textcolor{orangered}{0.70 (0.60-0.80)} \\ 
   & Costa (VM) & 0.90 (0.83-0.95) \\ 
   &  & \textcolor{orangered}{0.88 (0.80-0.95)} \\ 
  Preschoolers (3-6 years) & van Cauwenberghe & 0.22 (0.16-0.28) \\ 
   &  & \textcolor{orangered}{0.15 (0.10-0.21)} \\ 
   & Pate & 0.37 (0.30-0.45) \\ 
   &  & \textcolor{orangered}{0.30 (0.24-0.38)} \\ 
   & Butte (VA) & 0.75 (0.69-0.82) \\ 
   &  & \textcolor{orangered}{0.73 (0.67-0.80)} \\ 
   & Pate \& Pfeiffer & 0.84 (0.78-0.89) \\ 
   &  & \textcolor{orangered}{0.82 (0.76-0.87)} \\ 
   & Butte (VM) & 0.98 (0.96-1.00) \\ 
   &  & \textcolor{orangered}{0.98 (0.96-1.00)} \\ 
  School aged (5-6 years) & Puyau & 0.31 (0.17-0.45) \\ 
   &  & \textcolor{orangered}{0.32 (0.17-0.49)} \\ 
   & Evenson & 0.80 (0.68-0.91) \\ 
   &  & \textcolor{orangered}{0.79 (0.64-0.92)} \\ 
   \hline

\textbf{MASD: $0.700$; CIO: $0.814$}
\\
\hline
\caption{Percentage of sample meeting age-appropriate guidelines for physical activity according to age-appropriate ActiGraph cutpoints by age group. Percentile-based 95\% bootstrap confidence intervals are reported for original data results. Median and percentile-based 95\% confidence intervals of synthetic data results are printed in orange. PA, physical activity; VA, vertical axis; VM, vector magnitude; MASD, mean absolute standardized difference; CIO, confidence interval overlap}
\label{tab:Breau_fig3}
\end{longtable}

\begin{figure}[ht]
    \centering
    \includegraphics[width=\columnwidth]{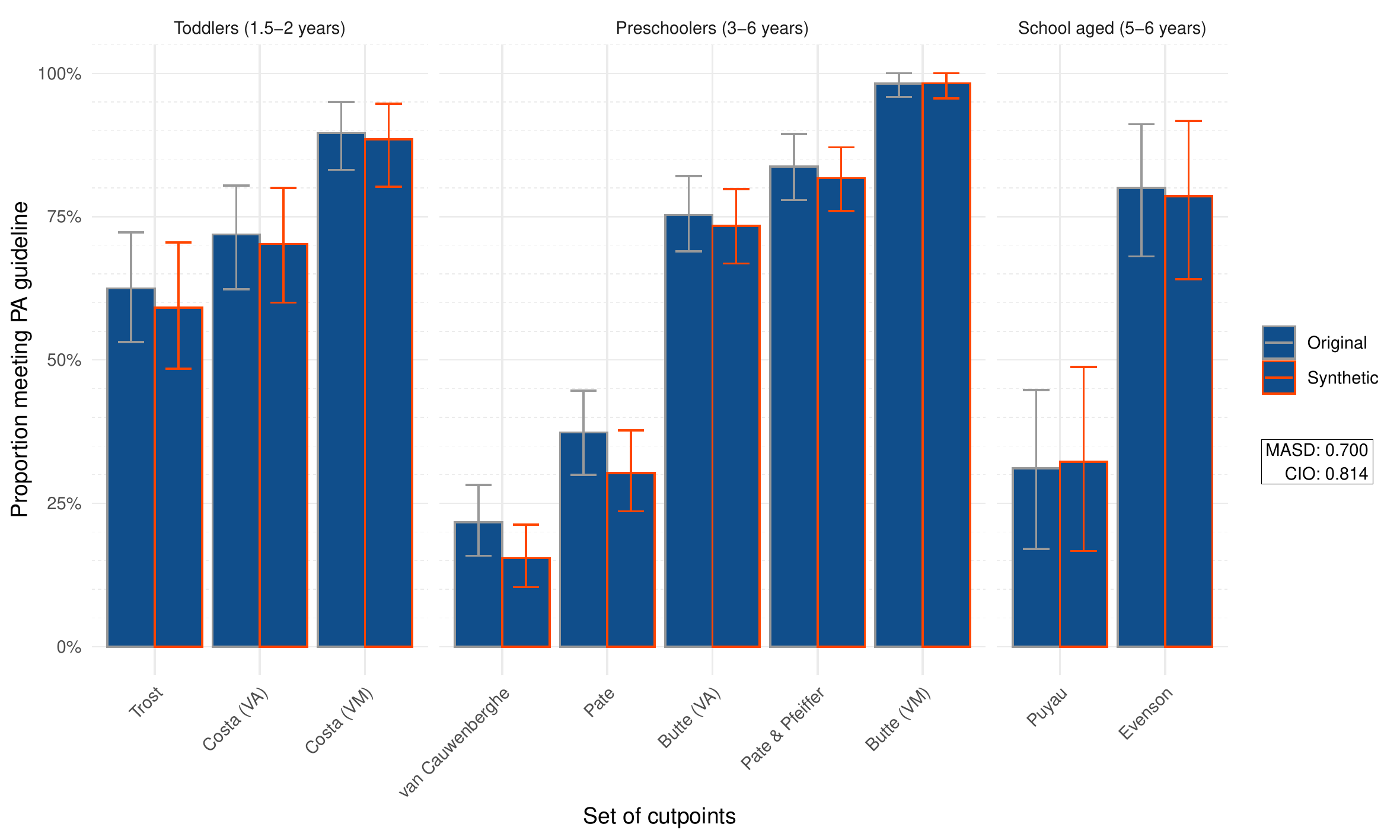}
    \caption{Percentage of sample meeting age-appropriate guidelines for physical activity according to age-appropriate ActiGraph cutpoints by age group. Percentile-based 95\% bootstrap confidence intervals are reported for original data results. Median and percentile-based 95\% confidence intervals of synthetic data results are reported. PA, physical activity; VA, vertical axis; VM, vector magnitude; MASD, mean absolute standardized difference; CIO, confidence interval overlap}
    \label{fig:Breau_fig3}
\end{figure}

\newpage


\subsection{Berger et al. \texorpdfstring{\cite{berger2021COVID}}{[Berger et al., 2021]}}
\label{appx:B5}

\subsubsection{Replication of Table 1 (general and task-specific synthesis)}

\begin{longtable}[l]{ll}
\hline
Variable & Value \\ 
  \hline
Number of participants (n) & 113527 (113527-113527) \\ 
   & \textcolor{orangered}{113527 (113527-113527)} \\ 
   & \textcolor{violetred4}{113527 (113527-113527)} \\ 
  Sex: Female (\%) & 51.75 (51.47-52.03) \\ 
   & \textcolor{orangered}{51.76 (51.45-52.05)} \\ 
   & \textcolor{violetred4}{51.76 (51.46-52.05)} \\ 
  Age (mean years) & 49.99 (49.92-50.07) \\ 
   & \textcolor{orangered}{49.69 (49.53-49.84)} \\ 
   & \textcolor{violetred4}{49.84 (49.74-49.94)} \\ 
  Age: 20-29 years (\%) & 9.08 (8.91-9.25) \\ 
   & \textcolor{orangered}{5.37 (4.98-6.08)} \\ 
   & \textcolor{violetred4}{7.98 (7.56-8.32)} \\ 
  Age: 60+ years (\%) & 26.76 (26.49-27.01) \\ 
   & \textcolor{orangered}{22.02 (21.36-22.83)} \\ 
   & \textcolor{violetred4}{25.50 (24.86-26.17)} \\ 
  Follow up time (mean years) & 2.66 (2.65-2.67) \\ 
   & \textcolor{orangered}{2.68 (2.66-2.70)} \\ 
   & \textcolor{violetred4}{2.68 (2.67-2.69)} \\ 
  Education: Hauptschule (\%) & 11.23 (11.04-11.42) \\ 
   & \textcolor{orangered}{11.23 (11.05-11.42)} \\ 
   & \textcolor{violetred4}{11.23 (11.04-11.41)} \\ 
  Education: Realschule (\%) & 28.62 (28.36-28.89) \\ 
   & \textcolor{orangered}{28.61 (28.34-28.89)} \\ 
   & \textcolor{violetred4}{28.62 (28.36-28.88)} \\ 
  Education: (Fach-)Abitur (\%) & 58.69 (58.41-58.98) \\ 
   & \textcolor{orangered}{58.69 (58.39-58.98)} \\ 
   & \textcolor{violetred4}{58.69 (58.41-58.97)} \\ 
  Education: none/other (\%) & 1.46 (1.39-1.53) \\ 
   & \textcolor{orangered}{1.46 (1.39-1.53)} \\ 
   & \textcolor{violetred4}{1.46 (1.38-1.52)} \\ 
  Living in relationship: Yes (\%) & 76.80 (76.55-77.04) \\ 
   & \textcolor{orangered}{76.80 (76.54-77.04)} \\ 
   & \textcolor{violetred4}{76.80 (76.55-77.04)} \\ 
  PHQ-9 score baseline examination woman (mean) & 4.14 (4.11-4.17) \\ 
   & \textcolor{orangered}{4.07 (4.02-4.12)} \\ 
   & \textcolor{violetred4}{4.13 (4.10-4.16)} \\ 
  PHQ-9 score baseline examination woman (SD) & 3.64 (3.60-3.67) \\ 
   & \textcolor{orangered}{3.28 (3.19-3.37)} \\ 
   & \textcolor{violetred4}{3.64 (3.60-3.68)} \\ 
  GAD-7 score baseline examination woman (mean) & 3.44 (3.41-3.47) \\ 
   & \textcolor{orangered}{3.37 (3.33-3.42)} \\ 
   & \textcolor{violetred4}{3.43 (3.40-3.46)} \\ 
  GAD-7 score baseline examination woman (SD) & 3.23 (3.20-3.26) \\ 
   & \textcolor{orangered}{2.89 (2.81-2.98)} \\ 
   & \textcolor{violetred4}{3.23 (3.19-3.26)} \\ 
  PHQ-9 score baseline examination man (mean) & 3.24 (3.21-3.27) \\ 
   & \textcolor{orangered}{3.27 (3.23-3.32)} \\ 
   & \textcolor{violetred4}{3.24 (3.21-3.27)} \\ 
  PHQ-9 score baseline examination man (SD) & 3.32 (3.27-3.36) \\ 
   & \textcolor{orangered}{3.02 (2.94-3.10)} \\ 
   & \textcolor{violetred4}{3.32 (3.28-3.36)} \\ 
  GAD-7 score baseline examination man (mean) & 2.58 (2.56-2.61) \\ 
   & \textcolor{orangered}{2.62 (2.59-2.66)} \\ 
   & \textcolor{violetred4}{2.59 (2.57-2.62)} \\ 
  GAD-7 score baseline examination man (SD) & 2.81 (2.78-2.85) \\ 
   & \textcolor{orangered}{2.59 (2.53-2.65)} \\ 
   & \textcolor{violetred4}{2.82 (2.79-2.86)} \\ 
   \hline

\textbf{General: \hspace{13pt} MASD: $9.183$; CIO: $0.438$}\\
\textbf{Task-specific: MASD: $1.780$; CIO: $0.747$}
\\
\hline
\caption{Characteristics of German National Cohort (NAKO Gesundheitsstude) participants taking part in the COVID survey in May 2020. Percentile-based 95\% bootstrap confidence intervals are reported for original data results. Median and percentile-based 95\% confidence intervals of synthetic data results are reported, general synthesis replication results are printed in orange, task-specific synthesis replication results in violet. PHQ-9, nine-item Patient Health Questionnaire; GAD-7, Generalized Anxiety Disorder seven-item scale; SD, standard deviation; MASD, mean absolute standardized difference; CIO, confidence interval overlap}
\label{tab:Berger_tab1_taskspec}
\end{longtable}

\newpage

\subsubsection{Replication of Table 2}

\begin{longtable}[l]{llll}
\hline
Variable & Value & Value woman & Value men \\ 
  \hline
Feeling of missing company of others (\%) & 79.88 (79.64-80.11) & 83.49 (83.17-83.79) & 76.00 (75.63-76.36) \\ 
   & \textcolor{orangered}{79.87 (79.63-80.10)} & \textcolor{orangered}{82.31 (81.91-82.72)} & \textcolor{orangered}{77.24 (76.79-77.69)} \\ 
  Feeling of being left out (\%) & 29.88 (29.62-30.16) & 35.09 (34.70-35.47) & 24.29 (23.92-24.67) \\ 
   & \textcolor{orangered}{29.88 (29.61-30.14)} & \textcolor{orangered}{34.11 (33.67-34.55)} & \textcolor{orangered}{25.34 (24.89-25.77)} \\ 
  Feeling of being socially isolated (\%) & 34.87 (34.61-35.15) & 40.48 (40.07-40.89) & 28.85 (28.47-29.22) \\ 
   & \textcolor{orangered}{34.87 (34.58-35.17)} & \textcolor{orangered}{39.28 (38.79-39.79)} & \textcolor{orangered}{30.14 (29.62-30.65)} \\ 
  Sum score loneliness (mean) & 4.93 (4.92-4.94) & 5.17 (5.15-5.18) & 4.68 (4.67-4.69) \\ 
   & \textcolor{orangered}{4.93 (4.92-4.94)} & \textcolor{orangered}{5.11 (5.09-5.13)} & \textcolor{orangered}{4.74 (4.72-4.76)} \\ 
  Sum score loneliness (SD) & 1.61 (1.60-1.61) & 1.67 (1.67-1.68) & 1.49 (1.48-1.50) \\ 
   & \textcolor{orangered}{1.44 (1.41-1.48)} & \textcolor{orangered}{1.49 (1.45-1.53)} & \textcolor{orangered}{1.37 (1.34-1.41)} \\ 
  Proportion of lonely participants (\%) & 31.67 (31.40-31.95) & 37.43 (37.04-37.83) & 25.49 (25.14-25.85) \\ 
   & \textcolor{orangered}{30.77 (30.15-31.32)} & \textcolor{orangered}{35.60 (34.81-36.35)} & \textcolor{orangered}{25.56 (24.93-26.16)} \\ 
  Subjective increase of loneliness during pandemic (\%) & 47.03 (46.74-47.34) & 51.00 (50.61-51.42) & 42.78 (42.36-43.21) \\ 
   & \textcolor{orangered}{47.04 (46.73-47.33)} & \textcolor{orangered}{50.48 (50.03-50.92)} & \textcolor{orangered}{43.35 (42.89-43.81)} \\ 
  Fear of natural disasters (\%) & 11.88 (11.69-12.07) & 15.76 (15.46-16.08) & 7.71 (7.49-7.93) \\ 
   & \textcolor{orangered}{11.88 (11.69-12.07)} & \textcolor{orangered}{14.39 (14.02-14.82)} & \textcolor{orangered}{9.19 (8.79-9.55)} \\ 
  Fear of cancer (\%) & 53.10 (52.81-53.39) & 58.91 (58.51-59.32) & 46.86 (46.41-47.27) \\ 
   & \textcolor{orangered}{53.10 (52.81-53.38)} & \textcolor{orangered}{57.29 (56.79-57.85)} & \textcolor{orangered}{48.60 (48.00-49.10)} \\ 
  Fear of cardiac infarction (\%) & 36.57 (36.28-36.85) & 36.91 (36.51-37.32) & 36.21 (35.80-36.62) \\ 
   & \textcolor{orangered}{36.57 (36.29-36.84)} & \textcolor{orangered}{37.69 (37.26-38.11)} & \textcolor{orangered}{35.36 (34.93-35.79)} \\ 
  Fear of COVID infection (\%) & 34.52 (34.24-34.80) & 38.56 (38.15-38.97) & 30.17 (29.79-30.56) \\ 
   & \textcolor{orangered}{34.51 (34.25-34.79)} & \textcolor{orangered}{37.63 (37.20-38.09)} & \textcolor{orangered}{31.17 (30.73-31.64)} \\ 
  PHQ-9 score COVID survey (mean) & 4.06 (4.04-4.09) & 4.58 (4.55-4.62) & 3.50 (3.47-3.54) \\ 
   & \textcolor{orangered}{4.04 (4.01-4.07)} & \textcolor{orangered}{4.52 (4.48-4.57)} & \textcolor{orangered}{3.53 (3.49-3.56)} \\ 
  PHQ-9 score COVID survey (SD) & 3.93 (3.90-3.95) & 4.07 (4.03-4.11) & 3.69 (3.65-3.73) \\ 
   & \textcolor{orangered}{3.59 (3.51-3.68)} & \textcolor{orangered}{3.71 (3.62-3.81)} & \textcolor{orangered}{3.38 (3.31-3.46)} \\ 
  GAD-7 score COVID survey (mean) & 3.37 (3.35-3.39) & 3.91 (3.88-3.94) & 2.78 (2.76-2.81) \\ 
   & \textcolor{orangered}{3.35 (3.32-3.38)} & \textcolor{orangered}{3.83 (3.79-3.87)} & \textcolor{orangered}{2.84 (2.81-2.87)} \\ 
  GAD-7 score COVID survey (SD) & 3.44 (3.41-3.46) & 3.62 (3.59-3.65) & 3.12 (3.09-3.16) \\ 
   & \textcolor{orangered}{3.14 (3.07-3.22)} & \textcolor{orangered}{3.27 (3.19-3.37)} & \textcolor{orangered}{2.91 (2.85-2.97)} \\ 
   \hline

\textbf{MASD: $8.491$; CIO: $0.281$}
\\
\hline
\caption{Relative frequency of loneliness and anxiety as well as depression and anxiety scores during pandemic measures in May 2020. Percentile-based 95\% bootstrap confidence intervals are reported for original data results. Median and percentile-based 95\% confidence intervals of synthetic data results are printed in orange. PHQ-9, nine-item Patient Health Questionnaire; GAD-7, Generalized Anxiety Disorder seven-item scale; SD, standard deviation; MASD, mean absolute standardized difference; CIO, confidence interval overlap}
\label{tab:Berger_tab2}
\end{longtable}

\newpage

\subsubsection{Replication of multivariable linear regression - Table 3 (general and task-specific synthesis)}

\begin{longtable}[l]{llll}
\hline
Variable & beta coefficient & 2.5\% & 97.5\% \\ 
  \hline
(Intercept) & 4.34 (4.27-4.41) & 4.27 (4.20-4.34) & 4.41 (4.34-4.48) \\ 
   & \textcolor{orangered}{4.30 (4.23-4.37)} & \textcolor{orangered}{4.24 (4.16-4.31)} & \textcolor{orangered}{4.37 (4.29-4.44)} \\ 
   & \textcolor{violetred4}{4.40 (4.32-4.49)} & \textcolor{violetred4}{4.33 (4.25-4.41)} & \textcolor{violetred4}{4.47 (4.39-4.56)} \\ 
  Age & 0.00 (-0.00-0.00) & -0.00 (-0.00-0.00) & 0.00 (0.00-0.00) \\ 
   & \textcolor{orangered}{-0.00 (-0.00-0.00)} & \textcolor{orangered}{-0.00 (-0.00--0.00)} & \textcolor{orangered}{0.00 (-0.00-0.00)} \\ 
   & \textcolor{violetred4}{0.00 (-0.00-0.00)} & \textcolor{violetred4}{-0.00 (-0.00-0.00)} & \textcolor{violetred4}{0.00 (-0.00-0.00)} \\ 
  Sex: Female & 0.28 (0.26-0.30) & 0.27 (0.25-0.28) & 0.30 (0.28-0.32) \\ 
   & \textcolor{orangered}{0.18 (0.16-0.21)} & \textcolor{orangered}{0.17 (0.15-0.20)} & \textcolor{orangered}{0.20 (0.18-0.23)} \\ 
   & \textcolor{violetred4}{0.27 (0.25-0.29)} & \textcolor{violetred4}{0.25 (0.23-0.27)} & \textcolor{violetred4}{0.28 (0.26-0.30)} \\ 
  Education level & -0.04 (-0.05--0.03) & -0.05 (-0.07--0.04) & -0.03 (-0.04--0.02) \\ 
   & \textcolor{orangered}{-0.02 (-0.04--0.01)} & \textcolor{orangered}{-0.03 (-0.05--0.02)} & \textcolor{orangered}{-0.01 (-0.02-0.00)} \\ 
   & \textcolor{violetred4}{-0.04 (-0.06--0.03)} & \textcolor{violetred4}{-0.06 (-0.07--0.04)} & \textcolor{violetred4}{-0.03 (-0.04--0.02)} \\ 
  In relationship: Yes & -0.13 (-0.15--0.11) & -0.15 (-0.18--0.13) & -0.11 (-0.13--0.09) \\ 
   & \textcolor{orangered}{-0.09 (-0.11--0.06)} & \textcolor{orangered}{-0.11 (-0.13--0.08)} & \textcolor{orangered}{-0.07 (-0.09--0.04)} \\ 
   & \textcolor{violetred4}{-0.13 (-0.16--0.11)} & \textcolor{violetred4}{-0.15 (-0.18--0.13)} & \textcolor{violetred4}{-0.11 (-0.14--0.09)} \\ 
  PHP-9 score COVID survey & 0.11 (0.10-0.11) & 0.10 (0.10-0.11) & 0.11 (0.11-0.11) \\ 
   & \textcolor{orangered}{0.10 (0.09-0.10)} & \textcolor{orangered}{0.09 (0.09-0.10)} & \textcolor{orangered}{0.10 (0.10-0.11)} \\ 
   & \textcolor{violetred4}{0.09 (0.08-0.10)} & \textcolor{violetred4}{0.09 (0.08-0.09)} & \textcolor{violetred4}{0.09 (0.09-0.10)} \\ 
  GAD-7 score COVID survey & 0.06 (0.06-0.07) & 0.06 (0.06-0.07) & 0.07 (0.06-0.07) \\ 
   & \textcolor{orangered}{0.08 (0.07-0.09)} & \textcolor{orangered}{0.08 (0.07-0.08)} & \textcolor{orangered}{0.08 (0.08-0.09)} \\ 
   & \textcolor{violetred4}{0.08 (0.07-0.08)} & \textcolor{violetred4}{0.07 (0.07-0.08)} & \textcolor{violetred4}{0.08 (0.08-0.09)} \\ 
  Fear of COVID infection: Yes & 0.10 (0.08-0.12) & 0.08 (0.06-0.10) & 0.12 (0.10-0.14) \\ 
   & \textcolor{orangered}{0.09 (0.07-0.11)} & \textcolor{orangered}{0.07 (0.05-0.09)} & \textcolor{orangered}{0.11 (0.09-0.13)} \\ 
   & \textcolor{violetred4}{0.10 (0.08-0.12)} & \textcolor{violetred4}{0.08 (0.06-0.10)} & \textcolor{violetred4}{0.12 (0.10-0.14)} \\ 
  Study center & -0.00 (-0.00--0.00) & -0.00 (-0.01--0.00) & -0.00 (-0.00-0.00) \\ 
   & \textcolor{orangered}{-0.00 (-0.00-0.00)} & \textcolor{orangered}{-0.00 (-0.00--0.00)} & \textcolor{orangered}{-0.00 (-0.00-0.00)} \\ 
   & \textcolor{violetred4}{-0.00 (-0.00--0.00)} & \textcolor{violetred4}{-0.00 (-0.01--0.00)} & \textcolor{violetred4}{-0.00 (-0.00-0.00)} \\ 
   \hline

\textbf{General: \hspace{13pt} MASD: $9.623$; CIO: $0.276$}\\
\textbf{Task-specific: MASD: $4.708$; CIO: $0.518$}
\\
\hline
\caption{Relationship between perceived loneliness and sociodemographic factors as well as symptoms of depression and anxiety among German National Cohort (NAKO Gesundheitsstudie) participants in May 2020. Percentile-based 95\% bootstrap confidence intervals are reported for original data results. Median and percentile-based 95\% confidence intervals of synthetic data results are reported, general synthesis replication results are printed in orange, task-specific synthesis replication results in violet. 2.5\% and 97.5\%, regression 95\% confidence interval limits; PHQ-9, nine-item Patient Health Questionnaire; GAD-7, Generalized Anxiety Disorder seven-item scale; MASD, mean absolute standardized difference; CIO, confidence interval overlap}
\label{tab:Berger_tab3_taskspec}
\end{longtable}

\newpage

\begin{figure}[ht]
    \centering
    \includegraphics[width=\columnwidth]{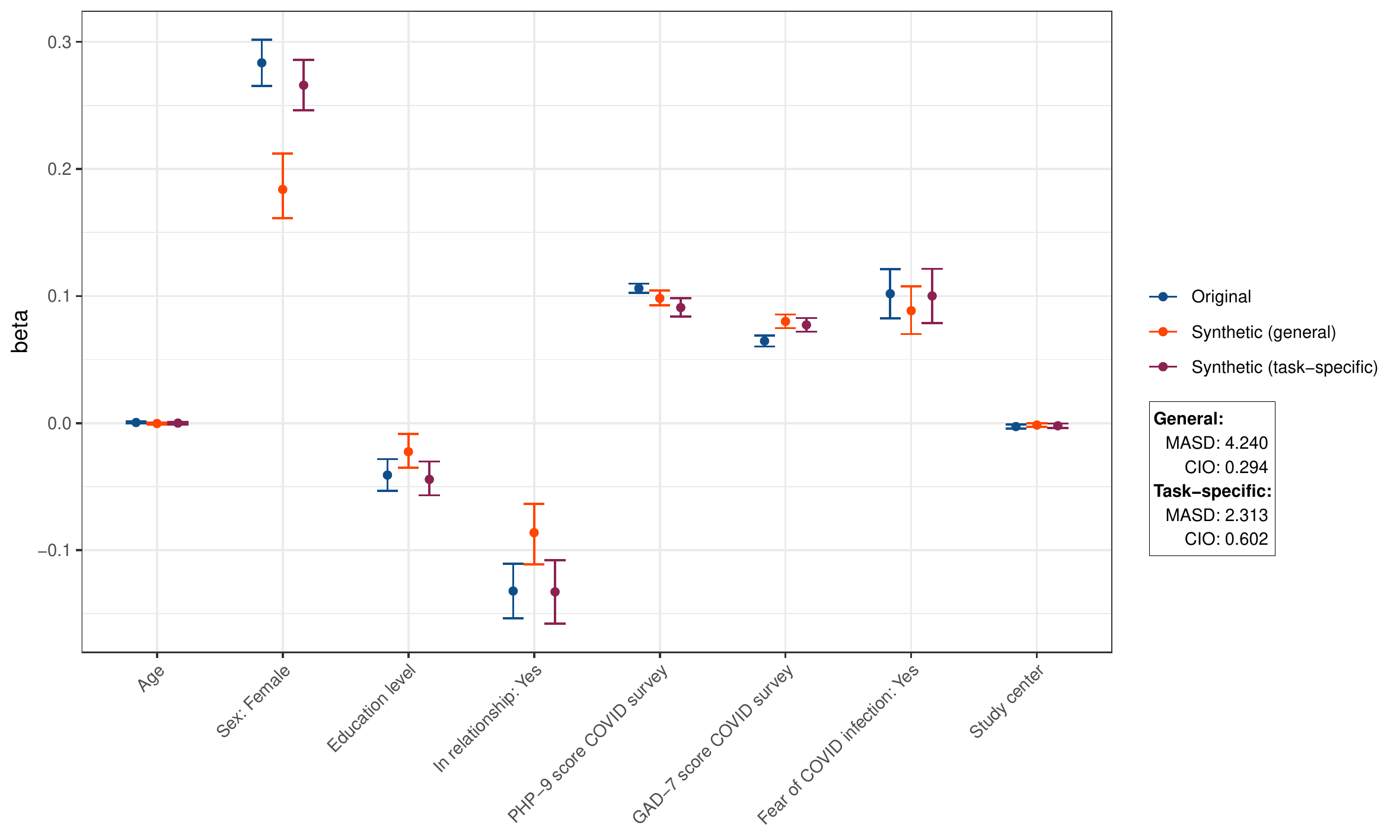}
    \caption{Relationship between perceived loneliness and sociodemographic factors as well as symptoms of depression and anxiety among German National Cohort (NAKO Gesundheitsstudie) participants in May 2020. Median beta estimates and percentile-based 95\% confidence intervals computed from the beta coefficient distribution across synthesis repetitions are reported for synthetic data. PHQ-9, nine-item Patient Health Questionnaire; GAD-7, Generalized Anxiety Disorder seven-item scale; MASD, mean absolute standardized difference; CIO, confidence interval overlap}
    \label{fig:Berger_reg}
\end{figure}

\newpage

\subsubsection{Replication of Figure 1}

\begin{longtable}[l]{llll}
\hline
Sex & Age group & Lonely participants (proportion) & Participants with increased loneliness (proportion) \\ 
  \hline
Male & 20-29 & 0.33 (0.32-0.35) & 0.58 (0.57-0.60) \\ 
   &  & \textcolor{orangered}{0.33 (0.31-0.35)} & \textcolor{orangered}{0.57 (0.55-0.59)} \\ 
   & 30-39 & 0.29 (0.28-0.30) & 0.50 (0.48-0.51) \\ 
   &  & \textcolor{orangered}{0.30 (0.29-0.31)} & \textcolor{orangered}{0.51 (0.50-0.53)} \\ 
   & 40-59 & 0.26 (0.25-0.26) & 0.45 (0.44-0.46) \\ 
   &  & \textcolor{orangered}{0.26 (0.25-0.27)} & \textcolor{orangered}{0.46 (0.45-0.47)} \\ 
   & 50-59 & 0.24 (0.23-0.25) & 0.41 (0.40-0.42) \\ 
   &  & \textcolor{orangered}{0.24 (0.24-0.25)} & \textcolor{orangered}{0.41 (0.41-0.42)} \\ 
   & 60-69 & 0.23 (0.23-0.24) & 0.36 (0.35-0.36) \\ 
   &  & \textcolor{orangered}{0.23 (0.22-0.23)} & \textcolor{orangered}{0.36 (0.35-0.37)} \\ 
   & 70+ & 0.23 (0.21-0.26) & 0.32 (0.29-0.34) \\ 
   &  & \textcolor{orangered}{0.20 (0.16-0.24)} & \textcolor{orangered}{0.30 (0.25-0.34)} \\ 
  Female & 20-29 & 0.47 (0.45-0.48) & 0.65 (0.64-0.66) \\ 
   &  & \textcolor{orangered}{0.43 (0.41-0.45)} & \textcolor{orangered}{0.63 (0.61-0.65)} \\ 
   & 30-39 & 0.44 (0.43-0.45) & 0.59 (0.58-0.61) \\ 
   &  & \textcolor{orangered}{0.40 (0.39-0.42)} & \textcolor{orangered}{0.58 (0.57-0.59)} \\ 
   & 40-59 & 0.35 (0.34-0.36) & 0.51 (0.50-0.51) \\ 
   &  & \textcolor{orangered}{0.36 (0.35-0.37)} & \textcolor{orangered}{0.52 (0.51-0.53)} \\ 
   & 50-59 & 0.35 (0.34-0.35) & 0.48 (0.47-0.49) \\ 
   &  & \textcolor{orangered}{0.34 (0.33-0.35)} & \textcolor{orangered}{0.48 (0.47-0.49)} \\ 
   & 60-69 & 0.37 (0.36-0.38) & 0.46 (0.45-0.47) \\ 
   &  & \textcolor{orangered}{0.33 (0.31-0.34)} & \textcolor{orangered}{0.44 (0.43-0.45)} \\ 
   & 70+ & 0.35 (0.32-0.38) & 0.40 (0.37-0.44) \\ 
   &  & \textcolor{orangered}{0.29 (0.24-0.35)} & \textcolor{orangered}{0.37 (0.31-0.42)} \\ 
   \hline

\textbf{MASD: $2.607$; CIO: $0.499$}
\\
\hline
\caption{Proportion of lonely participants and of those, who felt more lonely during the pandemic than before, by sex and age group (in years). Percentile-based 95\% bootstrap confidence intervals are reported for original data results. Median and percentile-based 95\% confidence intervals of synthetic data results are printed in orange; MASD, mean absolute standardized difference; CIO, confidence interval overlap}
\label{tab:Berger_fig1}
\end{longtable}

\begin{figure}[ht]
    \centering
    \includegraphics[width=\columnwidth]{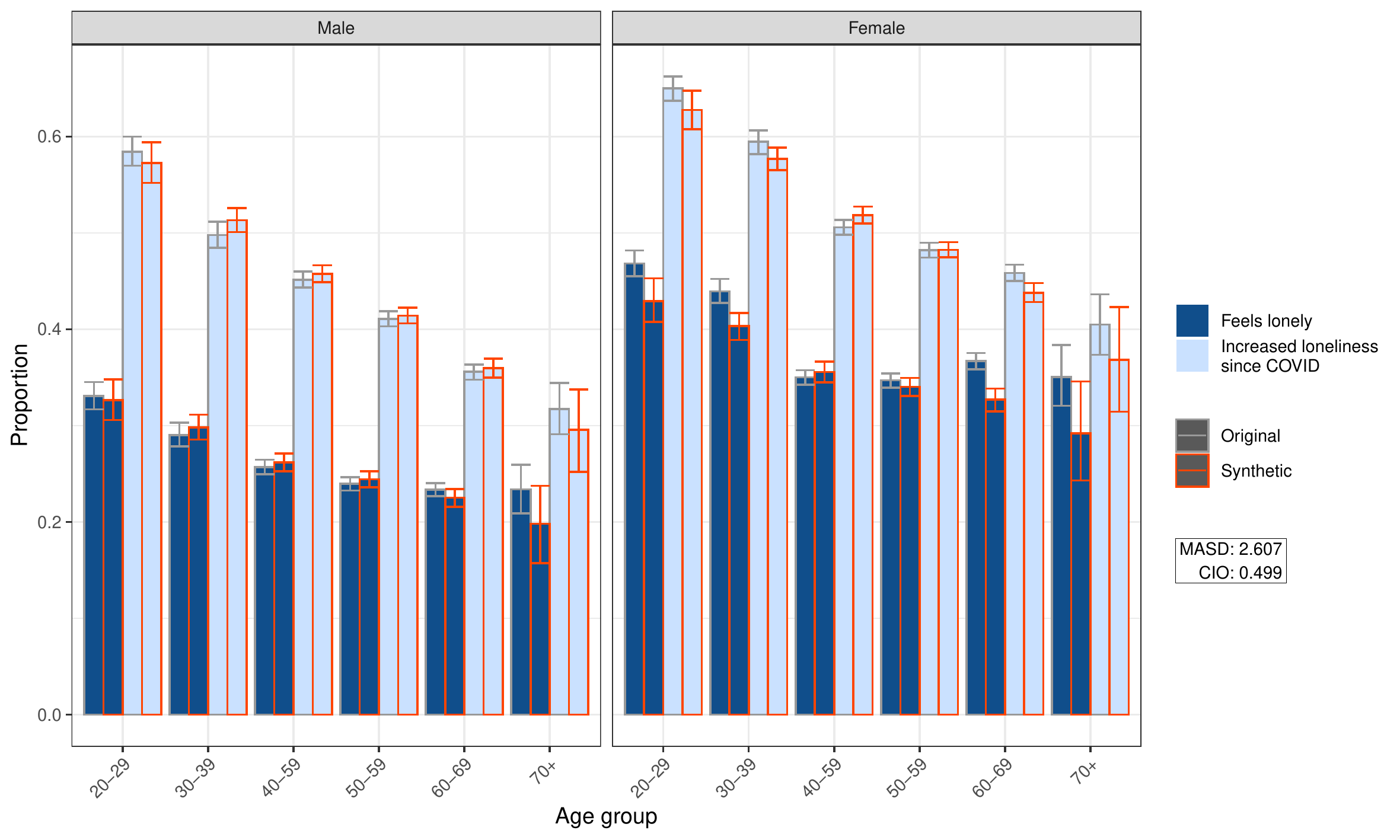}
    \caption{Proportion of lonely participants and of those, who felt more lonely during the pandemic than before, by sex and age group (years). Percentile-based 95\% bootstrap confidence intervals are reported for original data results. Median and percentile-based 95\% confidence intervals of synthetic data results are reported; MASD, mean absolute standardized difference; CIO, confidence interval overlap}
    \label{fig:Berger_fig1}
\end{figure}

\newpage

\subsubsection{Replication of Figure 2}

\begin{longtable}[l]{llll}
\hline
Sex & Loneliness score & PHQ-9 score COVID survey (mean) & GAD-7 score COVID survey (mean) \\ 
  \hline
Male & 3 & 2.56 (2.51-2.61) & 2.01 (1.96-2.05) \\ 
   &  & \textcolor{orangered}{2.38 (2.28-2.46)} & \textcolor{orangered}{1.87 (1.79-1.94)} \\ 
   & 4 & 2.63 (2.59-2.67) & 2.09 (2.06-2.13) \\ 
   &  & \textcolor{orangered}{2.65 (2.59-2.70)} & \textcolor{orangered}{2.11 (2.06-2.15)} \\ 
   & 5 & 3.46 (3.39-3.53) & 2.77 (2.71-2.83) \\ 
   &  & \textcolor{orangered}{3.62 (3.54-3.70)} & \textcolor{orangered}{2.93 (2.87-3.00)} \\ 
   & 6 & 4.77 (4.68-4.88) & 3.82 (3.74-3.90) \\ 
   &  & \textcolor{orangered}{4.69 (4.60-4.81)} & \textcolor{orangered}{3.83 (3.73-3.92)} \\ 
   & 7 & 5.44 (5.31-5.58) & 4.31 (4.21-4.43) \\ 
   &  & \textcolor{orangered}{5.69 (5.52-5.89)} & \textcolor{orangered}{4.63 (4.49-4.78)} \\ 
   & 8 & 6.39 (6.17-6.64) & 5.09 (4.90-5.28) \\ 
   &  & \textcolor{orangered}{6.82 (6.52-7.20)} & \textcolor{orangered}{5.54 (5.29-5.85)} \\ 
   & 9 & 9.03 (8.66-9.42) & 7.15 (6.84-7.47) \\ 
   &  & \textcolor{orangered}{8.49 (7.93-9.11)} & \textcolor{orangered}{6.87 (6.43-7.36)} \\ 
  Female & 3 & 3.34 (3.26-3.41) & 2.89 (2.82-2.96) \\ 
   &  & \textcolor{orangered}{3.03 (2.93-3.13)} & \textcolor{orangered}{2.55 (2.45-2.64)} \\ 
   & 4 & 3.23 (3.18-3.27) & 2.80 (2.76-2.84) \\ 
   &  & \textcolor{orangered}{3.27 (3.21-3.33)} & \textcolor{orangered}{2.76 (2.71-2.81)} \\ 
   & 5 & 4.22 (4.15-4.29) & 3.62 (3.55-3.68) \\ 
   &  & \textcolor{orangered}{4.32 (4.24-4.40)} & \textcolor{orangered}{3.67 (3.60-3.73)} \\ 
   & 6 & 5.24 (5.15-5.32) & 4.47 (4.39-4.54) \\ 
   &  & \textcolor{orangered}{5.35 (5.25-5.45)} & \textcolor{orangered}{4.54 (4.45-4.63)} \\ 
   & 7 & 6.12 (6.02-6.23) & 5.14 (5.04-5.23) \\ 
   &  & \textcolor{orangered}{6.33 (6.19-6.51)} & \textcolor{orangered}{5.35 (5.23-5.50)} \\ 
   & 8 & 7.24 (7.08-7.40) & 6.11 (5.97-6.26) \\ 
   &  & \textcolor{orangered}{7.51 (7.28-7.79)} & \textcolor{orangered}{6.31 (6.11-6.56)} \\ 
   & 9 & 9.22 (9.01-9.42) & 7.62 (7.44-7.80) \\ 
   &  & \textcolor{orangered}{8.94 (8.53-9.39)} & \textcolor{orangered}{7.45 (7.09-7.82)} \\ 
   \hline

\textbf{MASD: $3.464$; CIO: $0.368$}
\\
\hline
\caption{Relationship of drepression and anxiety symptoms with loneliness. Percentile-based 95\% bootstrap confidence intervals are reported for original data results. Median and percentile-based 95\% confidence intervals of synthetic data results are printed in orange. PHQ-9, nine-item Patient Health Questionnaire; GAD-7, Generalized Anxiety Disorder seven-item scale; MASD, mean absolute standardized difference; CIO, confidence interval overlap}
\label{tab:Berger_fig2}
\end{longtable}

\begin{figure}[ht]
    \centering
    \includegraphics[width=\columnwidth]{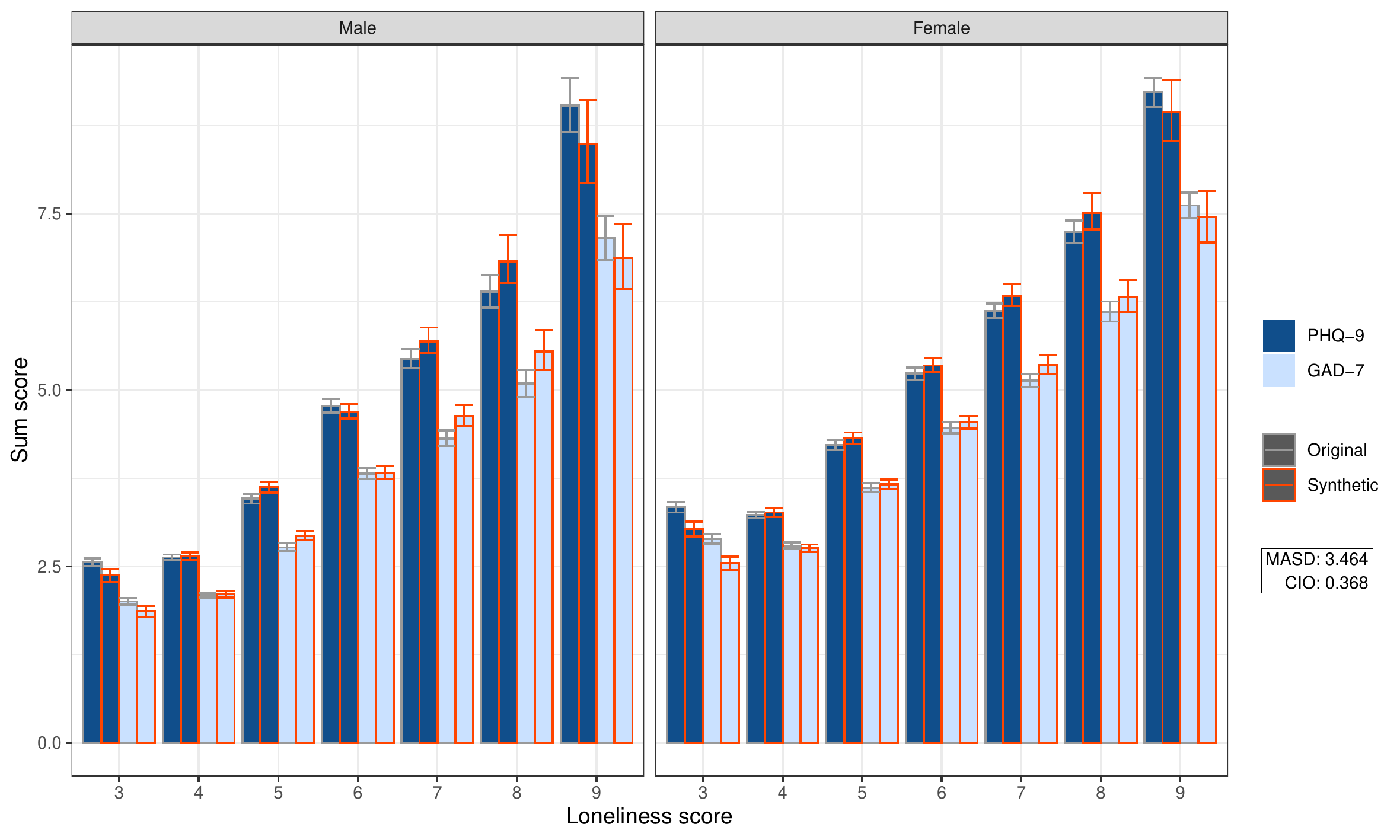}
    \caption{Relationship of depression and anxiety symptoms with loneliness. Percentile-based 95\% bootstrap confidence intervals are reported for original data results. Median and percentile-based 95\% confidence intervals of synthetic data results are reported. PHQ-9, 9-item Patient Health Questionnaire; GAD-7, Generalized Anxiety Disorder seven-item scale; MASD, mean absolute standardized difference; CIO, confidence interval overlap}
    \label{fig:Berger_fig2}
\end{figure}

\newpage

\subsubsection{Replication of Figure 3}

\begin{longtable}[l]{llll}
\hline
Sex & Lonely & PHQ-9 score baseline examination (mean) & GAD-7 score baseline examination (mean) \\ 
  \hline
Male & Yes & 4.17 (4.10-4.24) & 3.37 (3.31-3.42) \\ 
   &  & \textcolor{orangered}{4.19 (4.11-4.28)} & \textcolor{orangered}{3.42 (3.35-3.49)} \\ 
   & No & 2.90 (2.87-2.93) & 2.30 (2.27-2.32) \\ 
   &  & \textcolor{orangered}{2.93 (2.89-2.97)} & \textcolor{orangered}{2.33 (2.30-2.36)} \\ 
  Female & Yes & 4.79 (4.73-4.84) & 4.01 (3.96-4.06) \\ 
   &  & \textcolor{orangered}{4.76 (4.68-4.85)} & \textcolor{orangered}{3.98 (3.91-4.06)} \\ 
   & No & 3.73 (3.69-3.76) & 3.08 (3.05-3.11) \\ 
   &  & \textcolor{orangered}{3.66 (3.61-3.71)} & \textcolor{orangered}{3.01 (2.96-3.05)} \\ 
   \hline

\textbf{MASD: $2.157$; CIO: $0.542$}
\\
\hline
\caption{Baseline PHQ-9 and GAD-7 sum scores for lonely and non-lonely participants during the pandemic. Percentile-based 95\% bootstrap confidence intervals are reported for original data results. Median and percentile-based 95\% confidence intervals of synthetic data results are printed in orange. PHQ-9, nine-item Patient Health Questionnaire; GAD-7, Generalized Anxiety Disorder seven-item scale; MASD, mean absolute standardized difference; CIO, confidence interval overlap}
\label{tab:Berger_fig3}
\end{longtable}

\begin{figure}[ht]
    \centering
    \includegraphics[width=\columnwidth]{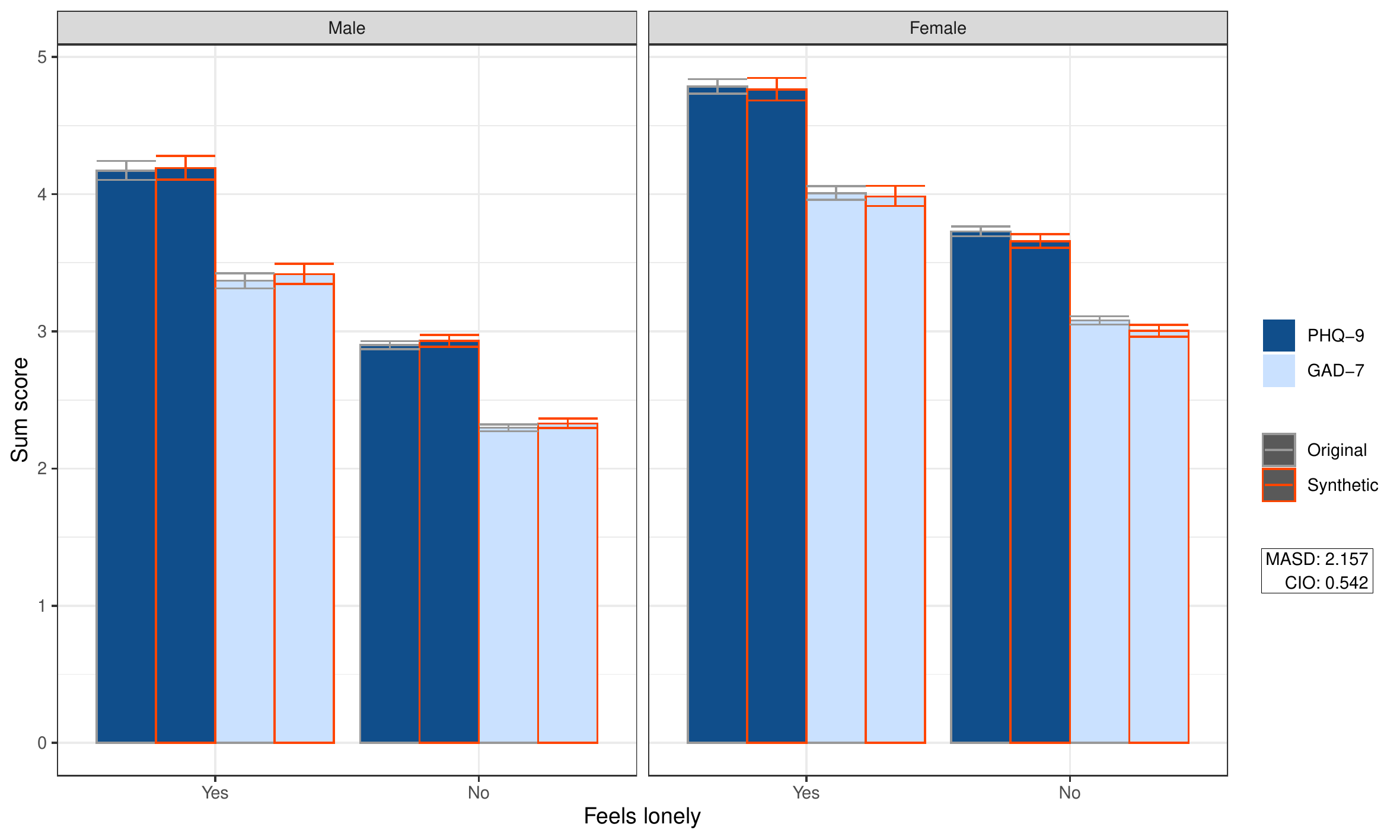}
    \caption{Baseline PHQ-9 and GAD-7 sum scores for lonely and non-lonely participants during the pandemic. Percentile-based 95\% bootstrap confidence intervals are reported for original data results. Median and percentile-based 95\% confidence intervals of synthetic data results are reported. PHQ-9, nine-item Patient Health Questionnaire; GAD-7, Generalized Anxiety Disorder seven-item scale; MASD, mean absolute standardized difference; CIO, confidence interval overlap}
    \label{fig:Berger_fig3}
\end{figure}

\newpage


\subsection{Tanoey et al. \texorpdfstring{\cite{tanoey2022diabetes}}{[Tanoey et al., 2022]}}
\label{appx:B6}

\subsubsection{Replication of Table 1 (general and task-specific synthesis)}

\begin{longtable}[l]{llllll}
\hline
Variable & Agg. & No diabetes & Diabetes type 1 & Diabetes type 2 & Total \\ 
  \hline
Number of participants (n) &  & 94408 (94252-94569) & 371 (331-410) & 6574 (6420-6724) & 101353 (101324-101381) \\ 
   &  & \textcolor{orangered}{94407 (94250-94562)} & \textcolor{orangered}{311 (270-351)} & \textcolor{orangered}{5847 (5477-6188)} & \textcolor{orangered}{100576 (100217-100884)} \\ 
   &  & \textcolor{violetred4}{94411 (94252-94563)} & \textcolor{violetred4}{371 (333-412)} & \textcolor{violetred4}{6571 (6429-6727)} & \textcolor{violetred4}{101353 (101353-101353)} \\ 
  Sex: Male & n & 43244 (42932-43555) & 204 (175-233) & 3602 (3489-3719) & 47050 (46735-47362) \\ 
   &  & \textcolor{orangered}{43267 (42956-43586)} & \textcolor{orangered}{156 (130-183)} & \textcolor{orangered}{3164 (2942-3397)} & \textcolor{orangered}{46587 (46222-46949)} \\ 
   &  & \textcolor{violetred4}{43264 (42953-43555)} & \textcolor{violetred4}{199 (171-227)} & \textcolor{violetred4}{3588 (3477-3707)} & \textcolor{violetred4}{47046 (46737-47346)} \\ 
   & Prop. & 0.46 (0.45-0.46) & 0.55 (0.50-0.60) & 0.55 (0.54-0.56) & 0.46 (0.46-0.47) \\
\end{longtable}

\newpage

\subsubsection{Replication of Table 2}

\begin{longtable}[l]{llllll}
\hline
Variable & Model & Age at diagnosis & Hazard ratio & 2.5\% & 97.5\% \\ 
  \hline
Birth order: First & Univariable & 0–40 & 0.83 (0.61-1.14) & 0.61 (0.45-0.82) & 1.12 (0.83-1.57) \\ 
   &  &  & \textcolor{orangered}{1.06 (0.72-1.63)} & \textcolor{orangered}{0.72 (0.49-1.04)} & \textcolor{orangered}{1.58 (1.05-2.56)} \\ 
  Birth order: Second &  &  & 0.69 (0.50-0.96) & 0.50 (0.36-0.69) & 0.94 (0.69-1.34) \\ 
   &  &  & \textcolor{orangered}{0.97 (0.65-1.51)} & \textcolor{orangered}{0.65 (0.44-0.98)} & \textcolor{orangered}{1.44 (0.95-2.37)} \\ 
  Birth order: $>$=Third &  &  & 0.66 (0.45-0.96) & 0.46 (0.30-0.67) & 0.95 (0.66-1.38) \\ 
   &  &  & \textcolor{orangered}{0.97 (0.63-1.51)} & \textcolor{orangered}{0.63 (0.41-0.96)} & \textcolor{orangered}{1.48 (0.95-2.39)} \\ 
  Birth order: Unknown &  &  & 1.07 (0.76-1.50) & 0.77 (0.54-1.07) & 1.49 (1.06-2.12) \\
\end{longtable}

\newpage

\subsubsection{Replication of univariable Cox regressions - Table 2 (general and task-specific synthesis)}

\begin{longtable}[l]{llll}
\hline
Variable & Hazard ratio & 2.5\% & 97.5\% \\ 
  \hline
Birth order: First & 0.83 (0.61-1.14) & 0.61 (0.45-0.82) & 1.12 (0.83-1.57) \\ 
   & \textcolor{orangered}{1.06 (0.72-1.63)} & \textcolor{orangered}{0.72 (0.49-1.04)} & \textcolor{orangered}{1.58 (1.05-2.56)} \\ 
   & \textcolor{violetred4}{0.84 (0.62-1.17)} & \textcolor{violetred4}{0.62 (0.46-0.85)} & \textcolor{violetred4}{1.14 (0.85-1.61)} \\ 
  Birth order: Second & 0.69 (0.49-0.96) & 0.50 (0.36-0.69) & 0.94 (0.68-1.35) \\ 
   & \textcolor{orangered}{0.97 (0.65-1.51)} & \textcolor{orangered}{0.65 (0.44-0.98)} & \textcolor{orangered}{1.44 (0.95-2.37)} \\ 
   & \textcolor{violetred4}{0.70 (0.51-0.97)} & \textcolor{violetred4}{0.51 (0.37-0.70)} & \textcolor{violetred4}{0.97 (0.71-1.35)} \\ 
  Birth order: $>$=Third & 0.66 (0.46-0.95) & 0.46 (0.31-0.66) & 0.95 (0.67-1.38) \\
\end{longtable}

\begin{figure}[ht]
    \centering
    \includegraphics[width=\columnwidth]{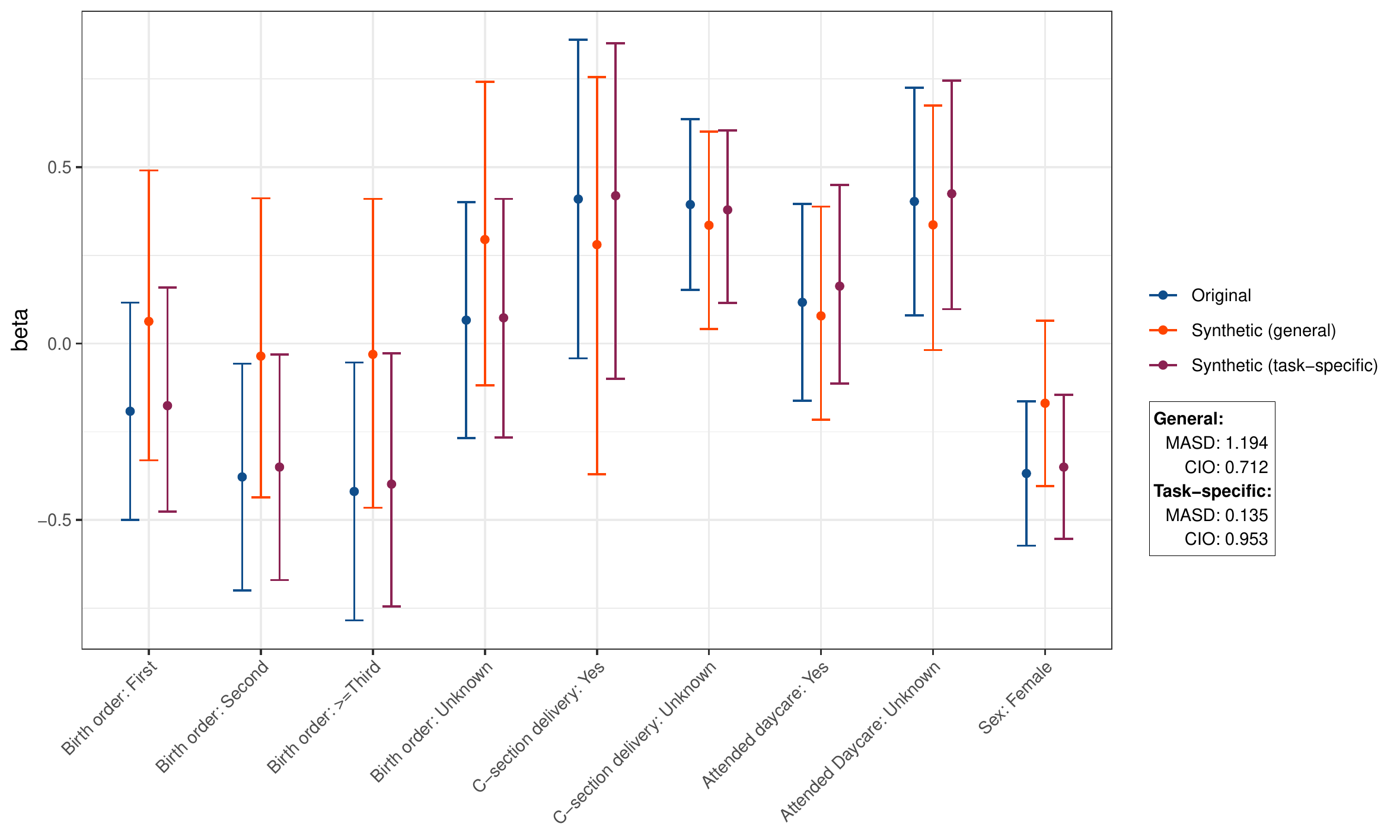}
    \caption{Type 1 diabetes univariable Cox regression estimates. Median beta estimates and percentile-based 95\% confidence intervals computed from the beta coefficient distribution across synthesis repetitions are reported for synthetic data. C-section, caesarean section; MASD, mean absolute standardized difference; CIO, confidence interval overlap}
    \label{fig:Tanoey_reg_taskspec}
\end{figure}

\newpage

\subsubsection{Replication of Figure 2 (general and task-specific synthesis)}

\begin{longtable}[l]{llll}
\hline
Sex & Age at diagnosis & Therapy & n \\ 
  \hline
Male & 0-15 & With insulin & 36 (25-48) \\ 
   &  &  & \textcolor{orangered}{15 (8-24)} \\ 
   &  &  & \textcolor{violetred4}{42 (28-56)} \\ 
   &  & Unknown & 13 (6-21) \\ 
   &  &  & \textcolor{orangered}{6 (2-12)} \\ 
   &  &  & \textcolor{violetred4}{12 (6-20)} \\ 
   & 16-30 & With insulin & 74 (58-91) \\ 
   &  &  & \textcolor{orangered}{44 (31-58)} \\ 
   &  &  & \textcolor{violetred4}{73 (56-92)} \\ 
   &  & Unknown & 6 (2-11) \\ 
   &  &  & \textcolor{orangered}{11 (5-18)} \\ 
   &  &  & \textcolor{violetred4}{7 (2-13)} \\ 
   & 31-40 & With insulin & 64 (49-80) \\ 
   &  &  & \textcolor{orangered}{62 (45-80)} \\ 
   &  &  & \textcolor{violetred4}{58 (44-76)} \\ 
   &  & Unknown & 11 (5-18) \\ 
   &  &  & \textcolor{orangered}{17 (9-27)} \\ 
   &  &  & \textcolor{violetred4}{10 (4-17)} \\ 
  Female & 0-15 & With insulin & 45 (32-59) \\ 
   &  &  & \textcolor{orangered}{18 (10-28)} \\ 
   &  &  & \textcolor{violetred4}{42 (29-55)} \\ 
   &  & Unknown & 6 (2-11) \\ 
   &  &  & \textcolor{orangered}{4 (1-8)} \\ 
   &  &  & \textcolor{violetred4}{6 (2-12)} \\ 
   & 16-30 & With insulin & 55 (41-70) \\ 
   &  &  & \textcolor{orangered}{42 (28-56)} \\ 
   &  &  & \textcolor{violetred4}{61 (46-80)} \\ 
   &  & Unknown & 5 (1-10) \\ 
   &  &  & \textcolor{orangered}{15 (8-24)} \\ 
   &  &  & \textcolor{violetred4}{5 (1-11)} \\ 
   & 31-40 & With insulin & 46 (33-60) \\ 
   &  &  & \textcolor{orangered}{51 (37-67)} \\ 
   &  &  & \textcolor{violetred4}{42 (30-56)} \\ 
   &  & Unknown & 10 (4-17) \\ 
   &  &  & \textcolor{orangered}{24 (14-38)} \\ 
   &  &  & \textcolor{violetred4}{9 (4-17)} \\ 
   \hline

\textbf{General: \hspace{13pt} MASD: $2.441$; CIO: $0.445$}\\
\textbf{Task-specific: MASD: $0.417$; CIO: $0.903$}
\\
\hline
\caption{Numbers of type 1 diabetes cases by sex, age at diagnosis (in years) and therapy. Percentile-based 95\% bootstrap confidence intervals are reported for original data results. Median and percentile-based 95\% confidence intervals of synthetic data results are reported, general synthesis replication results are printed in orange, task-specific synthesis replication results in violet. n, number of participants; MASD, mean absolute standardized difference; CIO, confidence interval overlap}
\label{tab:Tanoey_fig2_taskspec}
\end{longtable}

\newpage

\begin{figure}[ht]
    \centering
    \includegraphics[width=\columnwidth]{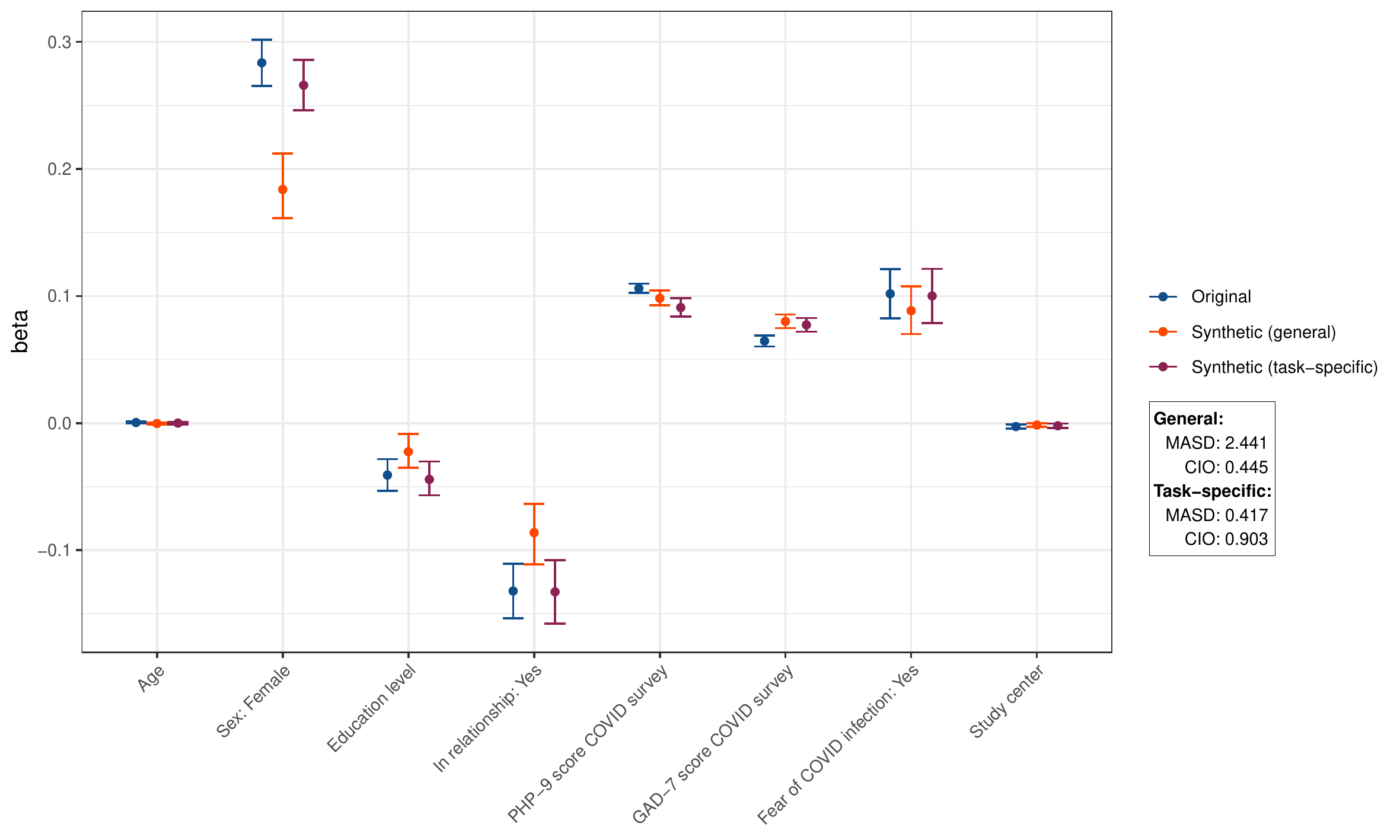}
    \caption{Numbers of participants with type 1 diabetes by sex, age at diagnosis (in years) and therapy. Percentile-based 95\% bootstrap confidence intervals are reported for original data results. Median and percentile-based 95\% confidence intervals of synthetic data results are reported. n, number of participants; MASD, mean absolute standardized difference; CIO, confidence interval overlap}
    \label{fig:Tanoey_fig2_taskspec}
\end{figure}

\newpage


\section{Comparative analysis of ARF performance across varying minimum node sizes and alternative synthesizers}
\label{appx:C}

{\normalsize

The evaluation in the main text focuses on the reproducibility of published epidemiological analyses using ARF-generated synthetic data, providing an interpretable and application-oriented assessment of utility. However, synthetic data quality is inherently multidimensional, and downstream utility represents only one aspect. Important additional dimensions include privacy risk and generalization to unseen data. In this section, we therefore provide a more comprehensive evaluation of the synthetic data generated by ARF, including an assessment of privacy risk (\cref{sec:appx_upto}) and generalization to out-of-sample data (\cref{sec:appx_gen}). In addition, we report a runtime analysis of the data synthesis process (\cref{sec:appx_runtime}).

In all experiments, we compared results across different values of the ARF minimum node size (2, 5, 10, 20, and 50), with 2 corresponding to the default setting. While random forests are generally considered relatively robust to hyperparameter choices in discriminative modelling \cite{probst2019RFhyperparameters}, this parameter controlling the granularity of the tree partitions directly influences the utility–privacy trade-off in generative modelling with ARF. In general, despite its potential benefits, we note that systematic hyperparameter optimization is often omitted in generative contexts due to the substantial computational overhead and the complexity of the evaluation process \cite{kindji2025tuning}. Results were further compared with five additional synthetic data generators to contextualize the findings and position ARF’s performance relative to alternative synthesis approaches. Specifically, we included \emph{synthpop} \cite{nowok2016synthpop}, \emph{Bayesian Networks} \cite{koller2009BN} and their differentially private \cite{dwork2014DP} variant \emph{PrivBayes} \cite{zhang2017PrivBayes}, as well as the deep learning–based approaches \emph{CTGAN} \cite{xu2019CTGAN} and \emph{TVAE} \cite{xu2019CTGAN}. These represent popular methods for tabular data synthesis, available in well-maintained packages on CRAN or PyPI and able to handle missing data. Recent promising deep learning approaches for tabular data synthesis, such as diffusion-based models \cite{zhang2024TabSyn}, do not currently meet these criteria. As both privacy and generalization must be evaluated on test data excluded from synthesizer training, all experiments in this section were conducted on five different 50/50 train-test splits.

}


\subsection{Utility-privacy trade-off}
\label{sec:appx_upto}

\begin{figure}[ht]
    \centering
    \includegraphics[width=\columnwidth]{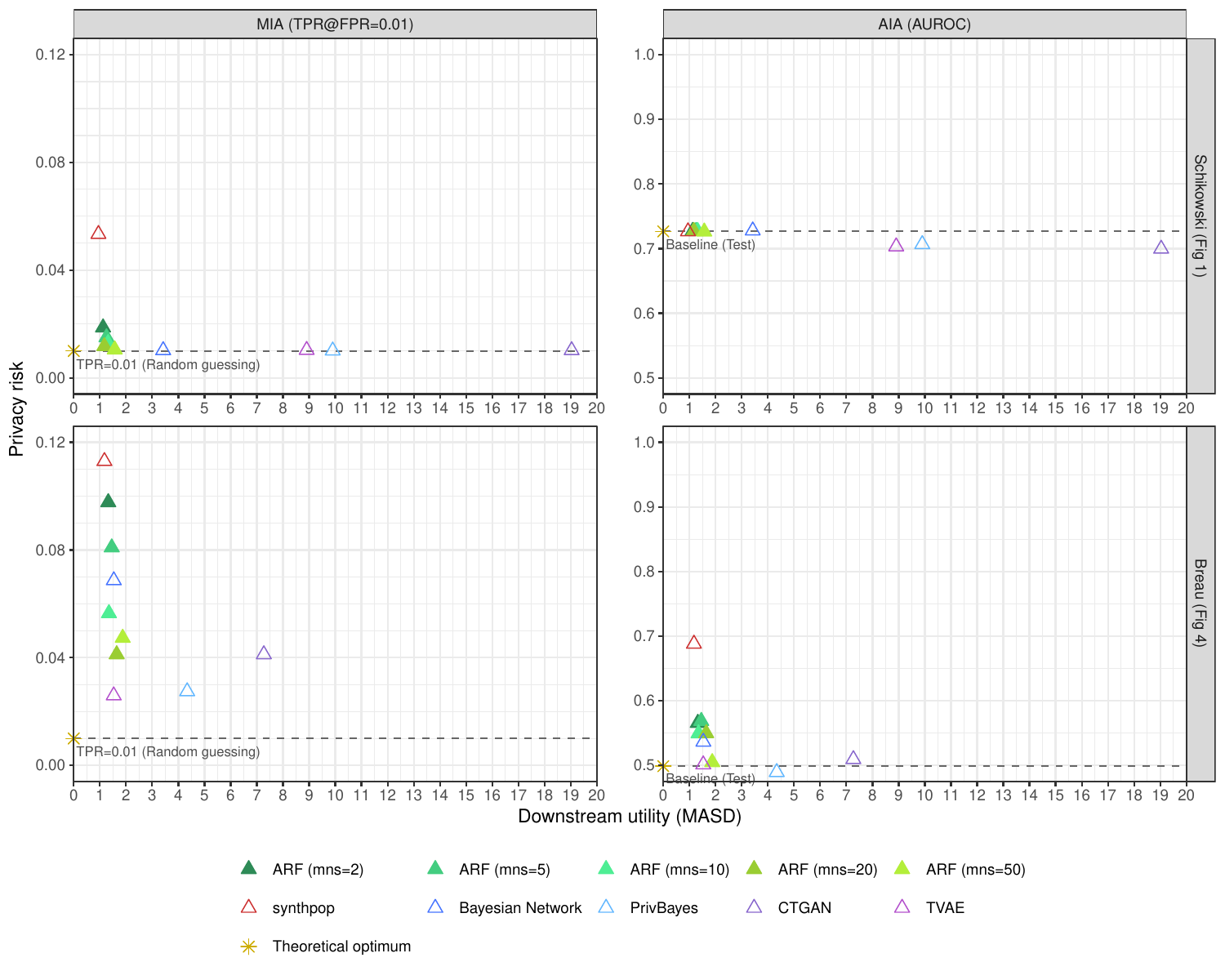}
    \caption{Utility (MASD) for downstream descriptive analyses from Schikowski et al.\@~\texorpdfstring{\cite{schikowski2020blutdruckmessung}}{[Schikowski et al., 2020]} (\texorpdfstring{\cref{fig:Schikwoski2}}{Figure 1}) and Breau et al.\@~\texorpdfstring{\cite{breau2022cutpoint}}{[Breau et al., 2022]} (\texorpdfstring{\cref{fig:Breau2}}{Figure 4}) vs. privacy risk MIA/AIA. Comparison of ARF performance (downstream utility: MASD; privacy risk: MIA \& AIA performance) across five minimum node sizes and five alternative synthesizers. Mean values over five synthesizer training repetitions are reported. ARF, adversarial random forests; MASD, mean absolute standardized difference; MIA, membership inference attack; TPR@FPR=0.01, attack true positive rate at a false positive rate of $0.01$; AIA, attribute inference attack; AUROC, area under the receiver operating characteristic curve, Fig, figure; mns, minimum node size of random forest}
    \label{fig:DU_MASD}
\end{figure}

\begin{figure}[ht]
    \centering
    \includegraphics[width=\columnwidth]{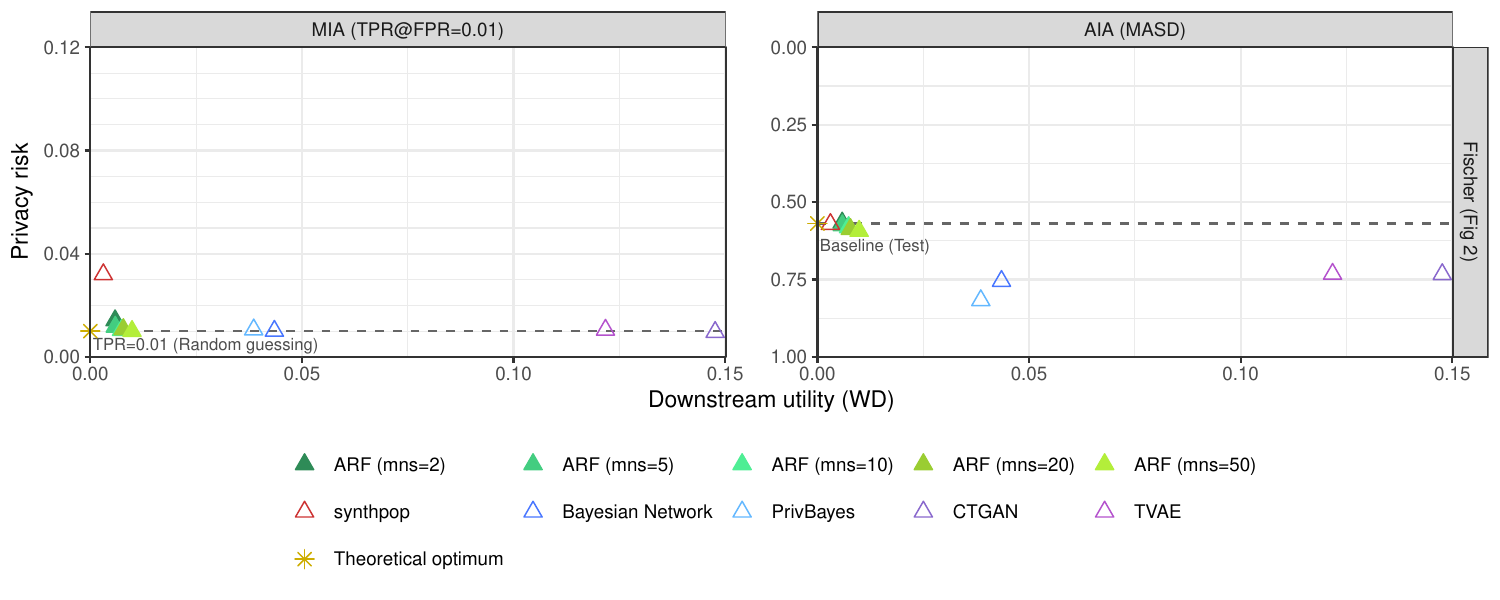}
    \caption{Utility (WD) for downstream multivariate distribution analysis from Fischer et al.\@~\texorpdfstring{\cite{fischer2020anthropometrisch}}{[Fischer et al., 2020]} (\texorpdfstring{\cref{fig:Fischer5}}{Figure 2}) vs. privacy risk MIA/AIA. Comparison of ARF performance (downstream utility: WD; privacy risk: MIA \& AIA performance) across five minimum node sizes and five alternative synthesizers. Mean values over five synthesizer training repetitions are reported. WD, multivariate Wasserstein distance; MASD, mean absolute standardized difference; MIA, membership inference attack; TPR@FPR=0.01, attack true positive rate at a false positive rate of $0.01$; AIA, attribute inference attack; AUROC, area under the receiver operating characteristic curve, Fig, figure; mns, minimum node size of random forest}
    \label{fig:DU_WD}
\end{figure}

\begin{figure}[h!]
    \centering
    \includegraphics[width=\columnwidth]{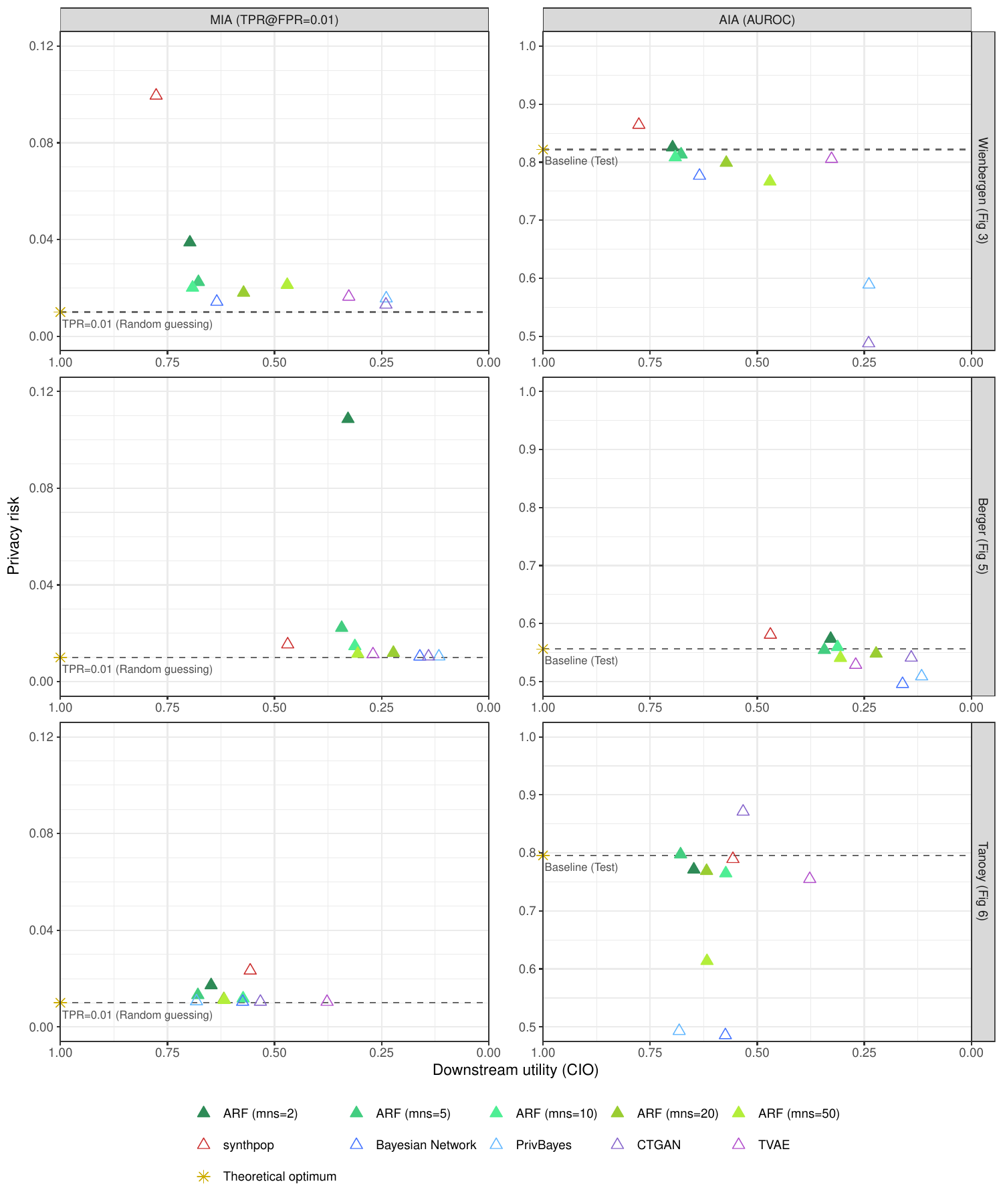}
    \caption{Utility (CIO) for downstream regression analyses from Wienbergen et al.\@~\texorpdfstring{\cite{wienbergen2022infarction}}{[Wienbergen et al., 2022]} (\texorpdfstring{\cref{fig:Wienbergen_reg}}{Figure 5}), Berger et al.\@~\texorpdfstring{\cite{berger2021COVID}}{[Berger et al., 2022]} (\texorpdfstring{\cref{fig:berger_reg}}{Figure 5}) and Tanoey et al.\@~\texorpdfstring{\cite{tanoey2022diabetes}}{[Tanoey et al., 2022]} (\texorpdfstring{\cref{fig:tanoey_reg}}{Figure 6}) vs. privacy risk MIA/AIA. Comparison of ARF performance (downstream utility: CIO; privacy risk: MIA \& AIA performance) across five minimum node sizes and five alternative synthesizers. Mean values over five synthesizer training repetitions are reported. CIO, mean confidence interval overlap; MIA, membership inference attack; TPR@FPR=0.01, attack true positive rate at a false positive rate of $0.01$; AIA, attribute inference attack; AUROC, area under the receiver operating characteristic curve, Fig, figure; mns, minimum node size of random forest}
    \label{fig:DU_CIO}
\end{figure}

{\normalsize

In this section, the utility dimension was again evaluated by replicating the downstream analyses in \crefrange{fig:Schikwoski2}{fig:berger_reg}. For descriptive analyses, we report the mean absolute standardized difference (MASD) and the the mean confidence interval overlap (CIO) for regression analyses. We refer to \cref{sec:appx_A} for a definition of these metrics. For MASD calcuation, we quantified the standard error of the original estimates again via bootrapping with 2\,000 repetitions. The calculation of the CIO was based on the confidence intervals obtained from the regression models. For the distribution analysis in \cref{fig:Fischer5}, we computed the mean of the Wasserstein distances (WD) \cite{villani2021WD} between training and synthetic data over all four subgroups. Average values over all five repetitions were reported.

To assess privacy risk, we performed distance-, local density-, and classifier-based membership inference attacks (MIAs) \cite{ward2026MIA}, as well as classifier-based attribute inference attacks (AIAs) \cite{stadler2022groundhog}. Membership inference attacks aim to determine whether a given data point was part of the training data used to fit the synthesizer. Attribute inference attacks attempt to infer the value of a sensitive target variable given a set of key variables, which are typically assumed to be publicly available or easily accessible. The selection of key and target variables for each dataset is provided in \cref{stab:AIA}.

\begin{table}[h]
\small
\centering
\begin{tabular}{lll}
\hline
\textbf{Publication} & \textbf{Key variables} & \textbf{Target variable} \\ \hline
Schikowski et al. \cite{schikowski2020blutdruckmessung} & \texttt{Age}, \texttt{Sex} & \texttt{Hypertension} \\
Fischer et al. \cite{fischer2020anthropometrisch} & \texttt{Age}, \texttt{Sex}, \texttt{Height}, \texttt{Weight}, \texttt{Study Center} & 	\texttt{VAT} \\
Wienbergen et al.\@~\cite{wienbergen2022infarction} & \texttt{Age}, \texttt{Sex}, \texttt{BMI}, \texttt{Education}, \texttt{Birth Germany} & \texttt{EOMI} \\
Breau et al. \cite{breau2022cutpoint} & \texttt{Age}, \texttt{Sex}, \texttt{Weight}, \texttt{Height}, \texttt{Ethnicity} & \texttt{Income Household} \\
Berger et al.\@~\cite{berger2021COVID} & \texttt{Age}, \texttt{Sex}, \texttt{Study Center}, \texttt{Education} & \texttt{Suicidal Ideation} \\
Tanoey et al. \cite{tanoey2022diabetes} & \texttt{Age}, \texttt{Sex}, \texttt{Height}, \texttt{Migration Background} & \texttt{Diabetes} \\
\hline \hline
\end{tabular}
\caption{Overview of key and target variables for each publication used for attribtute inference attacks. VAT, visceral adipose tissue; EOMI, early-onset myocardial infarction}
\label{stab:AIA}
\end{table}

The success of MIAs was evaluated using the true positive rate at a false positive rate of $0.01$ (TPR@FPR=0.01). A value of $0.01$ corresponds to performance equivalent to random guessing, while higher values indicate an increased privacy risk. In practice, the realistic privacy risk depends on the prevalence of training records in the underlying population and can therefore be substantially lower when the training dataset represents only a small fraction of that population. Furthermore, these attacks assume that an adversary has access to a reference dataset from the population. In our evaluation, this assumption was simulated using train–test splits, but such information may not always be available in practice. At the same time, the considered attacks represent only a subset of possible inference strategies. More sophisticated and computationally intensive approaches, such as shadow modelling attacks \cite{stadler2022groundhog}, could potentially achieve stronger results. However, unlike differentially private \cite{dwork2014DP} methods, empirical attacks can never provide formal guarantees, as stronger attacks may exist or emerge in the future.

AIAs success was assessed via the area under the receiver operating characteristic curve (AUROC) for categorical targets and MASD for continuous targets (Fischer et al. \cite{fischer2020anthropometrisch}), using a random forest model. Results were compared to a baseline model trained on real test data. Perfect privacy preservation is indicated when a model trained on synthetic data performs no better than the baseline trained on test data.

The results are presented in \crefrange{fig:DU_MASD}{fig:DU_CIO}. The ARF models consistently ranked among the synthesizers with the highest utility. Lower values of the minimum node size generally achieved higher utility, with only minor differences across minimum node sizes of 2, 5, and 10. We note that the results for ARF with a minimum node size of 2 were lower than those reported in the main text, as expected given the reduced training sample size in this evaluation. While utility fluctuated across most of the alternative synthesizers, synthpop was the most consistent competitor, showing better utility values than the ARF models for Wienbergen et al.\@~\cite{wienbergen2022infarction} (synthpop: CIO $=$ 0.776; ARF: CIO $\leq$ 0.698) and Berger et al. \cite{berger2021COVID} (synthpop: CIO $=$ 0.469; ARF: CIO $\leq$ 0.343), lower utility for Tanoey et al.\@~\cite{tanoey2022diabetes} (synthpop: CIO $=$ 0.557; ARF: CIO $\leq$ 0.679), and comparable performance on the remaining datasets.

However, ARF consistently exhibited lower MIA risks than synthpop, which showed increased risks particularly for Wienbergen et al. (synthpop: TPR@FPR=0.01 $=$ 0.100; ARF: TPR@FPR=0.01 $\leq$ 0.0388) and Schikowski et al.\@~\cite{schikowski2020blutdruckmessung} (synthpop: TPR@FPR=0.01 $=$ 0.053; ARF: TPR@FPR=0.01 $\leq$ 0.019). An exception was Berger et al.\@~\cite{berger2021COVID}, where all models exhibited comparably low utility (CIO $\leq$ 0.5) and synthpop demonstrated the overall best utility–privacy trade-off. The elevated MIA risk for ARF with a minimum node size of 2 in Berger et al.\@~\cite{berger2021COVID} can be attributed to the dataset consisting almost entirely of categorical variables, where the limited combination space leads to overly close reproduction of feature combinations. Substantially lower risk values were achieved with minimum node sizes greater than 2. For Breau et al.\@~\cite{breau2022cutpoint}, MIA risk values were elevated for all synthesizers, presumably due to the small dataset size, with TVAE achieving the most favourable utility–privacy trade-off.

While ARF demonstrated negligible attribute inference attack (AIA) risks across experiments, synthpop substantially increased prediction performance for the target variables in Breau et al.\@~\cite{breau2022cutpoint} (synthpop: AUROC $=$ 0.688; test baseline: AUROC $=$ 0.5) and further increased the already high baseline AUROC for Wienbergen et al.\@~\cite{wienbergen2022infarction} (synthpop: AUROC $=$ 0.864; test baseline: AUROC $=$ 0.822).

Overall, the ARF models demonstrated the most favourable balance between utility and privacy for Schikowski et al.\@~\cite{schikowski2020blutdruckmessung}, Fischer et al.\@~\cite{fischer2020anthropometrisch}, Wienbergen et al.\@~\cite{wienbergen2022infarction} (together with the Bayesian Network), and Tanoey et al.\@~\cite{tanoey2022diabetes} (together with PrivBayes). They showed the second-best trade-off behind TVAE for Breau et al.\@~\cite{breau2022cutpoint} and behind synthpop for Berger et al.\@~\cite{berger2021COVID}.

}

\newpage


\newpage

\newpage

\newpage


\subsection{Generalization to unseen test data}
\label{sec:appx_gen}

\begin{figure}[h!]
    \centering
    \includegraphics[width=\columnwidth]{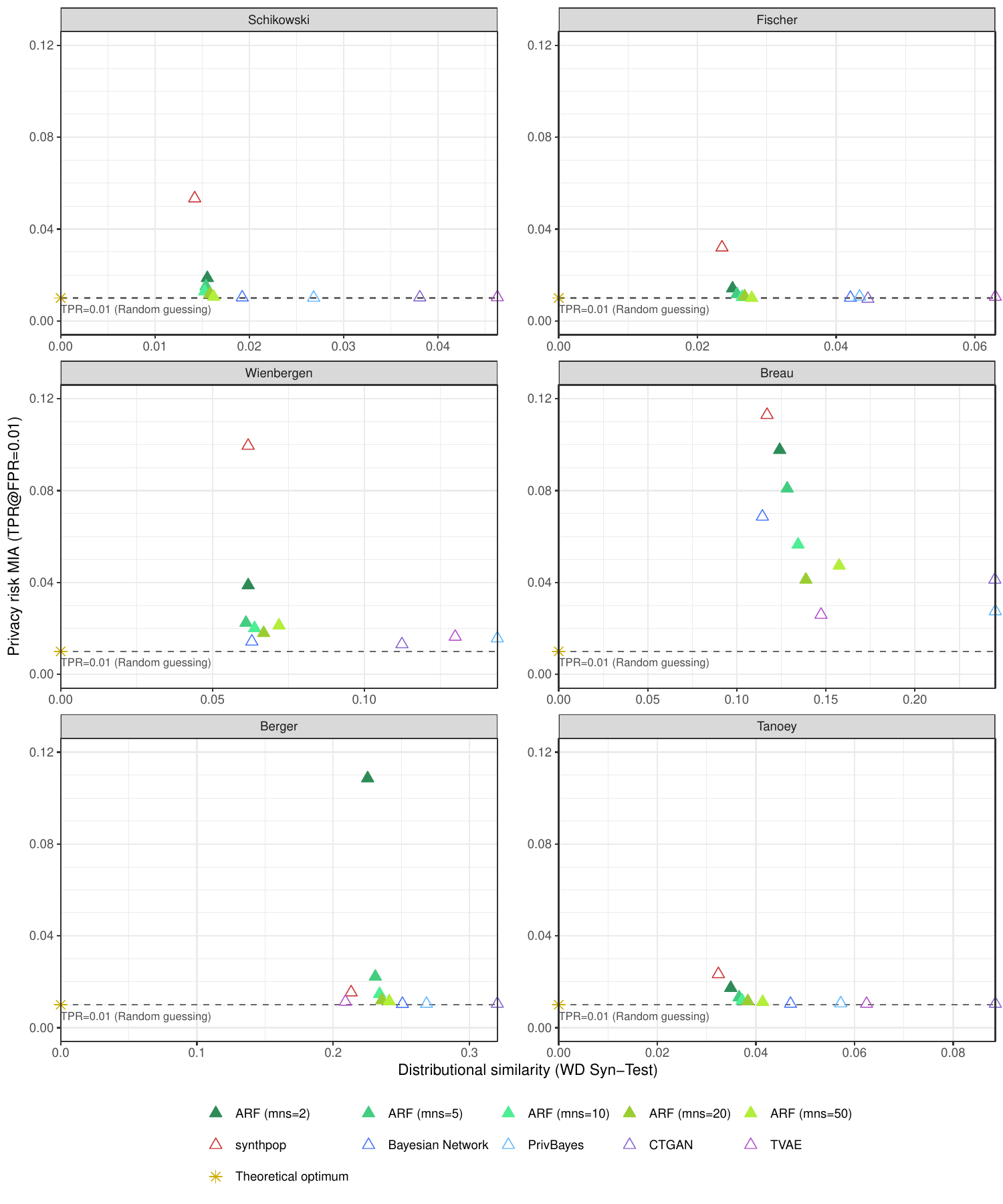}
    \caption{Distributional similarity (WD Syn-Test) vs. privacy risk MIA. Comparison of ARF performance (distributional similarity: WD Syn-Test; privacy risk: MIA performance) across five minimum node sizes and five alternative synthesizers. Mean values over five synthesizer training repetitions are reported. WD Syn-Test, multivariate Wasserstein distance between synthetic and test data not used for synthesizer training; MIA, membership inference attack; TPR@FPR=0.01, attack true positive rate at a false positive rate of $0.01$; mns, minimum node size of random forest}
    \label{fig:Appx_generalization}
\end{figure}

{\normalsize

For many applications, such as the use of synthetic data as test data or for data augmentation, it is important that the learned distributions generalize to the underlying population or unseen data rather than merely reproducing the training sample distribution. To assess this, we measured the multivariate Wasserstein distance (WD) \cite{villani2021WD} between synthetic data and test data not used for synthesizer training. Gower’s distance \cite{gower1971gower} was used as the ground metric, enabling the application of WD to mixed-type data.

The results are shown in \cref{fig:Appx_generalization} and plotted against the MIA privacy risks previously presented in \cref{sec:appx_A}. For most datasets, the findings were consistent with the downstream utility results reported in \cref{sec:appx_upto}. Some differences were observed: TVAE now showed the highest distributional similarity to the test data for Berger et al.\@~\cite{berger2021COVID}, while synthpop achieved the highest distributional similarity for Tanoey et al.\@~\cite{tanoey2022diabetes}, albeit at the cost of the highest MIA risk.

Overall, the results indicate that ARF did not merely reproduce the training distribution but achieves competitive generalization to unseen data while often maintaining a favourable privacy risk level in comparison to the alternative synthesizers.

}

\newpage

\subsection{Runtime}
\label{sec:appx_runtime}

\begin{figure}[ht]
    \centering
    \includegraphics[width=\columnwidth]{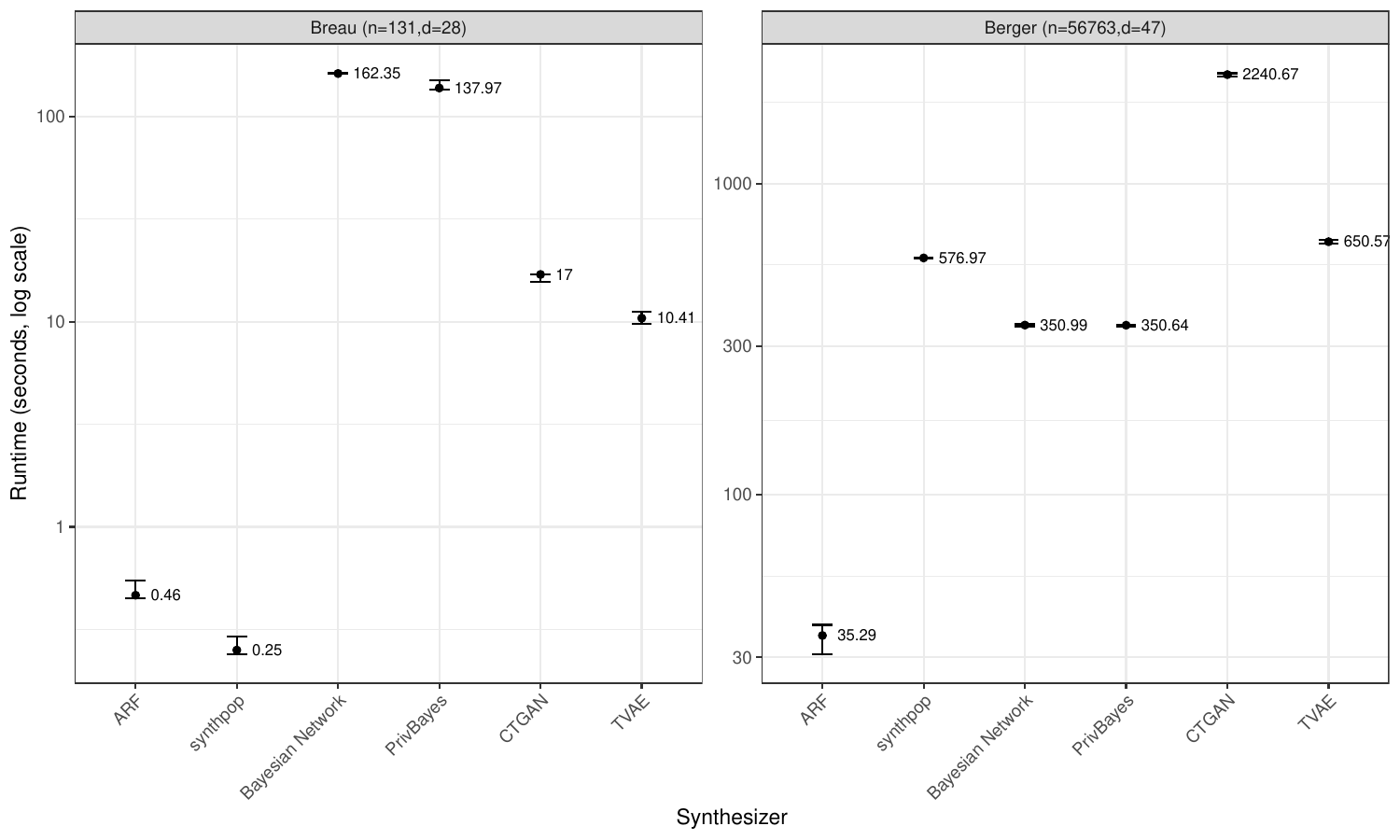}
    \caption{Synthesis runtime for Breau et al.\@~\texorpdfstring{\cite{breau2022cutpoint}}{[Breau et al., 2022]} and Berger et al.\@~\texorpdfstring{\cite{berger2021COVID}}{[Berger et al., 2022]}. Synthesis runtime comparison (synthesizer training and sampling in seconds) for ARF and five alternative synthesizers. Median, minimum, and maximum values over five synthesizer repetitions are reported. ARF, adversarial random forests with minumum node size of two (default); n, number of training samples; d, dimensionality}
    \label{fig:Appx_runtime}
\end{figure}

{\normalsize

\cref{fig:Appx_runtime} shows runtime results for the smallest dataset (Breau et al.\@~\cite{breau2022cutpoint}) and the largest dataset (Berger et al.\@~\cite{berger2021COVID}). For Breau et al.\@~\cite{breau2022cutpoint}, ARF and synthpop finished training and sampling in under 0.5 seconds, while the deep learning–based approaches required more than ten seconds and the Bayesian network–based approaches more than two minutes. For Berger et al.\@~\cite{berger2021COVID}, ARF (median runtime: 35.29 seconds) performed orders of magnitude faster than all other synthesizers.

}

\end{document}